\documentclass[aps,prd,preprint,superscriptaddress,showkeys,nofootinbib,nobibnotes]{revtex4-1}
\usepackage{amssymb}
\usepackage{amscd}
\usepackage{latexsym}
\usepackage{amsmath}
\usepackage{hyperref}
\usepackage{multirow}
\usepackage{graphicx}
\usepackage{pstricks}
\usepackage{color}
\usepackage{epstopdf}
\usepackage{setspace}

\begin{document}

%
\preprint{HUPD1708}
%
\title{A new mechanism for generating particle number asymmetry through interactions}
%

\author{Takuya Morozumi}
\email[]{morozumi@hiroshima-u.ac.jp}
\affiliation{Graduate School of Science, Hiroshima University, Higashi-Hiroshima, 739-8526, Japan\\ }%
\affiliation{Core of Research for Energetic Universe, Hiroshima University, Higashi-Hiroshima, 739-8526, Japan}%
\author{Keiko I. Nagao}
\email[]{nagao@dap.ous.ac.jp}
\affiliation{National Institute of Technology, Niihama College, Ehime 792-8580, Japan \\ }
\affiliation{Okayama University of Science, Ridaicho, Kita-ku, Okayama-shi 700-0005 Japan }
\author{Apriadi Salim Adam}
\email[]{apriadiadam@hiroshima-u.ac.jp}
\affiliation{Graduate School of Science, Hiroshima University, Higashi-Hiroshima, 739-8526, Japan }%
%
\author{Hiroyuki Takata}
\email[]{takata@tspu.edu.ru}
\affiliation{Tomsk State Pedagogical University, Tomsk, 634061, Russia}
\date{\today}
\begin{abstract}
A new mechanism for generating particle number asymmetry (PNA) has been developed. This mechanism is realized with a Lagrangian including a complex scalar field and a neutral scalar field. The complex scalar carries U(1) charge which is associated with the PNA. It is written in terms of the condensation and Green's function, which is obtained with two-particle irreducible (2PI) closed time path (CTP) effective action (EA). In the spatially flat universe with a time-dependent scale factor, the time evolution of the PNA is computed. We start with an initial condition where only the condensation of the neutral scalar is non-zero. The initial condition for the fields is specified by a density operator parameterized by the temperature of the universe. With the above initial conditions, the PNA vanishes at the initial time and later it is generated through the interaction between the complex scalar and the condensation of the neutral scalar. We investigate the case that both the interaction and the expansion rate of the universe are small and include their effects up to the first order of the perturbation. The expanding universe causes the effects of the dilution of the PNA, freezing interaction and the redshift of the particle energy. As for the time dependence of the PNA, we found that PNA oscillates at the early time and it begins to dump at the later time. The period and the amplitude of the oscillation depend on the mass spectrum of the model, the temperature and the expansion rate of the universe.  
\end{abstract}
\maketitle

\tableofcontents

\newpage
\section{Introduction}
\label{sec1}
The origin of BAU has long been a question of great interest in explaining why there is more baryon than anti-baryon in nature. Big bang nucleosynthesis (BBN) \cite{Kneller2004} and cosmic microwave background \cite{Ade:2015xua} measurements give the BAU as $\eta \equiv n_{B}/s \backsimeq 10^{-10}$, where $n_{B }$ is the baryon number density and $s$ is the entropy density. In order to address this issue, many different models and mechanisms have been proposed \cite{Sakharov:1967dj,Yoshi_PhysRevLett.41.281,FUKUGITA198645,AFFLECK1985361,COHEN1988913}.  The mechanisms discussed in the literature satisfy the three Sakharov conditions \cite{Sakharov:1967dj}, namely, (i) baryon number ($ B $) violation, (ii) charge ($C$) and charge-parity ($CP$) violations, and (iii) a departure from the thermal equilibrium. For reviews of different types of models and mechanisms, see, for example, \cite{DAVIDSON2008105,RevModPhys.71.1463,ZUREK201491}. Recently, the variety of the method for the calculation of BAU has been also developed \cite{PhysRevD.91.123529,PhysRevD.94.035007,PhysRevD.95.095014}. 

In the present paper, we further extend the model of scalar fields \cite{Hotta2014} so that it generates the PNA through interactions. In many of previous works, the mechanism generating BAU relies on the heavy particle decays. Another mechanism uses  U(1) phase of the complex scalar field \cite{AFFLECK1985361}. In this work, we develop a new mechanism to generate PNA. The new feature of our approach is briefly explained below. 

The model which we have proposed \cite{Morozumi2017} consists of a complex scalar field and a neutral scalar field. The PNA is related to the U(1) current of the complex field. In our model, the neutral scalar field has a time-dependent expectation value which is called condensation. In the new mechanism, the oscillating condensation of the neutral scalar interacts with the complex scalar field. Since the complex scalar field carries U(1) charge, the interactions with the condensation of the neutral scalar generate PNA. The interactions break U(1) symmetry as well as charge conjugation symmetry. At the initial time, the condensation of the neutral scalar is non-zero. We propose a way which realizes such initial condition.  

As for the computation of the PNA, we use 2PI formalism combined with density operator formulation of quantum field theory \cite{Calzetta1988}. 
The initial conditions of the quantum fields are specified with the density operator. The density operator is parameterized by the temperature of the universe at the initial time. We also include the effect of the expansion of the universe. It is treated perturbatively and the leading order term which is proportional to the Hubble parameter at the initial time is considered. With this method, the time dependence of the PNA is computed and the numerical analysis is carried out. Especially, the dependence on the various parameters of the model such as masses and strength of interactions is investigated. We also study the dependence on the temperature and the Hubble parameter at the initial time. We first carry out the numerical simulation without specifying the unit of parameter sets. Later, in a radiation dominated era, we specify the unit of the parameters and estimate the numerical value of the PNA over entropy density.

This paper is organized as follows. In section \ref{sec2}, we introduce our model with $ CP $ and particle number violating interactions. We also specify the density operator as the initial state. In section \ref{sec3}, we derive the equation of motion for Green's function and field by using 2PI CTP EA formalism. We also provide the initial condition for Green's function and field. In section \ref{sec4}, using the solution of Green's function and field, we compute the expectation value of the PNA. Section \ref{sec5} provides the numerical study of the time dependence of the PNA. We will also discuss the dependence on the parameters of the model. Section \ref{sec6} is devoted to conclusion and discussion. In Appendix \ref{sec_apply}, we introduce a differential equation which is a prototype for Green's function and field equations. Applying the solutions of the prototype, we obtain the solutions for both Green's function and field equations. In Appendices \ref{f_and_g}-\ref{sec_calJ}, the useful formulas to obtain the PNA for non-vanishing Hubble parameter case are derived. 

\section{A model with CP and particle number violating interaction}
\label{sec2}

In this section, we present a model which consists of scalar fields \cite{Morozumi2017}. It has both $ CP $ and particle number violating  features. As an initial statistical state for scalar fields, we employ the density operator for thermal equilibrium.

Let us start by introducing a model consists of a neutral scalar, $N$, and a complex scalar, $\phi$. 
The action is given by,
\begin{equation}
\label{Lageq}
	\begin{aligned}
S =& \int d^{4}x \sqrt{-g} \left(\mathcal{L}_{\text{free}}+\mathcal{L}_{\text{int}}\right),  \\
\mathcal{L}_{\text{free}} =& g^{\mu\nu} \nabla_{\mu} \phi^{\dagger}
\nabla_{\nu} \phi - m_{\phi}^{2} | \phi |^{2} + \frac{1}{2} \nabla_{\mu} N \nabla^{\mu} N  - \frac{M_N^2}{2} N^{2}+ \frac{B^2}{2}  (\phi^2 + \phi^{\dagger 2})    \\
&+ \left(\frac{\alpha_2}{2} \phi^2 + h.c.\right)R+ \alpha_{3} |\phi|^{2}R , \\
\mathcal{L}_{\text{int}} =& 
A\phi^2 N + A^{\ast} \phi^{\dagger 2} N + A_{0}|\phi|^{2}N ,
  \end{aligned}
\end{equation}
where $g_{\mu\nu}$ is the metric and $R$ is the Riemann curvature. With this Lagrangian, we aim to produce the PNA through the soft-breaking terms of U(1) symmetry whose coefficients are denoted by $A$ and $B^2$. One may add the quartic terms to the Lagrangian which are invariant under the U(1) symmetry. Though those terms  preserve the stability of the potential for large field configuration and are also important for the renormalizability, we assume they do not lead to the leading contribution for the generation of the PNA. We also set the coefficients  of the odd power terms for $N^n (n=1,3)$ zero in order to obtain a simple oscillating behavior for the time dependence of the condensation of $N$.
We assume that our universe is homogeneous for space and employ the Friedmann-Lema\^itre-Robertson-Walker metric, 
\begin{align}
g_{\mu \nu} &= (1, -a^2(x^0), -a^2(x^0), -a^2(x^0)), \label{metric_g}
\end{align}
where $a(x^0)$ is the scale factor at time $x^0$. Correspondingly the Riemann curvature is given by,
\begin{align}
R(x^0) &= 6 \left[\frac{\ddot{a}(x^0)}{a(x^0)} + \left(\frac{\dot{a}(x^0)}{a(x^0)} \right)^2\right] \label{Rie}.
\end{align}
In Eq.\eqref{Lageq}, the terms proportional to $A$, $B$ and $\alpha_{2}$ are the particle number violating interactions. In general, only one of the phases of those parameters can be rotated away. Throughout this paper, we study the special case that $B$ and $\alpha_{2}$ are real numbers and $A$ is a complex number. Since only $A$ is a complex number, it is a unique source of the $ CP $ violation.

We rewrite all the fields in terms of real scalar fields, $\phi_i$ ($ i = 1, 2, 3$), defined as,
\begin{eqnarray}
\phi=\frac{\phi_1+i \phi_2}{\sqrt{2}}, \quad N=\phi_3. 
\end{eqnarray}
With these definitions, the free part of the Lagrangian is rewritten as,
\begin{eqnarray}
\mathcal{L}_{\text{free}} 
=  \frac{1}{2}  \sqrt{- g}  [ g^{\mu \nu} \nabla_{\mu}
\phi_i \nabla_{\nu}^{} \phi_i - \tilde{m}_{i}^2 (x^0) \phi_i^2 ], 
\end{eqnarray}
where the  kinetic term is given by,
\begin{align}
g^{\mu \nu} \nabla_{\mu} \phi_{i} \nabla_{\nu}^{} \phi_{i} = \frac{\partial \phi_i}{\partial x^0}\frac{\partial \phi_i}{\partial x^0} - \frac{1}{a(x^0)^2} \frac{\partial \phi_{i}}{\partial x^j}  \frac{\partial \phi_{i}}{\partial x^j},
\end{align}
and their effective masses are given as follows,
\begin{eqnarray}
\tilde{m}_1^2 (x^0) & = & m_{\phi}^2 - B^2 - (\alpha_2 + \alpha_3) R (x^0),\label{eq:3} \\
\tilde{m}_2^2 (x^0) & = & m_{\phi}^2 + B^2 + (\alpha_2 - \alpha_3 ^{}) R(x^0),\label{eq:4}  \\
\tilde{m}_3^2 (x^0) & = & m_N^2.  \label{eq:5}
\end{eqnarray}
Non-zero $B^2$ or $\alpha_2$ leads to the non-degenerate mass spectrum for $\phi_1$ and $\phi_2$. The interaction Lagrangian is rewritten with a totally symmetric coefficient  $A_{ijk}$,
\begin{eqnarray}
{\cal L}_{\text{int}}&=&
\sum_{ijk=1}^3 \frac{1}{3} A_{ij k} \phi_i \phi_j \phi_k 
\end{eqnarray}
with $i,j,k=1,2,3$. The non-zero components of $A_{ijk}$ are written with the couplings for cubic interaction, $A$ and $A_{0}$, as shown in Table \ref{tb:1}. We also summarize the qubic interactions and their properties according to U(1) symmetry and $ CP $ symmetry. 
\begin{table}[t]
	\begin{center}
		\caption{The cubic interactions and their properties}
		\begin{tabular}{|c|c|} \hline \hline 
			Cubic interaction coupling & Property \\ \hline \hline
			$A_{113}=\frac{A_0}{2}+{\rm Re}.(A)$ & -- \\ \hline 
			$A_{223}=\frac{A_0}{2}-{\rm Re}.(A)$ & -- \\ \hline
			$A_{113}-A_{223}=2 {\rm Re.}(A)$  &   U(1) violation \\ \hline
			$A_{123}=-{\rm Im}.(A)$ &\small{U(1), $ CP $ violation} \\ \hline \hline 
		\end{tabular}
		\label{tb:1}
	\end{center}
\end{table} 

N\"{o}ether current related to the U(1) transformation is,
\begin{eqnarray}
j_\mu(x)&=& \frac{i}{2} \left(  \phi^{\dagger} \overset{\leftrightarrow}{\partial_{\mu}}
\phi - \phi \overset{\leftrightarrow}{\partial_{\mu}} \phi^{\dagger} \right)  . \label{neother}
\end{eqnarray}
In terms of real scalar fields, the N\"{o}ether current alters into,
\begin{eqnarray}
j_{\mu}  & = & \frac{1}{2} \left(\phi_2 \overset{\leftrightarrow}{\partial_{\mu}} \phi_1 - 
\phi_1 \overset{\leftrightarrow}{\partial_{\mu}} \phi_2\right). \label{defjnew}
\end{eqnarray}
The ordering of the operators in Eq.\eqref{neother} is arranged so that it is Hermite and the particle number operator, 
\begin{eqnarray}
Q(x^0) = \int d^3 \mathbf{x} \sqrt{-g}\ j_0(x),
\end{eqnarray}
has a normal ordered expression. Then, in the vanishing limit of interaction terms and particle number violating terms, the vacuum expectation value of the particle number vanishes. With the above definition, $j_0(x)$ is the PNA per unit comoving volume. The expectation value of the PNA is written with a density operator,
\begin{eqnarray}
\langle j_0(x) \rangle={\rm Tr}(j_0(x)\rho(t_{0})). \label{jvalue}
\end{eqnarray}
Note that, the PNA is a Heisenberg operator and $\rho(t_0)$ is a density operator  which specifies the  state at the initial time $x^0=t_0$. In this work, we use the density operator 
with zero chemical potential. It is specifically given by,
\begin{eqnarray}
\rho(t_{0})=\frac{e^{-\beta H}}{{\rm Tr}(e^{-\beta H})}, \label{rhospe}
\end{eqnarray}
where $\beta$ denotes inverse temperature, $1/T$, and $H$ is a Hamiltonian which includes linear term of fields,
\begin{eqnarray}
H & = & \frac{1}{2}\sum_{i = 1}^3 \int d^3 {\bf x}\ a(t_0)^3  \left[  \pi_{\phi_i}
\pi_{\phi_i} + \frac{\nabla \phi_i \cdot \nabla \phi_i}{a (t_0)^2} + \tilde{m}_{i}^2 
\left( \phi_i- v_{i}\right)^2  \right] , \label{Hnew}
\end{eqnarray}
where $v_i$ is a constant. The linear term of fields in Eq.\eqref{Hnew} is prepared for the non-zero expectation value of fields. Note that the density operator in Eq.\eqref{rhospe} is not exactly the
same as the thermal equilibrium one since in the Hamiltonian, the interaction part are not included. 
Since we assume three dimensional space is translational invariant, then the expectation value of the PNA depends on time $x^0$ and the initial time $t_0$.
As we will show later, the non-zero expectation value for the field $\phi_3$ leads to the time dependent 
condensation which is the origin of the non-equilibrium time evolution of the system.

Below we consider the matrix element of the density operator given in Eq.\eqref{rhospe}. We start with the following density operator for one component real scalar field as an example, 
\begin{eqnarray}
 \rho(t_0)  &=& \frac{e^{-\beta H_{\text{example}}}}{\text{Tr}(e^{-\beta H_{\text{example}}})}, \label{rhospe2} \\
H_{\text{example}} & = & \frac{1}{2} \int d^3 {\bf x}\ a(t_0)^3  \left[  \pi_{\phi}
\pi_{\phi} + \frac{\nabla \phi \cdot \nabla \phi}{a (t_0)^2} + \tilde{m}^2 
\left( \phi - v\right)^2  \right]. \label{Hnew2}
\end{eqnarray}
The above Hamiltonian is obtained  from that of Eq.\eqref{Hnew}  by keeping only one of the real scalar fields. The matrix element of the initial density operator in Eq.\eqref{rhospe2} is 
written in terms of the path integral form of the imaginary time formalism given as,
\begin{eqnarray}
\langle \phi^1 | \rho(t_0) | \phi^2 \rangle  &=& \frac{\int_{\phi(0)=\phi^2, \phi(\beta)=\phi^1} 
d \phi
e^{- S^{\text{example}}_E[\phi]}}{\int d\phi^1  \int_{\phi(0)=\phi^1, \phi(\beta)=\phi^1} d\phi e^{- S^{\text{example}}_E[\phi]} },
\end{eqnarray}
where $S^{\text{example}}_E$ is an Euclidean action which corresponds to the Hamiltonian in Eq.\eqref{Hnew2} and it is given by,
\begin{eqnarray}
S^{\text{example}}_E [\phi (\mathbf{x}, u)]= \frac{1}{2}  \int_0^{\beta} du \int d^3 \mathbf{x} \left\{ \left( \frac{\partial
\phi}{\partial u} \right)^2 +\frac{\nabla \phi \cdot \nabla \phi}{a(t_{0})^2} + \tilde{m}^2 (\phi - v)^2 \right\}.
\end{eqnarray}
After carring out the path integral, the density matrix is written with 
$S^{\text{example}}_E [\phi_{\text{cl}}(\mathbf{x},u)]$
which is the action for the
classical orbit $\phi_{\text{cl}}$ satisfying the boundary conditions, $\phi_{\text{cl}}(u=0)=\phi^2, \phi_{\text{cl}}(u=\beta)=\phi^1$.  
It is given as the functional of the boundary fields $\phi^i (i=1,2)$ and vacuum 
expectation value $v$ as,
 \begin{eqnarray}
 \langle \phi^1 | \rho(t_0) | \phi^2 \rangle  &=&  \frac{\exp \left[-S^{\text{example}}_{\text{Ecl}} [\phi^1, \phi^2] \right]}{\int d \phi^1 
 \exp [- S^{\text{example}}_{\text{Ecl}} [\phi^1, \phi^1]]}, \label{FR_2}
 \end{eqnarray}
where $S^{\text{example}}_{\text{Ecl}} [\phi^1, \phi^2]\simeq S^{\text{example}}_E [\phi_{\text{cl}}(\mathbf{x},u)]$ is given by,
\begin{align}
S^{\text{example}}_{\text{Ecl}} [\phi^1, \phi^2] =& -\frac{a(t_{0})^6}{2}   \int
\frac{d^3 \mathbf{k}}{(2 \pi)^3} \left\{ \sum_{i=1}^{2} \phi^i ({\bf k}) \kappa^{i i} (-
{\bf k})  \phi^i (- {\bf k}) \right. \nonumber \\
& \left.  -\phi^1 ({\bf k})\kappa^{1 2}(- {\bf k}) \phi^2 (- {\bf k}) -\phi^2 ({\bf k}) \kappa^{2 1}(- {\bf k}) \phi^1 (- {\bf k}) \right\} \nonumber \\
& + a(t_{0})^6 \left\{ \sum_{i=1}^2 \phi^i (0)   \kappa^{ii } (0) -\phi^1 (0) \kappa^{12 } (0) -\phi^2 (0) \kappa^{21 } (0) \right\} v \label{SEcl_new_2}. 
\end{align}
In the above expression, we drop the terms which are proportional to $v^2$ because they do not contribute to the normalized density matrix.
$\kappa^{b d} ({\bf k}) $ is defined as \cite{Hotta2014},
\begin{equation}
\label{kappij}
\begin{aligned}
\kappa^{11} ({\bf  k}) =  \kappa^{22} ({\bf  k})
:=& - \frac{1}{a(t_{0})^3}  \frac{\omega ({\bf k}) \cosh \beta \omega ({\bf k})}{\sinh
	\beta \omega ({\bf k})}, \\
\kappa^{12} ({\bf  k}) =  \kappa^{21} ({\bf  k})
:=& - \frac{1}{a(t_{0})^3}  \frac{\omega ({\bf k})}{\sinh \beta \omega ({\bf k})}, \\
\omega ({\bf k}) :=& \sqrt{\frac{{\bf k}^2}{a(t_{0})^2} + \tilde{m}^2}.
\end{aligned}
\end{equation}
Using the above definitions, one can write the density matrix in Eq.\eqref{FR_2} as the following form,
\begin{eqnarray}
\langle \phi^1 | \rho(t_0) | \phi^2 \rangle &=&N
\exp \left[  \int \sqrt{- g (x)}
d^4 xJ^a(x) c^{a b} \phi^b(x) \right. \nonumber \\
&& \left.  + \frac{1}{2}  \int d^4 xd^4 y \sqrt{- g (x)} \phi^a (x) c^{a b} K^{bd} (x, y) c^{de} \phi^e (y) \sqrt{- g (y)} \right],
\label{eq:dme}
\end{eqnarray}
with $\phi^1(t_0)=\phi^1$ and $ \phi^2(t_0)=\phi^2$. The upper indices $a$ and $b$ are $1$ or $2$.
$c^{ab}$ is the metric of CTP formalism \cite{Calzetta1988} and $c^{11} = - c^{22} = 1$ and $c^{12} = c^{21} = 0$.
In the above expression, the source terms $J$ and $K$ do not vanish only at the initial time $t_0$ and they are given by,
\begin{align}
J^b (x) :=&  - i \delta (x^0 - t_0) j^b ,\quad  j^b  =  - a_{t_0}^3  \kappa^{bd}_{} (\mathbf{k}=0) c^{d e} v^e 
,   \label{NewJ} \\
K^{b d} (x, y) :=&  - i \delta (x^0 - t_0) \delta (y^0 - t_0) \kappa^{b d}({\bf x - y}), \quad \kappa^{ab} ({\bf x}) = 
\int \frac{d^3 k}{(2 \pi)^3} \kappa^{ab} ({\bf k}) e^{- i {\bf k \cdot x}}, \label{eq:47} 
\end{align}
where $\kappa^{ab}({\bf k})$ is given in Eq.\eqref{kappij}. 
In Eq.(\ref{eq:dme}), $N$ is a normalization constant  which is given as,
\begin{eqnarray}
\frac{1}{N} &=& \int d\phi^1 \int d\phi^2 \delta(\phi^1-\phi^2) \exp \left[  \int \sqrt{- g (x)}
d^4 xJ^a(x) c^{a b} \phi^b(x)  \right. \nonumber \\
&& \left.   +\frac{1}{2}  \int d^4 xd^4 y \sqrt{- g (x)} \phi^a (x) c^{a b} K^{bd} (x, y) c^{de} \phi^e (y) \sqrt{- g (y)} \right].
\end{eqnarray}
\section{Two Particle Irreducible Closed Time Path Effective Action}
\label{sec3}


In this section, we derive the equations of motion, i.e., the Schwinger-Dyson equations (SDEs) for both Green's function and field. SDEs are obtained by taking the variation of 2PI EA with respect to fields and Green's functions, respectively. In addition, we also provide the initial condition for Green's function and field to solve SDEs.


\subsection{2PI Formalism in Curved Space-Time}
2PI CTP EA in curved space-time has been investigated in \cite{PhysRevD.56.661} and their formulations can be applied to the present model. In 2PI formalism, one introduces non-local source term denoted as $K$ and local source term denoted as $J$,
\begin{eqnarray}
e^{iW [J, K]} &=&  \int d \phi \exp \left( i \left[ S[\phi,g] + \int \sqrt{- g (x)}
d^4 xJ_i^a(x) c^{a b} \phi_i^b(x) + \frac{1}{2}  \int d^4 xd^4 y \sqrt{- g (x)}  \right. \right. \nonumber \\
& & \left. \left. \times \phi_i^a (x) c^{a b} K^{bd}_{ij} (x, y) c^{de} \phi_j^e (y) \sqrt{- g (y)} \right]
\right), \label{funtional_W}
\end{eqnarray}
where $i,j=1$ or $2$ and $S[\phi,g] $ is given by,
\begin{eqnarray}
 S[\phi,g] = \int d^{4}x \sqrt{-g(x)}  \left[\frac{1}{2} c^{ab} (g^{\mu \nu} \nabla_{\mu}
\phi_i^{a} \nabla_{\nu}^{} \phi_i^b - \tilde{m}_{i i}^2 \phi_{i}^{a} \phi^b_i )   + \frac{1}{3} D_{abc} A_{ijk} \phi_{i}^{a} \phi_j^b \phi_{k}^{c}  \right] ,
\end{eqnarray}
where  $D_{111}=-D_{222}=1$ and the other components are zero. 
The upper indices of the field and the source terms distinguish two different time paths in closed time path formalism \cite{Calzetta1988}. 
One can define the mean fields $\bar{\phi}^a_i$ and  Green's function by taking the
functional derivative with respect to the source terms $J$ and $K$, respectively,
\begin{eqnarray}
\bar{\phi}^a_{i}(x)&=&\frac{c^{ab}}{\sqrt{-g(x)}} \frac{\delta W[J,K]}{\delta J_{i}^{b}(x) },  \\
\bar{\phi}_i^a (x) \bar{\phi}_j^e (y)  + G^{ a e}_{i j} (x,y) &=& 
 2 \frac{c^{a b}}{ \sqrt{- g (x)}} \frac{\delta W[J,K]}{\delta K^{bd}_{i j} (x,y) } \frac{c^{de}}{\sqrt{- g (y)}}.
\end{eqnarray}

If  one sets 
the source terms to be the ones given in Eqs.\eqref{NewJ} and \eqref{eq:47}, one can show 
that the 
expectation value of the product of the field operators with the initial density operator is related 
to the Green function and mean fields.  
Definitely, we can prove the following relations,
\begin{align}
\bar{\phi}^a_i (x) =&  \frac{\int d \phi^1 \int d
  \phi^2 \langle \phi^2 | \phi_i (x^{}) | \phi^1 \rangle \exp (- S_{E \text{cl}} [\phi^1,
  \phi^2])}{\int d \phi^1 \int d \phi^2 \delta (\phi^1 - \phi^2) \exp (- S_{E
  \text{cl}} [\phi^1, \phi^2])} \label{phiH} \\ 
=& \mathrm{Tr} [\phi_i (x) \rho (t_0)], \\
G^{12}_{ij} (x, y) =& \frac{\int \int d \phi^1 d \phi^2 \langle \phi^2 | \Phi^{}_j (y)
  \Phi^{}_i (x) | \phi^1  \rangle e^{- S_{E \text{cl}} [\phi^1, \phi^2]} }{\int
  \int d \phi^1 d \phi^2 \delta (\phi^2 - \phi^1) e^{- S_{E \text{cl}}
  [\phi^1, \phi^2]}} \label{2phiH}\\
 =& \mathrm{Tr} [\phi_j (y) \phi_i (x) \rho (t_0)] -\bar{\phi}^2_j (y) \bar{\phi}^1_i (x) ,
\label{def_trac_jrho}
\end{align}
with $\bar{\phi}^a (x)=\bar{\phi} (x)$ 
and $\Phi_{}(x)$ is a Heisenberg operator which has form as $\Phi (x^0, \mathbf{x}) \equiv  \phi (x^0, \mathbf{x}) - \bar{\phi} (x^0) $. 
With Eq.\eqref{def_trac_jrho}, one can write the expectation value of the current as the sum of the 
contribution from Green function and the current of the mean fields. 
Then Eq.\eqref{jvalue} alters into,
\begin{align}
\langle j_{0}(x)\rangle = \text{Re}\left[\left( \frac{\partial}{\partial x^{0}} -
\frac{\partial}{\partial y^{0}} \right)G_{12}^{12} (x, y) \big|_{y
	\rightarrow x} + \bar{\phi}^{2}_2 (x)\overset{\leftrightarrow}{\partial_{0}}  \bar{\phi}^{1}_1 (x)  \right], \label{orGenj}
\end{align}
where we have used Eq.\eqref{defjnew} and the following relations, 
\begin{align}
G_{i j}^{a b \ast} (x,y)  & =  \tau^{1 a c} G^{c d}_{i j} (x,y)  \tau^{1 d b},\\
\bar{\phi}^{a \ast}_{i} (x)  & =  \tau^{1 a b} \bar{\phi}^b_{i}(x),
\end{align}
where $\tau$ is the Pauli matrix.

The Green functions and expectation value of fields  are derived as solutions of the
SDEs which are obtained with 2PI EA.
The 2PI EA is related  to the generating functional $W[J, K]$  by Legendre
transfomation as \cite{Calzetta2008a,Cornwall:1974vz}, 
\begin{align}
\Gamma_2 [G, \bar{\phi},g] =&  W[J,K] - \int 
d^4 x  \sqrt{- g (x)} J_i^a(x) c^{a b} \bar{\phi}_i^b(x)   \nonumber \\
& 
- \frac{1}{2}  \int d^4 x \int d^4 y \sqrt{- g (x)}  c^{a b} K^{bd}_{ij} (x, y) c^{de} \{ \bar{\phi}_i^a (x) \bar{\phi}_j^e (y)  + G^{ a e}_{i j} (x,y)\} \sqrt{- g (y)} .
\label{Gamma_2b}
\end{align}

Let us write the 2PI EA $\Gamma_{2}$ in our model, in which 
%
we only keep the interaction term up to the first order of cubic interaction, $A_{ijk}$. It is given as,
\begin{eqnarray}
\Gamma_2 [G, \bar{\phi}, g] &=& S[\bar{\phi}, g] + \frac{1}{2}  \int d^4 x \int d^4 y
\frac{\delta^2 S[\bar{\phi}, g]}{\delta \bar{\phi}_i^a (x) \delta \bar{\phi}_j^b (y)}  G^{ab}_{ij} (x, y)  
 + \frac{i}{2}
\text{TrLn}\ G^{- 1}  ,  \label{Gamma_2}
\end{eqnarray}
where $S [\bar{\phi},g]$ is the action written in terms of mean fields as,
\begin{align}
S [\bar{\phi},g] =& \int d^{4}x \sqrt{-g(x)}  \left[-\bar{\phi}_{i}^{a} \frac{1}{2} c^{ab} ( \Box + \tilde{m}_{i i}^2) \bar{\phi}^b_i  + \frac{1}{3} D_{abc} A_{ijk} \bar{\phi}_{i}^{a} \bar{\phi}_j^b \bar{\phi}_{k}^{c}  
\right] \nonumber \\
&  + \frac{1}{2} \int d^4 x \sqrt{- g (x)} [\delta (x^0 - T) - \delta
(x^0-t_{0})] \bar{\phi}^a_{i} c^{ab}  \dot{\bar{\phi}}^b_{i}.  \label{eq:17} 
\end{align}
In Eq.\eqref{Gamma_2}, the interactions are included in the first term as well as in the second term. 
In the action above, we have also taken into account the surface term at the boundary which corresponds to the last term of Eq.\eqref{eq:17}. 
$T$ and $t_0$ in Eq.\eqref{eq:17} are the upper bound and the lower bound of the time integration, respectively. 

\subsection{Schwinger Dyson Equations}
Now let us derive SDEs for both Green's function and field. These equations can be obtained by taking the variation of the 2PI EA, $\Gamma_2$, with respect to the scalar field $\bar{\phi}$ and Green's function $G$. 


In the following, we first derive SDEs for the field. The variation of the 2PI EA in Eq.\eqref{Gamma_2b} with respect to the scalar field $\bar{\phi}$ leads to,
\begin{eqnarray}
\frac{1}{\sqrt{-g(x)}} \frac{\delta\Gamma_2}{\delta \bar{\phi}^{a}_{i}(x)} = - c^{ab}  J_{i}^{b} (x)  - \int d^{4}z\  c^{ab} K^{bc}_{ij} (x,z) c^{cd}  \sqrt{-g(z)} \bar{\phi}^{d}_{j}(z).  \label{eq:19}
\end{eqnarray}
Using Eqs.\eqref{NewJ} and \eqref{eq:47}, one computes the right hand side of the above equation as,
\begin{align}
 & c^{ab} J_i^b (x) + \int d^4 z c^{ab} K_{ij}^{bc}
  (x, z) c^{cd} \sqrt{- g (z)}  \bar{\phi}^d_j (z) \nonumber \\
   = & - i \delta
  (x^0 - t_0)
     \left( v_i  \tilde{m}_i \tanh \frac{\beta \tilde{m}_i}{2} +
  c^{ab} \kappa^{bc}_{ii} (\mathbf{k}= 0) c^{cd} a
  (t_{ 0})^3 v_i^d \right) \nonumber\\
   = & 0 ,
\end{align}
where we have used $\kappa^{ab}({\bf k})$ given in Eq.\eqref{kappij}. 
The left hand side of Eq.\eqref{eq:19} is computed using Eq.\eqref{Gamma_2} 
and one obtains the following equation of motion of the scalar field $\bar{\phi}$,
\begin{eqnarray}
(\delta_{i j} \Box + \tilde{m}_{i j}^2)
\bar{\phi}^d_j (x) &=& 
c^{da} D_{a b c} A_{i j k} \left\{ \bar{\phi}_j^b (x)
\bar{\phi}_k^c (x) + G^{b c}_{jk} (x, x)\right\} 
, \label{eomPHI}
\end{eqnarray}
where the Laplacian of Friedman-Lema\^itre-Robertson-Walker metric is given by,
\begin{eqnarray}
\Box = \nabla_{\mu} \nabla^{\mu}  =  \frac{\partial^2}{\partial x^{0 2}} -
\frac{1}{a (x^0)^2} \nabla \cdot \nabla + 3 \frac{\dot{a}}{a} 
\frac{\partial}{\partial x^0}.  \label{lapc}
\end{eqnarray}

Next, the equation of motion for Green's function is derived in the following way. The variation of the 2PI EA in Eq.\eqref{Gamma_2b} with respect to Green's function $G$ leads to,
\begin{eqnarray}
\frac{\delta \Gamma_2}{\delta G^{ab}_{ij} (x, y)} = - \frac{1}{2} c^{ac}
\sqrt{- g (x)} K^{cd}_{ij} (x, y) \sqrt{-g(y)} c^{db}.  \label{G2GF01}
\end{eqnarray}
The left hand side of the above equation is obtained by taking variation of Eq.\eqref{Gamma_2} with respect to Green's function as, 
\begin{eqnarray}
\frac{\delta \Gamma_2}{\delta G^{ab}_{ij} (x, y)} & = & - \frac{i}{2} (G^{- 1})^{b a}_{j i} (y, x) + \frac{1}{2} 
\frac{\delta^2 S [\bar{\phi},g] }{\delta \bar{\phi}_i^a (x) \delta \bar{\phi}_j^b (y)} 
, \label{G2GF02} 
\end{eqnarray}
where the second term of above expression is computed using action in Eq.\eqref{eq:17}. 
Taking all together Eqs.\eqref{G2GF01} and \eqref{G2GF02}, one obtains the following two differential equations for Green's function,
\begin{eqnarray}
(\overset{\rightarrow}{\Box}_x + \tilde{m}^2_i)_{} G^{a b}_{i j}(x,y)  & = & - i
\delta_{i j} \frac{c^{a b}}{\sqrt{- g(x)}} \delta (x - y) + 2 c^{a d} D_{d c e} A_{i k l} \bar{\phi}^e_{l, x} G^{cb}_{kj, xy} \nonumber\\
& & + \int d^4 z K^{a
	e}_{i k} (x, z) \sqrt{- g(z)} c^{e f} G^{f b}_{k j} (z, y) 
	, \label{eomxGF} 
\end{eqnarray}
\begin{eqnarray}
(\overset{\rightarrow}{\Box}_y + \tilde{m}^2_j) G^{a b}_{i j} (x, y)  & = & -
i \delta_{_{i j} } \delta (x - y) \frac{c^{a b}}{\sqrt{- g(y)}} + 2 G^{ac}_{ik, xy} D_{c e f} A_{k j l} \bar{\phi}^f_{l, y} c^{e b} \nonumber \\
& & + \int d^4 z
G^{a e}_{i k} (x, z) c^{e f} \sqrt{- g (z)} K^{f b}_{k j} (z, y) 
, \label{eomyGF}
\end{eqnarray}
where $\Box_x = \nabla_x^{\mu} \nabla_{\mu}^x$ and $\Box_y = \nabla_y^{\mu} \nabla_{\mu}^y$.

Next, we rescale Green's function, field and coupling constant of interaction as follows,
\begin{eqnarray}
\bar{\phi} (x^0) & =: & \left( \frac{a_{t_0}}{a (x^0)}
\right)^{3 / 2} \hat{\varphi} (x^0) \label{rdefphi},\\
G (x^0, y^0, {\bf k}) & =: & \left( \frac{a_{t_0}}{a (x^0)}  \right)^{3 / 2} \hat{G} (x^0, y^0, {\bf k}) \left( \frac{a_{t_0}}{a
	(y^0)} \right)^{3 / 2} \label{rdefG}, \\
\hat{A} (x^0) & := & \left( \frac{a_{t_0}}{a
	(x^0)} \right)^{3 / 2} A, \label{rdefA}
\end{eqnarray}
where $a_{t_0}$ stands for the initial value for the scale factor and we have defined $a_{t_0}:=a(t_0)$ and we have used Fourier transformation for Green's function as,
\begin{eqnarray}
  G (x^0, y^0, \mathbf{k}) & = & \int d^3 \mathbf{r} G (x^0, \mathbf{r}, y^0,
  0) e^{i \mathbf{k} \cdot \mathbf{r}} \label{G00}.
\end{eqnarray}
By using these new definitions, SDEs in Eqs.\eqref{eomPHI} is written  as,
\begin{align}
\left[ \frac{\partial^2}{{\partial x^{0}}^{2}} + \bar{m}_{i}^2 (x^0) \right]_{
	} \hat{\varphi}^d_i (x^0) =& 
	c^{da} D_{a b c}  \hat{A}_{i j k} (x^0) \{ \hat{\varphi}_j^b (x^0)
\hat{\varphi}_k^c (x^0) + \hat{G}^{bc}_{jk} (x, x) \} .
\label{diffphi01}
\end{align}

Next SDEs for the rescaled Green's function in Eqs.\eqref{eomxGF} and \eqref{eomyGF} are written as,
\begin{align}
\left[ \frac{\partial^2}{{\partial x^{0}}^{2}} + \Omega^{2}_{i, {\bf k}} (x^0) \right] \hat{G}^{a b}_{i j, x^0 y^0}
({\bf k}) =&  2 c^{a d} D_{d c e}  \hat{A}_{i k l, x^0} 
\hat{\varphi}^e_{l, x^0} 
\hat{G}^{cb}_{kj, x^0 y^0} ({\bf k})
- i \delta_{i j} \delta_{x^0 y^0} \frac{c^{a b}}{a^3_{t_0}}  \nonumber \\
& - i
\delta_{t_0 x^0} \kappa_{i k}^{a e} ({\bf k}) a^3_{t_0} c^{e f} \hat{G}^{f
	b}_{k j, t_0 y^0}  ({\bf k}) 
, \label{diffGx01} \\
\left[ \frac{\partial^2}{{\partial y^{0}}^{2}} + \Omega^{2}_{i, {\bf k}} (y^0) \right] \hat{G}_{i j, x^0 y^0}^{a b}
({\bf k}) = &  2 \hat{G}^{ac}_{ik, x^0 y^0} ({\bf k}) D_{c e f}  \hat{A}_{k j l, y^0}
\hat{\varphi}^f_{l, y^0} 
	c^{e b}
- i \delta_{i j} \delta_{x^0 y^0} \frac{c^{a b}}{a^3_{t_0}}  \nonumber \\
& - i \hat{G}^{ae}_{ik,
	x^0 t_0} ({\bf k}) c^{e f} a^3_{t_0} \kappa_{k j}^{f b} ({\bf k})
\delta_{t_0 y^0} , \label{diffGy01}
\end{align}
where we have defined,
\begin{eqnarray}
\Omega^{2}_{i, {\bf k}} (x^0) &:=& \frac{{\bf k}^{2}}{a (x^0)^2} + \bar{m}_i^2 (x^0), \label{Ome} \\
\bar{m}_i^2 (x^0) &:=& \tilde{m}_i^2 (x^0) -
\frac{3}{2} \left( \frac{\ddot{a} (x^0)}{a (x^0)} \right) - \frac{3}{4}
\left( \frac{\dot{a} (x^0)}{a (x^0)} \right)^2. \label{defmass}
\end{eqnarray}
Note that the first derivative with respect to time which is originally presented in the expression of Laplacian, Eq.\eqref{lapc}, is now absent in the expression of SDEs for the rescaled fields and Green's functions.


\subsection{The initial condition for Green's function and field}

In this subsection, the initial conditions for Green's function and field are determined. For simplicity, let us look back to example model for one real scalar field. 
We first compute the initial condensation of the field $\bar{\phi}(t_0)=\hat{\varphi}(t_0)$ (see Eq.\eqref{rdefphi}). Using Eq.\eqref{phiH} and setting $x^0=t_0$, we compute it as follows,
\begin{eqnarray}
\bar{\phi}(t_0)\equiv \langle \phi (t_0, {\bf x}) \rangle & := & \frac{ \int  d \phi\  \phi (t_0,{\bf x}) \exp \left[-S^{\text{example}}_{\text{Ecl}} [\phi, \phi] \right]}{\int  d \phi \exp \left[ - S^{\text{example}}_{\text{Ecl}} [\phi, \phi] \right] }
\nonumber\\
& = & \frac{\int d \phi\  \phi ({\bf x}) \exp \left[ -
	\frac{1}{2} \int d^3 {\bf x} d^3 {\bf y} \phi ({\bf x}) D
	({\bf x} - {\bf y}) \phi ({\bf y}) + 2 a_{t_0}^3 \int d^3
	{\bf x} \phi ({\bf x}) j^1 \right]}{\int d \phi \
	\exp \left[ - \frac{1}{2} \int d^3 {\bf x} d^3 {\bf y} \phi
	({\bf x}) D ({\bf x} - {\bf y}) \phi ({\bf y}) + 2
	a_{t_0}^3 \int d^3 {\bf x} \phi ({\bf x}) j^1 \right]},\nonumber \\ \label{phi_t0}
\end{eqnarray}
where we have computed the last term of Eq.\eqref{SEcl_new_2} using Eq.\eqref{NewJ} and 
$D ({\bf r})$ is defined as \cite{Hotta2014},
\begin{eqnarray}
D ({\bf r}) & = & 2 a_{t_0}^3  \int \frac{d^3 \textbf{k}}{(2 \pi)^3} 
\frac{\omega ({\bf k})  (\cosh \beta \omega ({\bf k}) -
	1)}{\sinh \beta \omega ({\bf k})} e^{- i\ {\bf r \cdot k}}.  \label{Dr}
\end{eqnarray}
To proceed the calculation, we denote $J ({\bf x})$ as $J ({\bf x}) = 2 a_{t_0}^3 j^1$. Then the initial condensation of field $\langle \phi (t_0, {\bf x}) \rangle$ is given by,
\begin{eqnarray}
\langle \phi (t_0, {\bf x}) \rangle & = & \frac{\int d \phi'
	({\bf x}) \left\{ \phi' ({\bf x}) + \int d^3 {\bf z} D^{- 1}
	({\bf x} - {\bf z}) J ({\bf z}) \right\} \exp \left[ -
	\frac{1}{2} \int d^3 {\bf x} d^3 {\bf y} \phi' ({\bf x}) D
	({\bf x} - {\bf y}) \phi' ({\bf y}) \right]}{\int d \phi'
	({\bf x}) \exp \left[ - \frac{1}{2} \int d^3 {\bf x} d^3
	{\bf y} \phi' ({\bf x}) D ({\bf x} - {\bf y}) \phi'
	({\bf y}) \right]} \nonumber\\
& = & v, \label{phi_t0_2}
\end{eqnarray}
where we have defined $\phi' ({\bf x}) = \phi ({\bf x}) - \int d^3
{\bf z} D^{- 1} ({\bf z} - {\bf x}) J ({\bf z})$ and $D^{-	1} ({\bf z} - {\bf x})$ satisfies,
\begin{eqnarray}
	\int d^3 {\bf x} D^{- 1} ({\bf z} - {\bf x}) D ({\bf x}
	- {\bf y}) =  \delta ({\bf z} - {\bf y}) .
\end{eqnarray}

Next we will compute the initial condition for Green's function $G (t_0, t_0, \mathbf{ k})=\hat{G} (t_0, t_0, \mathbf{ k})$ (see Eq.\eqref{rdefG}). By using Eq.\eqref{2phiH} and setting $x^0=t_0$ and $y^0=t_0$, one computes it as follows,
\begin{eqnarray}
  G^{ab} (t_0, {\bf x},t_0, {\bf y})  & = & \frac{ \int \int  d \phi^{1} d \phi^{2}  \langle \phi^2 | \Phi (t_0,{\bf y})
 \Phi (t_0,{\bf x}) | \phi^1  \rangle   \exp \left[-S^{\text{example}}_{\text{Ecl}} [\phi^1, \phi^2] \right]}{\int  d \phi \exp \left[ - S^{\text{example}}_{\text{Ecl}} [\phi, \phi] \right] }
\nonumber\\
& = & \frac{\int d \phi' \  \phi' ({\bf y}) \phi' ({\bf x}) \exp \left[ -
	\frac{1}{2} \int d^3 {\bf x} d^3 {\bf y} \phi' ({\bf x}) D
	({\bf x} - {\bf y}) \phi' ({\bf y})  \right]}{\int d \phi' \
	\exp \left[ - \frac{1}{2} \int d^3 {\bf x} d^3 {\bf y} \phi'
	({\bf x}) D ({\bf x} - {\bf y}) \phi' ({\bf y})  \right]},\nonumber \\
	&=& D^{-1}(\mathbf{y}- \mathbf{x}). \label{Gt0t0}
\end{eqnarray}
Using Eqs.\eqref{G00} and \eqref{Dr}, Eq.\eqref{Gt0t0} becomes,
\begin{eqnarray}
G^{ab} (t_0, t_0, \mathbf{ k})  & = & D^{-1} (\mathbf{k}) \nonumber \\
& = & \frac{1}{2 \omega
	({\bf k}) a_{t_0}^3} \left[ \frac{\sinh \beta \omega ({\bf
		k})}{\cosh \beta \omega ({\bf k}) - 1} \right].
\end{eqnarray}


The above results with the example model can be extended to our model and  we summarize them as,
\begin{eqnarray}
\hat{G}^{a b}_{i j, t_0 t_0} ({\bf k}) &=&  \delta_{ij}  \frac{1}{2 \omega_i
	({\bf k}) a_{t_0}^3} \left[ \frac{\sinh \beta \omega_i ({\bf
		k})}{\cosh \beta \omega_i ({\bf k}) - 1} \right] ,\\
\hat{\varphi}^{a}_{i}(t_0)&=&v_i.
\label{eq:incon}
\end{eqnarray}

Next we derive the time derivative of the field and Green's function at the initial time $t_0$. First we integrate the field equation in Eq.\eqref{diffphi01} with respect to time. By setting $x^0=t_0$, we obtain,
\begin{align}
	\frac{\partial \hat{\varphi}_{i, x^0}}{\partial x^0}\Big{|}_{x^0=t_0} = 0 . \label{1TD_field}
\end{align} 
Similarly, we integrate Eq.\eqref{diffGx01} with respect to time $x^0$. By setting both $x^0$ and $y^0$ equal to $t_0$, we obtain the following initial condition, 
\begin{eqnarray}
	\lim_{x^{0}\rightarrow t_0} \frac{\partial }{\partial x^{0}} \hat{G}^{a b}_{i j, x^0 t_0}
	({\bf k}) = - i \delta_{i j} \frac{c^{a b}}{a^3_{t_0}} - i
	\kappa_{i k}^{a e} ({\bf k}) a^3_{t_0} c^{e f} \hat{G}^{f
		b}_{k j, t_0 t_0}  ({\bf k}) .
\end{eqnarray}
Finally, we integrate Eq.\eqref{diffGy01} with respect to time $y^0$. By setting both $x^0$ and $y^0$ equal to $t_0$, we obtain another initial condition, 
\begin{eqnarray}
	\lim_{y^{0}\rightarrow t_0} \frac{\partial }{\partial y^{0}} \hat{G}_{i j, t_0 y^0}^{a b}
	({\bf k})  = - i
	\delta_{i j}  \frac{c^{a b}}{a^3_{t_0}} - i \hat{G}^{ae}_{ik,
		t_0 t_0} ({\bf k}) c^{e f} a^3_{t_0} \kappa_{k j}^{f b} ({\bf k}) .
\end{eqnarray}

\section{The expectation value of PNA} 
\label{sec4}

The SDEs obtained in the previous section allow us to write the solutions for both Green's functions and fields in the form of integral equations. 
In this section, we present the correction to the expectation value of the PNA up to the first order contribution with respect to the cubic interaction. For this purpose, in subsection \ref{subsec1}, we show how one analytically obtains the solutions of SDEs. We write down the solutions up to the first order of the cubic interaction. In the subsection \ref{subsec2}, we also write the expectation value of the PNA up to the first order of the cubic interaction and investigate it by taking into account of the time dependence of the scale factor.

\subsection{The solution of Green's function and fields including $o(A)$ corrections}
\label{subsec1}

The SDEs in present work are inhomogeneous differential equations of the second order. To solve the differential equation, the variation of constants method is used. With the method, the solutions of SDEs are written in the form of integral equations. We solve the integral equation pertubatively and the solutions up to the first order of the cubic interaction are obtained. 
We first write the solutions of fields as,
\begin{eqnarray}
\hat{\varphi}^d_{ i, x^0} & = & \hat{\varphi}^{d,\text{free}}_{i, x^0} + \hat{\varphi}^{d, o (A)}_{i, x^0}, \label{phigen}  \\
\hat{\varphi}^{d,\text{free}}_{ i, x^0} & = & - \bar{K}'_{i, x^0 t_0}
\hat{\varphi}^d_{i, t_0},\label{phifree}  \\
\hat{\varphi}^{d, o (A)}_{i, x^0} & = & \int_{t_0}^{x^0} \bar{K}_{i, x^0 t}  
c^{da} D_{a b c} \hat{A}_{i j k}
(t)  \left\{ \hat{\varphi}^{b, \text{free}}_{j} (t) \hat{\varphi}^{c,
	\text{free}}_{k} (t) + \int \frac{d^3 k}{(2 \pi)^3} \hat{G}^{b c,
	\text{free}}_{j k, t t} ({\bf k}) \right\} 
	dt
, 
\label{phihatOA}
\end{eqnarray}
where $\hat{\varphi}^{\text{free}}$ denotes the free part contribution while $\hat{\varphi}^{o (A)}$ is the contribution due to the first order of the cubic interaction. In Appendix \ref{App2}, Eqs.\eqref{phigen}-\eqref{phihatOA} are derived in details.
$\bar{K}_{i,x^0 y^0}:=\bar{K}_{i,x^0 y^0,{\bf k=0}}$ and $\bar{K}_{i,x^0 y^0,{\bf k}}$
is defined by,
\begin{eqnarray}
\bar{K}_{i, x^0 y^0, {\bf k}} & : = & \frac{1}{W_{i,{\bf k}}} \{ f_{i, {\bf k}} (x^0) g_{i, {\bf k}} (y^0) -
g_{i, {\bf k}} (x^0) f_{i, {\bf k}} (y^0) \}, \label{defKbar2}
\end{eqnarray}
where $W_{i,{\bf k}}$ is defined as,
\begin{eqnarray}
W_{i,{\bf k}}:=\dot{f}_{i, {\bf k}}(x^{0})g_{i, {\bf k}}(x^{0})-f_{i, {\bf k}}(x^{0})\dot{g}_{i, {\bf k}}(x^{0}). \label{defW2}
\end{eqnarray}
$f_{i, {\bf k}}$ and $g_{i, {\bf k}}$ are the solutions which satisfy the following homogeneous differential equations,
\begin{eqnarray}
\left[ \frac{\partial^2}{{\partial x^{0}}^{2}} + \Omega^2_{i, {\bf k}} (x^0) \right] f_{i,{\bf k}}
(x^0) & = & 0,\label{fx0} \\
\left[ \frac{\partial^2}{{\partial x^{0}}^{2}} + \Omega^2_{i, {\bf k}} (x^0) \right] g_{i,{\bf k}}
(x^0) & = & 0 , \label{gx0}
\end{eqnarray} 
where $\Omega^2_{i, {\bf k}} (x^0)$ is given in Eq.\eqref{Ome}. In Appendix \ref{f_and_g}, $f_{i, {\bf k}}$ and $g_{i, {\bf k}}$ are derived in details. 
$\bar{K'}_{i,x^0 y^0}:=\bar{K'}_{i,x^0 y^0,{\bf k=0}}$ and $\bar{K'}_{i,x^0 y^0,{\bf k}}$ is also defined as follows,
\begin{align}
\bar{K'}_{i,x^0 y^0,{\bf k}}:=\frac{\partial \bar{K}_{i,x^0 y^0,{\bf k}}}{\partial y^0}. \label{kprime_bar2}
\end{align}

Next we write down the solution of Green's function as follows,
\begin{eqnarray}
\hat{G}^{a b}_{i j, x^0 y^0} ({\bf k}) & = & \hat{G}^{a b, \text{free}}_{i j, x^0
	y^0} ({\bf k}) + \hat{G}_{i j, x^0 y^0}^{a b, o (A)}
({\bf k}), \label{solG}  \\
\hat{G}^{a b, \text{free}}_{i j, x^0 y^0} ({\bf k}) & = &
\frac{\delta_{ij}}{2 \omega_{i, {\bf k}} a_{t_0}^3} \coth \frac{\beta
	\omega_{i, {\bf k}}}{2}  \left(\begin{array}{cc}
1 & 1\\
1 & 1
\end{array}\right)^{a b}    \left[  \bar{K}'_{i, x^0 t_0,{\bf k}}  \bar{K}'_{i, y^0 t_0,{\bf k}}   + \omega_{i, {\bf k}}^2 
\bar{K}_{i, x^0 t_0,{\bf k}} \bar{K}_{y^0 t_0,{\bf k}} \right] \nonumber\\
&  & + \frac{i \delta_{ij}}{2 a_{t_0}^3} \bar{K}_{i, x^0 y^0,{\bf k}}  \{
\epsilon^{a b} + c^{a b}  (\theta (y^0 - x^0) - \theta (x^0 - y^0)) \}, \label{Ghatfree}
\end{eqnarray}
where $\epsilon^{a b}$ is an anti-symmetric tensor and its non-zero components are given as $\epsilon^{12}=1$ while $\theta(t)$ denotes a unit step function,
\begin{eqnarray}
\hat{G}_{i j, x^0 y^0}^{a b, o (A)} ({\bf k}) & = & \int_{t_0}^{y^0} R^{a b, o (A)}_{i j, x^0 t, \mathbf{k}}  \bar{K}_{j, y^0 t,
\mathbf{k}} dt  \nonumber \\ & & + \int_{t_0}^{x^0} \bar{K}_{i, x^0 t, \mathbf{k}} Q^{a c, o (A)}_{i j, tt_0,
\mathbf{k}}  (E_{j j, \mathbf{k}}^{T, c b}  \bar{K}_{j, y^0 t_0, \mathbf{k}} -
\bar{K}'_{j, y^0 t_0, \mathbf{k}} \delta^{c b}) dt , \label{GhatOA} 
\end{eqnarray}
where  $Q^{o(A)}$, $R^{o(A)}$ and $E_{{\bf k}}$ are given as,
\begin{align}
Q^{a b,o(A)}_{ij,x^0 y^0,{\bf k}}   = & 2 c^{a d} D_{d c e}  \hat{A}_{i k l, x^0} 
\hat{\varphi}^{e,
	\text{free}}_{l, x^0} 
\hat{G}^{c b,\text{free}}_{k j, x^0 y^0} ({\bf k}), 
\\
R^{a b ,o(A)}_{ij,x^0 y^0,{\bf k}}   = & 2 \hat{G}^{ac,\text{free}}_{ik, x^0 y^0} ({\bf k}) D_{c e f}  \hat{A}_{k j l, y^0} 
\hat{\varphi}^{f,\text{free}}_{l, y^0}
c^{e b}
, \\
E^{ac}_{ik,{\bf k}} =& -i\kappa^{ae}_{ik}({\bf k}) a_{t_{0}}^{3} c^{ec} , \label{defE}
\end{align}
and $\kappa_{ij}^{ab}({\bf k})$ is given in Eq.\eqref{kappij}. In Appendices \ref{app_a4} and \ref{app_a5}, we derive Eqs.\eqref{Ghatfree} and \eqref{GhatOA} in detail respectively.

\subsection{ The expectation value of PNA including $o(A)$ corrections and the first order of the Hubble parameter}
\label{subsec2}
Next we compute the PNA in Eq.\eqref{orGenj} including the first order correction with respect to $A$ ($o(A)$) and the effect of expansion up to the first order of the Hubble parameter. By using rescaled fields, Green's function and coupling constant in Eqs.\eqref{rdefphi}-\eqref{rdefA}, one can write down total contribution to the expectation value of PNA with order $o(A)$ corrections as,
\begin{eqnarray}
\left( \frac{a (x^0)}{a_{t_0}} \right)^3 \langle j_0 (x^0) \rangle & = & \text{ Re} \left[  \hat{\varphi}^{1,\text{free}}_{2} (x^0)
\dot{\hat{\varphi}}^{1 \ast,\text{free}}_{1} (x^0) - \hat{\varphi}^{1,\text{free}}_{1} (x^0) \dot{\hat{\varphi}}^{1 \ast,\text{free}}_{2} (x^0) \right]   \nonumber\\
& & +\int
\frac{d^3 \mathbf{k}}{(2 \pi)^3} \left.  \left( \frac{\partial}{\partial x^0} -
\frac{\partial}{\partial y^0} \right) \text{ Re} \left[ \hat{G}^{12, o(A)}_{12}
(x^0, y^0, {\bf k})\right]  \right|_{y^{0} \rightarrow x^0} \nonumber\\
&  & + \text{ Re} \left[  \hat{\varphi}^{1,\text{free}}_{2} (x^0)
\dot{\hat{\varphi}}^{1 \ast, o(A)}_{1} (x^0) - \hat{\varphi}^{1,\text{free}}_{1} (x^0) \dot{\hat{\varphi}}^{1 \ast, o(A)}_{2} (x^0) \right] 
\nonumber\\
&  & + \text{ Re} \left[  \hat{\varphi}^{1, o(A)}_{2} (x^0)
\dot{\hat{\varphi}}^{1 \ast,\text{free}}_{1} (x^0) - \hat{\varphi}^{1, o(A)}_{1} (x^0) \dot{\hat{\varphi}}^{1 \ast,\text{free}}_{2} (x^0) \right] 
 . 
\label{OAcor}
\end{eqnarray}
The first line of the above equation is the zeroth order of the cubic interaction while the next three terms are the first order. 

As was indicated previously, we will further investigate the expectation value of the PNA for the case of time-dependent scale factor.  
For that purpose, one can expand scale factor around $t_0$ for $0< t_0 \leqslant x^0$ as follows,
\begin{eqnarray*}
	a (x^0) & = & a (t_0) + (x^0 - t_0) \dot{a} (t_0) + \frac{1}{2} (x^0 -
	t_0)^2 \ddot{a} (t_0) + \ldots\\
	& = & a^{(0)} + a^{(1)} (x^0) + a^{(2)} (x^0)  + \ldots .
\end{eqnarray*}
We first assume that $a^{(n + 1)} (x^0) < a^{(n)} (x^0)$ when $x^0$ is near
$t_0$. Then one can keep only the following terms,
\begin{eqnarray}
a (x^0) & \simeq & a^{(0)} + a^{(1)} (x^0), \label{sctu_scale}
\end{eqnarray}
and $a^{(n)} (x^0)$ for ($n \geqslant 2$) are set to be zero. $a^{(0)}$
corresponds to the  constant scale factor and $a^{(1)} (x^0)$ corresponds to linear Hubble parameter
$H (t_0)$. Thus it can be written as,
\begin{eqnarray}
\frac{a (x^0)}{a (t_0)} & = & 1 + (x^0 - t_0) H (t_0),  \label{SF}
\end{eqnarray}
where $H (t_0) $ is given by,
\begin{align}
	H (t_0)= \frac{\dot{a}(t_0)}{a(t_0)}, \label{Ht0}
\end{align}
and $t_0 > 0$. Throughout this study, we only keep first order of $H (t_0)$ as the first non-trivial approximation. For the case that Hubble parameter is positive, it corresponds to the case for the expanding universe. Under this situation, $\dot{a} (x^0)  =  a (t_0) H (t_0)$ and $\ddot{a} (x^0) = 0 $.

Now let us briefly go back to Eq.\eqref{defmass}. With these approximations, the second term of Eq.\eqref{defmass} is apparently vanished. Since $\dot{a} (x^0)$ is proportional to linear $H (t_0)$, the third term of Eq.\eqref{defmass} involves second order of $H(t_0)$. Hence, one can neglect it and the Riemann curvature $R (x^0)$ in Eq.\eqref{Rie} is also vanished.
Therefore, $\bar{m}_i^2 (x^0)$ is
simply written as $\tilde{m}^2_i$. Now $\tilde{m}^2_i$ are given as,
\begin{eqnarray}
\tilde{m}_1^2 & = & m_{\phi}^2 - B^2,\\
\tilde{m}_2^2 & = & m_{\phi}^2 + B^2,\\
\tilde{m}_3^2 & = & m_N^2.
\end{eqnarray}
Next we define $\omega_{i, {\bf k}}$ as,
\begin{align}
 \omega_{i, {\bf k}} :=\sqrt{\frac{{\bf k}^2}{a_{t_0}^{2}}+\tilde{m}_i^2}.
\end{align}
We consider $\Omega_{i, {\bf k}} (x^0)$ defined in Eq.\eqref{Ome}. One can expand it around time $t_0$ as,
\begin{eqnarray}
\Omega_{i, {\bf k}} (x^0) 
& \simeq & \omega_{i, {\bf k}}  + (x^0 - t_0) \left.
\frac{\partial}{\partial x^0} \Omega_{i,{\bf k}} (x^0) \right|_{x^0 = t_0}  \nonumber \\
& = & \omega_{i, {\bf k}}  \left\{ 1 - H (t_0) (x^0 -
t_0) \frac{{\bf  k}^2}{[a (t_0) \omega_{i, {\bf k}} (t_0)]^2} \right\}. \label{OmeB2}
\end{eqnarray}

Now let us investigate the expectation value of PNA under these approximations. For the case that $\hat{\varphi}_{1, t_0} = \hat{\varphi}_{2, t_0} = 0$ and $\hat{\varphi}_{3, t_0} \neq 0$, the non-zero contribution to the expectation value of PNA comes only from $o (A)$ corrections to Green's function. From Eq.\eqref{OAcor}, we can obtain,
\begin{eqnarray}
\langle j_0 (x^0) \rangle & =& 
\frac{2  }{a (x^0)^3}\hat{\varphi}_{3, t_0} \int \frac{d^3
	{\bf k}}{(2 \pi)^3}  \int_{t_0}^{x^0} \hat{A}_{123,t} (- \bar{K}'_{3, tt_0, {\bf 0}})
\left[ \left\{ \frac{1}{2 \omega_{2, {\bf k}} (t_0)} \right.
\right.  \nonumber\\
&  & \times \coth \frac{\beta \omega_{2, {\bf k}} (t_0)}{2}
[(\dot{\bar{K}}_{1, x^0 t, {\bf k}}  \bar{K}'_{2, x^0 t_0,
	{\bf k}} - \bar{K}_{1, x^0 t, {\bf k}} 
\dot{\bar{K}}'_{2, x^0 t_0, {\bf k}} ) \bar{K}'_{2, tt_0,
	{\bf k}}  \nonumber\\
&  & + \omega^2_{2, {\bf k}} (t_0) (\dot{\bar{K}}_{1, x^0 t,
	{\bf k}}  \bar{K}_{2, x^0 t_0, {\bf k}} - \bar{K}_{1, x^0 t,
	{\bf k}}  \dot{\bar{K}}_{2, x^0 t_0, {\bf k}}) \bar{K}_{2, tt_0,
	{\bf k}}]\}  \nonumber\\
&  & - \{1 \leftrightarrow 2 \text{for lower indices} \}] dt
, \label{OAcor_B}
\end{eqnarray}
where we have used Eqs.\eqref{phifree} and \eqref{GhatOA}. Following the expression of the scale factor in Eq.\eqref{SF}, $\bar{K}$ is also divided into the part of the constant scale factor and the part which is proportional to $H(t_0)$.  In Appendix \ref{sec_kbar}, $\bar{K}$ and its derivative are derived  in details.
In the above expression, $H(t_0)$ is also included in $\hat{A}(t)$. Since we are interested in the PNA up to the first order of $H(t_0)$, we expand it as follows,
\begin{eqnarray}
\hat{A} (t) & \simeq &   A
\left\{ 1 - \frac{3}{2} (t - t_0) H (t_0) \right\}. \label{rdefA_B}
\end{eqnarray}
Furthermore, substituting Eqs.\eqref{SF}, \eqref{rdefA_B} and $\bar{K}$ and its derivative in Eqs.\eqref{k0}, \eqref{k1}, \eqref{kpri0}-\eqref{kdotpri1} into Eq.\eqref{OAcor_B}, one can divide the PNA into two parts,
\begin{eqnarray}
\langle j_0 (x^0) \rangle & = & \langle j_0 (x^0) \rangle_{ 
	\text{1st} } + \langle j_0 (x^0) \rangle_{ \text{2nd}
} , 
\label{NTC}
\end{eqnarray}
\begin{eqnarray}
\langle j_0 (x^0) \rangle_{  \text{1st} } & = & \frac{2
	\hat{\varphi}_{3, t_0} A_{1 2 3}}{a_{t_0}^3} \int \frac{d^3 
	{\bf k}}{(2 \pi)^3} \int_{t_0}^{x^0} \left\{ 1 - 3 (x^0 - t_0) H (t_0)  -
\frac{3}{2} (t - t_0) H (t_0) \right\} \nonumber\\
&  & \times \left[ \left\{ \frac{(- \bar{K}^{(0)\prime}_{3, t t_0,
		{\bf 0}} )}{2 \omega_{2, {\bf k}} (t_0) } \coth \frac{\beta
	\omega_{2, {\bf k}} (t_0)}{2} \left[ \left( \bar{K}^{(0)\prime}_{2,^{}
	x^0 t_0, {\bf k}} \overset{\leftrightarrow}{\partial\ \dot{}}
\bar{K}^{(0)}_{1, x^0 t, {\bf k}} \right) \bar{K}^{(0)\prime}_{2, t
	t_0, {\bf k}} \right. \right. \right. \nonumber\\
&  & \left. \left. \left. + \omega^2_{2, {\bf k}} (t_0) \left(
\bar{K}^{(0)}_{2, x^0 t_0, {\bf k}} \overset{\leftrightarrow}{\partial\
	\dot{}} \bar{K}^{(0)}_{1, x^0 t, {\bf k}} \right) \bar{K}^{(0)}_{2, t
	t_0, {\bf k}} \right] \right\} - \{ 1 \leftrightarrow 2 \text{ for}
\text{ lower} \text{ indices} \} \right] d t, \nonumber \\
\label{J01st} \\
\langle j_0 (x^0) \rangle_{ \text{2nd} } & = & \frac{2
	\hat{\varphi}_{3, t_0} A_{1 2 3}}{a_{t_0}^3} \int \frac{d^3 
	{\bf k}}{(2 \pi)^3} \int_{t_0}^{x^0}  \left[ \left\{
\frac{(- \bar{K}^{(0)\prime}_{3, t t_0, {\bf 0}} )}{2 \omega_{2,
		{\bf k}} (t_0) } \coth \frac{\beta \omega_{2, {\bf k}} (t_0)}{2}
\right. \right. \nonumber\\
&  & \times \left[ \left( \bar{K}^{(0)\prime}_{2,^{} x^0 t_0, {\bf k}}
\overset{\leftrightarrow}{\partial\ \dot{}} \bar{K}^{(0)}_{1, x^0 t,
	{\bf k}} \right) \bar{K}^{(1)\prime}_{2, t t_0, {\bf k}} \right.
\nonumber\\
&  & + \left( \bar{K}^{(1)\prime}_{2,^{} x^0 t_0, {\bf k}}
\overset{\leftrightarrow}{\partial\ \dot{}} \bar{K}^{(0)}_{1, x^0 t,
	{\bf k}} + \bar{K}^{(0)\prime}_{2,^{} x^0 t_0, {\bf k}}
\overset{\leftrightarrow}{\partial\ \dot{}} \bar{K}^{(1)}_{1, x^0 t,
	{\bf k}} \right) \bar{K}^{(0)\prime}_{2, t t_0, {\bf k}}
\nonumber\\
&  & + \omega^2_{2, {\bf k}} (t_0) \left[ \left( \bar{K}^{(0)}_{2, x^0
	t_0, {\bf k}} \overset{\leftrightarrow}{\partial\ \dot{}}
\bar{K}^{(0)}_{1, x^0 t, {\bf k}} \right) \bar{K}^{(1)}_{2, t t_0,
	{\bf k}} \right. \nonumber\\
&  & \left. \left. \left. + \left( \bar{K}^{(1)}_{2, x^0 t_0, {\bf k}}
\overset{\leftrightarrow}{\partial\ \dot{}} \bar{K}^{(0)}_{1, x^0 t,
	{\bf k}} + \bar{K}^{(0)}_{2, x^0 t_0, {\bf k}}
\overset{\leftrightarrow}{\partial\ \dot{}} \bar{K}^{(1)}_{1, x^0 t,
	{\bf k}} \right) \bar{K}^{(0)}_{2, t t_0, {\bf k}} \right] \right]
\right\} \nonumber\\
&  & \left. - \{ 1 \leftrightarrow 2 \text{ for} \text{ lower} \text{ indices} \}  \right] d t  \label{J02nd},
\end{eqnarray}
and the derivative $\overset{\leftrightarrow}{\partial\ \dot{}}$ acts on the first argument of $\bar{K}$ and defined as follows,
\begin{eqnarray}
	\bar{K}_{2, x^0 t, {\bf k}} \overset{\leftrightarrow}{\partial\ \dot{}}
	\bar{K}_{1, x^0 t, {\bf k}} & = & \bar{K}_{2, x^0 t, {\bf k}} \left( \frac{\partial}{\partial x^0}
	\bar{K}_{1, x^0 t, {\bf k}} \right)  -
	 \left( \frac{\partial}{\partial x^0}
	\bar{K}_{2, x^0 t, {\bf k}} \right)\bar{K}_{1, x^0 t, {\bf k}}.
\end{eqnarray}
Each term of the PNA shown in Eqs.\eqref{J01st} and \eqref{J02nd} can be understood as follows. The first term is the PNA with the constant scale factor. The second term with a prefactor $-3H(t_0)(x^0 - t_0)\frac{1}{a_{t_0}^{3}}\simeq\frac{1}{a(x^0)^{3}}-\frac{1}{a_{t_0}^{3}}$ is called the dilution effect. The third term with a prefactor $-\frac{3}{2}A_{123}(t-t_0)H(t_0) \simeq \hat{A}_{123}(t)-A_{123}$ is called the freezing interaction effect. The fourth term which corresponds to $\langle j_0 (x^0) \rangle_{ \text{2nd} }$ is called the redshift effect. Below we explain their physical origins. The dilution of the PNA is caused by the increase of the volume of the universe. The origin of the freezing interaction effect can be understood with Eq.\eqref{rdefA_B}. It implies that the strength of the cubic interaction $\hat{A}(t)$ controlling the size of PNA, decreases as the scale factor grows. The origin of the redshift can be explained as follows. As shown in Eq.\eqref{Ome}, as the scale factor grows, the physical wavelength becomes large. Therefore, the momentum and the energy of the particles becomes small. Note that this effect does not apply to the zero-mode such as condensate which is homogeneous and is a constant in the space. 

Before closing this section,
we compute the production rate of the PNA per unit time which is a useful expression
when we understand the numerical results of the PNA. We compute the time derivative of the PNA for
the case of the constant scale factor $H_{t_0}=0$.
By setting $H_{t_0}=0$, one obtains it at the initial time $x^1=x^0-t_0=0$,
\begin{eqnarray}
\frac{\partial}{\partial x^1}\langle j_0 (x^1+t_0) \rangle|_{x^1=0} & = & \frac{ v_3 A_{123}}{a_{t_0}^3}\int^{\infty}_{0}
\frac{k^2 dk}{2 \pi^2}\frac{1}{\omega_{1 ,{\bf k}}\omega_{2 ,{\bf k}}} 
\nonumber \\
& & \times[(n_2-n_1) (\omega_{1, {\bf k}}+ \omega_{2,{\bf k}}) +(n_2+n_1+1) 
(\omega_{1, {\bf k}}- \omega_{2,{\bf k}})]  , \label{tder}
\end{eqnarray}
where $n_i$ is the distribution functions for the Bose particles,
\begin{eqnarray}
n_i=\frac{e^{-\beta \omega_{i, {\bf k}}}}{1-e^{-\beta \omega_{i ,{\bf k}}}}, (i=1,2).
\end{eqnarray}
In Appendix \ref{sec_calJ}, we derive Eq.\eqref{tder} in detail.
Because we assume $\tilde{m}_1 < \tilde{m}_2$, one obtains inequality
$n_2 < n_1$. From the expression above, the production rate of PNA at the initial time
is negative for $v_3 A_{123}>0$. One also finds the rate is logarithmically divergent for
the momentum ($\mathbf{k}$) integration,
\begin{eqnarray}
\frac{\partial}{\partial x^1}\langle j_0 (x^1+t_0) \rangle|_{x^1=0}^{\text{divergent}}\simeq
\frac{ v_3 A_{123}}{2 \pi^2 a_{t_0}^3} 
 \frac{\tilde{m}_1^2-\tilde{m}_2^2}{2} \text{log}\left( \frac{k_{\text{max}}}{\mu} \right)  ,
\end{eqnarray}
where $\mu=O(\tilde{m}_i)$ ($i=1,2$) and $k_{\text{max}}$ is an ultraviolet cut off for the momentum integration. 
With the expression, one expects that for  the positive $v_3 A_{123}$, 
the PNA becomes negative from zero just after the
initial time and the behavior will be confirmed in the numerical
simulation.

\section{Numerical results}
\label{sec5}

In this section, we numerically study the time dependence of the PNA. The PNA depends on the parameters of the model such as masses and coupling constants. It also depends on the initial conditions and the expansion rates of the universe. Since the PNA is linearly proportional to the coupling constant $A_{123}$ and the initial value of the field $\hat{\varphi}_{3, t_0}$, we can set these parameters as unity in the unit of energy and later on one can multiply their values. As for the initial scale factor $a_{t_0}$, without loss of generality, one can set this dimensionless factor is as unity. For the other parameters of the model, we choose $\tilde{m}_2,B$ and $\omega_{3,{\bf 0}}=\tilde{m}_3$ as independent parameters since the mass $\tilde{m}_1$ is written as,
\begin{align}
\tilde{m}_1^2=\tilde{m}_2^2-2 B^2. \label{def_B}
\end{align}
The temperature $T$ and the expansion rate $H(t_0)$ determine the environment for the universe. The former determines the thermal distribution of the scalar fields. 
Within the approximation for the time dependence of the scale factor in Eq.\eqref{SF}, $H(t_0)$ is the only parameter which controls the expansion rate of the universe. The approximation is good for the time range which satisfies the following inequality,
\begin{align}
x^0-t_0 \ll \frac{1}{3 H(t_0)}. \label{tr}
\end{align}
The time dependence of PNA is plotted as a function of the dimensionless time defined as,
\begin{align}
t=\omega_{3,{\bf 0}}^r (x^0-t_0), \label{dimt}
\end{align}
where $\omega^r_{3,{\bf 0}}$ is a reference frequency. In terms of the dimensionless time, the condition of Eq.\eqref{tr} is written as,
\begin{align}
  t \ll t_{\text{max}} \equiv  \frac{\omega_{3,{\bf 0}}^r}{3 H(t_0)}. \label{tmax}
\end{align}

How the PNA behaves with respect to time is discussed in the following subsection (\ref{sec5A}-\ref{sec5C}). The results, as will be shown later, revealed that the PNA has an oscillatory behavior. We also investigate the parameter dependence for two typical cases, one of which corresponds to the longer period and the other corresponds to the shorter period.  
In the numerical simulation, we do not specify the unit of parameters.  Note that the numerical values for the dimensionless quantities such as ratio of masses do not depend on the choice of the unit as far as the quantities in the ratio are given in the same unit. In subsection \ref{sec5D}, we assign the unit for the parameters and estimate the ratio of the PNA over entropy density.

\subsection{The PNA with the longer period}
\label{sec5A}

\begin{figure}[htbp]
	\centering
	\includegraphics[width=.7\textwidth]{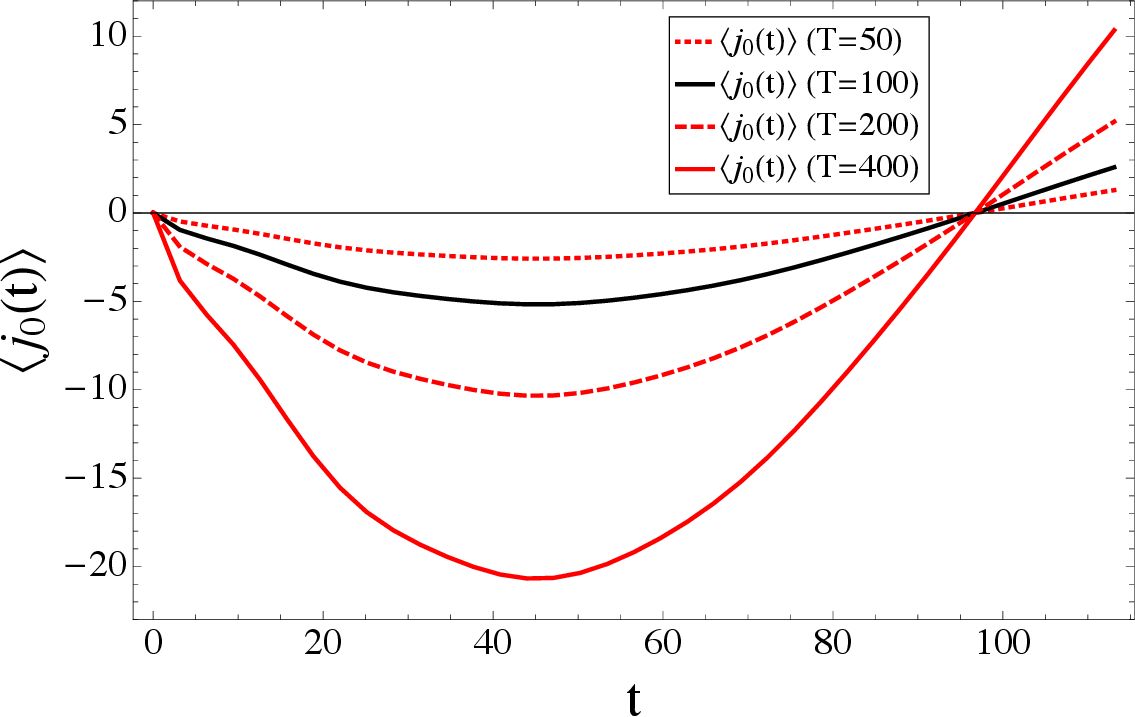}
	\caption{\label{figure2} Dependence on temperature $T$ of the time evolution of PNA. In horizontal axis, we use the dimensionless time $t = \omega_{3,{\bf 0}}^r (x^0 - t_0)$ where we choose $\omega_{3, {\bf 0}}^r = 0.35$. We fix a set of parameters as ($\tilde{m}_1, \tilde{m}_2, B,H_{t_0}, \omega_{3, {\bf 0}}$)=($0.04, 0.05, 0.021,10^{- 3}, 0.0035$) for all of the lines. The dotted red, black, dashed red and red lines show the cases $T=50, 100, 200$ and $400$, respectively. }
\end{figure}
Let us now consider the PNA which has the longer period. While we investigate the dependence of several parameters, we fix two parameters as $(\tilde{m}_2,H_{t_0})=(0.05,10^{-3})$. In Fig. \ref{figure2}, the temperature ($T$) dependence of PNA is shown. It depends on the temperature only through hyperbolic function as shown in Eq.\eqref{OAcor_B}. In this figure, $t_{\text{max}}$ in Eq.\eqref{tmax} is around $110$. What stands out of this figure is the change of the amplitude for PNA among the three curves. As the temperature increases, the amplitude of the oscillation becomes larger. 
\begin{figure}[htbp]
	\centering
	\includegraphics[width=.7\textwidth]{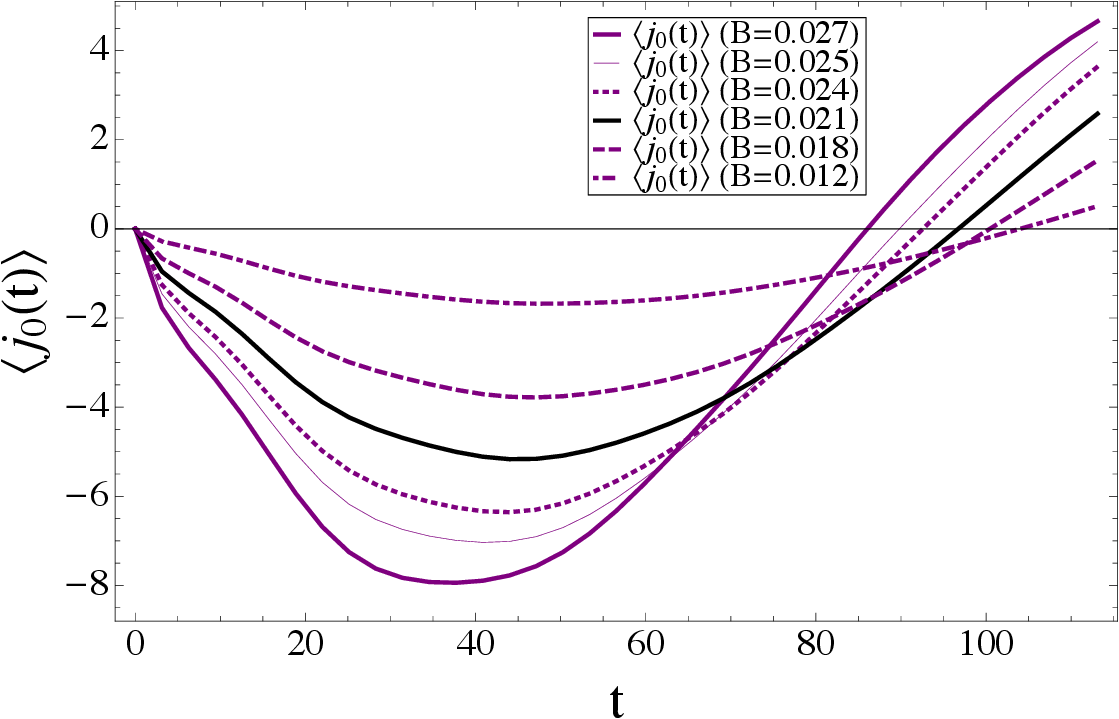}
	\caption{\label{figure3} Dependence on parameter $B$ of the time evolution of PNA. The horizontal axis is the dimensionless time defined as $t = \omega_{3, {\bf 0}}^r (x^0 -t_0)$. As a reference angular frequency, we choose $\omega_{3, {\bf 0}}^r = 0.35$. We fix a set of
		parameters as ($\tilde{m}_2, T, H_{t_0}, \omega_{3, {\bf 0}}$)=($0.05,100,10^{- 3},0.0035$) for all of the lines. The purple, thin purple, dotted purple, black, dashed purple and dot-dashed purple lines show the cases $B=0.027, 0.025, 0.024, 0.021, 0.018$ and $0.012$, respectively. }
\end{figure}
In Fig. \ref{figure3}, we show the $B$ dependence. Interestingly, both of the amplitude and the period of the oscillation change when we alter the parameter $B$. As it increases, the amplitude becomes larger and its period becomes shorter. 
\begin{figure}[htbp]
	\centering
	\includegraphics[width=.7\textwidth]{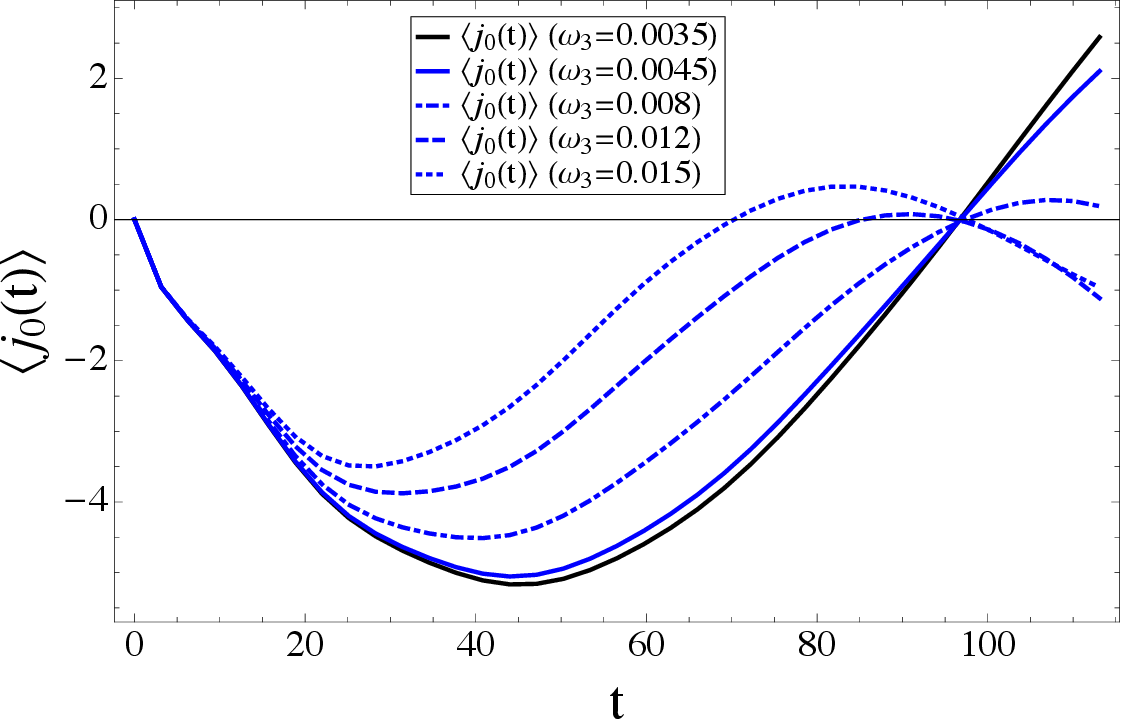}
	\caption{\label{figure4} The $\omega_{3, {\bf 0}}$ dependence of the time evolution of PNA. In horizontal axis, we use the dimensionless time $t =\omega_{3, {\bf 0}}^r (x^0 - t_0)$ where we choose $\omega_{3, {\bf 0}}^r = 0.35$. We use a set of parameters as ($\tilde{m}_1, \tilde{m}_2, B,T, H_{t_0}$)=($0.04,0.05,0.021,100,10^{- 3}$) for all of the lines. The black, blue, dot-dashed blue, dashed blue and dotted blue lines show the cases $\omega_{3, {\bf 0}}= 0.0035, 0.0045, 0.008, 0.012$ and $0.015$, respectively. }
\end{figure}
Fig. \ref{figure4} shows the dependence of the PNA on $\omega_{3, {\bf 0}}$. As shown in the black, blue and dot-dashed blue lines, the position of the first node does not change when $\omega_{3, {\bf 0}}$ takes its value within the difference of $\tilde{m}_1$ and $\tilde{m}_2$. However, the amplitude of oscillation gradually decreases as $\omega_{3, {\bf 0}}$ increases up to the mass difference. The more interesting findings were observed when $\omega_{3, {\bf 0}}$ becomes larger than the mass difference. As $\omega_{3, {\bf 0}}$ becomes larger, the amplitude decreases and the new node is formed at once. The dashed and dotted blue lines show this behavior.    
\begin{figure}[htbp]
	\centering
	\includegraphics[width=.70\textwidth]{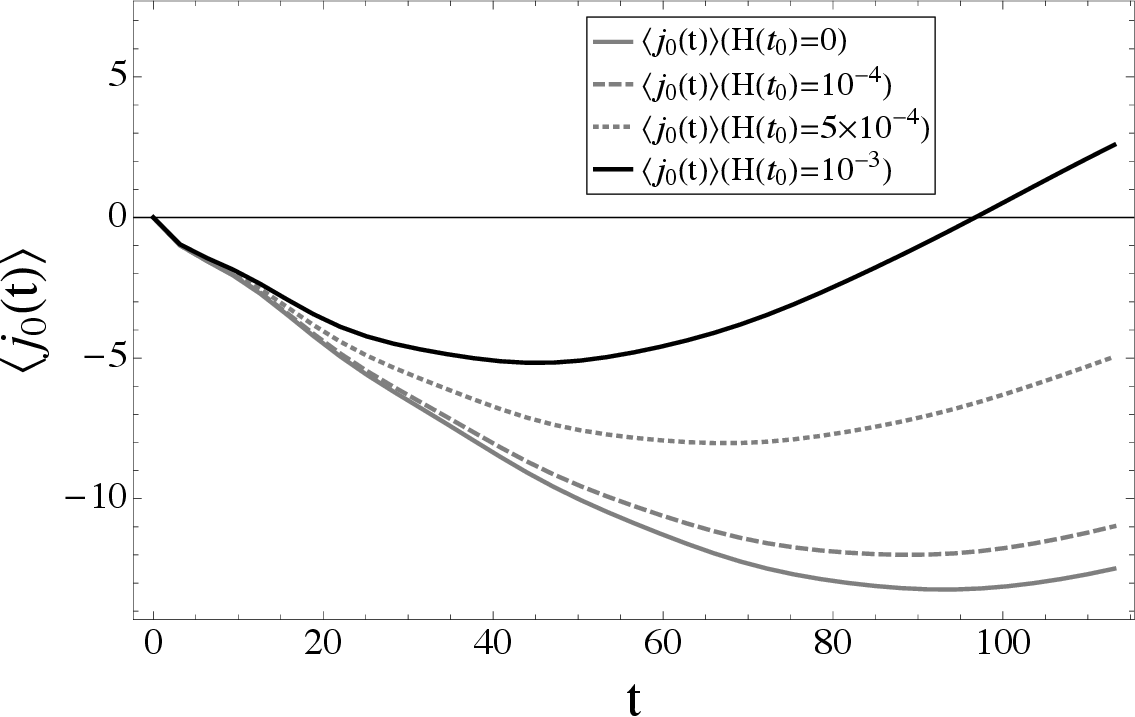}
	\caption{\label{figure5} Dependence on the expansion rate $H_{t_0}$ of the time evolution of PNA. In the horizontal axis, we use the dimensionless time $t = \omega_{3,{\bf 0}}^r (x^0 - t_0)$ where we choose $\omega_{3, {\bf 0}}^r = 0.35$. We fix a set of parameters as ($\tilde{m}_1, \tilde{m}_2, B,T, \omega_{3, {\bf 0}}$)=($0.04,0.05,0.021,100,0.0035$) for all of the lines. The gray, dashed gray, dotted gray and black lines show the cases $H_{t_0}=0, 10^{- 4}, 5\times 10^{- 4}$ and $10^{- 3}$, respectively.}
\end{figure}
The dependence on the expansion rate ($H_{t_0}$) is shown in Fig. \ref{figure5}. There is an interesting aspect of this figure at the fixed time $t$. As the expansion rate becomes larger, the size of PNA becomes smaller. 

\subsection{The PNA with the shorter period}
\label{sec5B}
\begin{figure}[t]
	\centering
	\includegraphics[width=.70\textwidth]{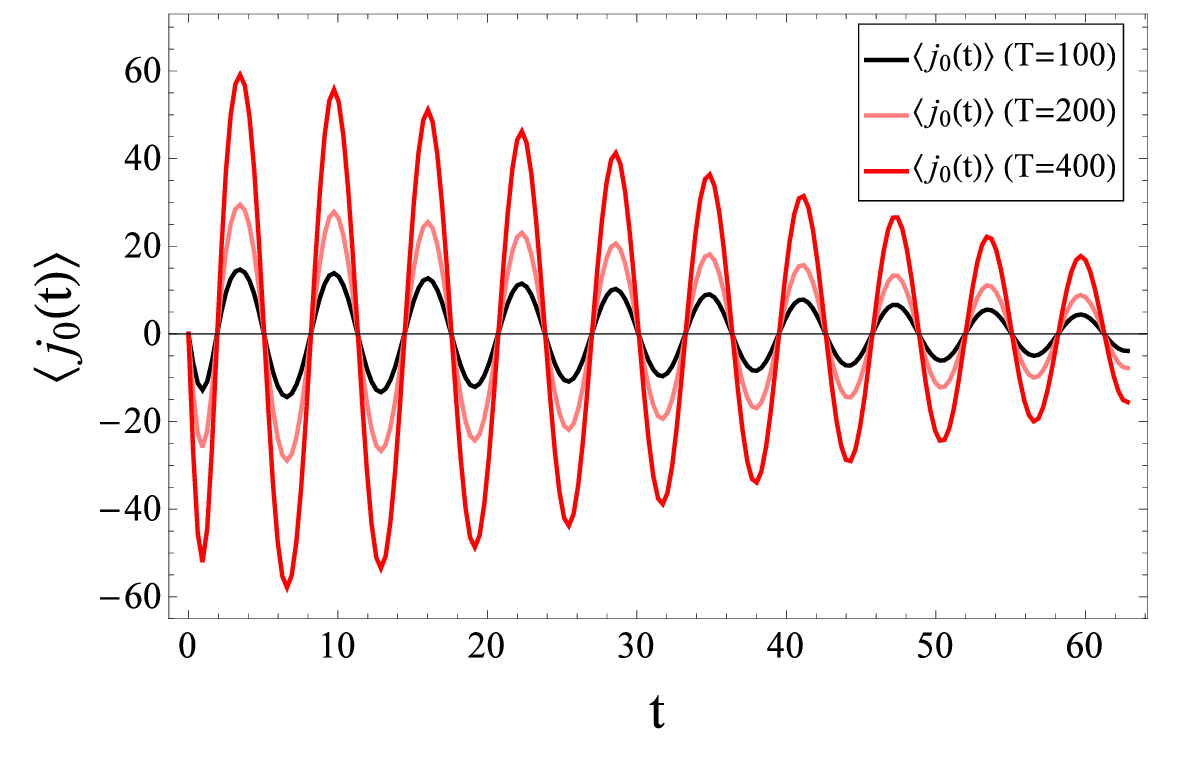}
	\caption{\label{figure6} Dependence on temperature $T$ of the time evolution of PNA. As for horizontal axis, we use the dimensionless time $t = \omega_{3,{\bf 0}}^r (x^0 - t_0)$ where we choose $\omega_{3, {\bf 0}}^r = 0.35$. We fix a set of parameters as $(\tilde{m}_1,\tilde{m}_2,B,\omega_3,H_{t_0})=(2,3,1.58,0.35,10^{-3})$ for all the lines. The black, light red and red lines show the cases $T=100,200$ and $400$, respectively. }
\end{figure}

\begin{figure}[t]
	\centering
	\begin{tabular}{c c}
		\includegraphics[width=.5\textwidth]{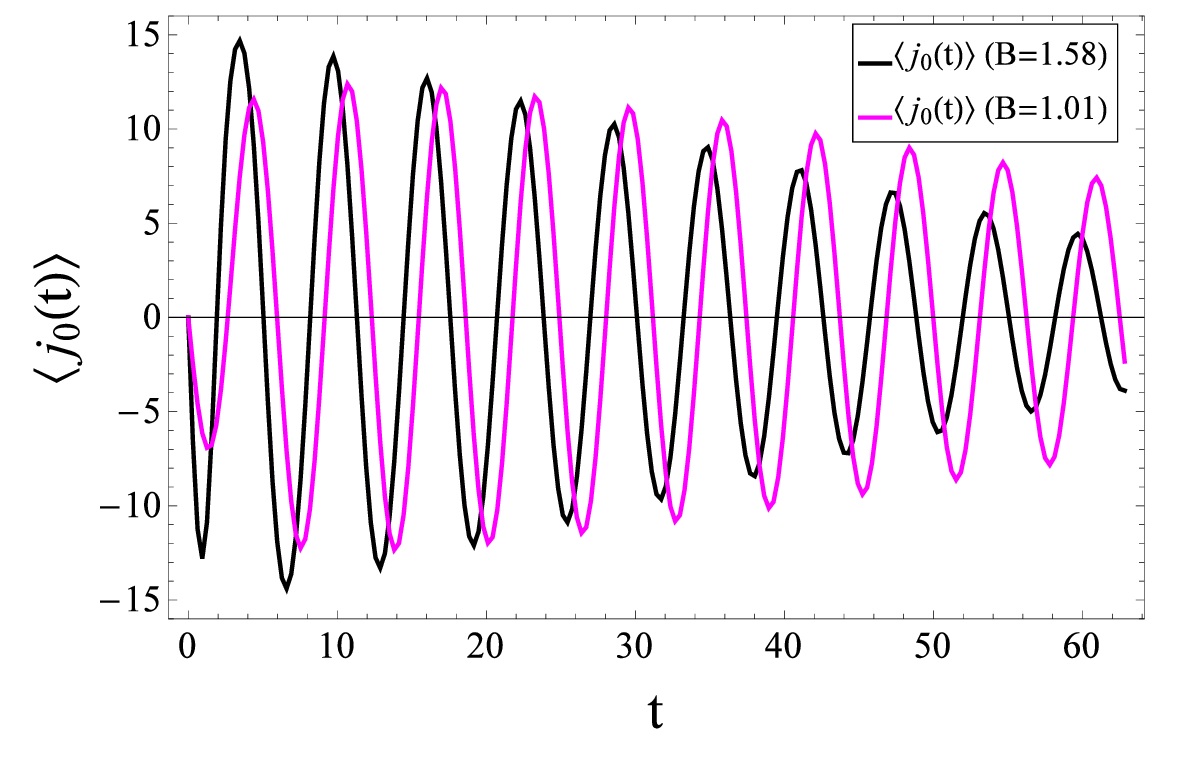}
		&
		\includegraphics[width=.5\textwidth]{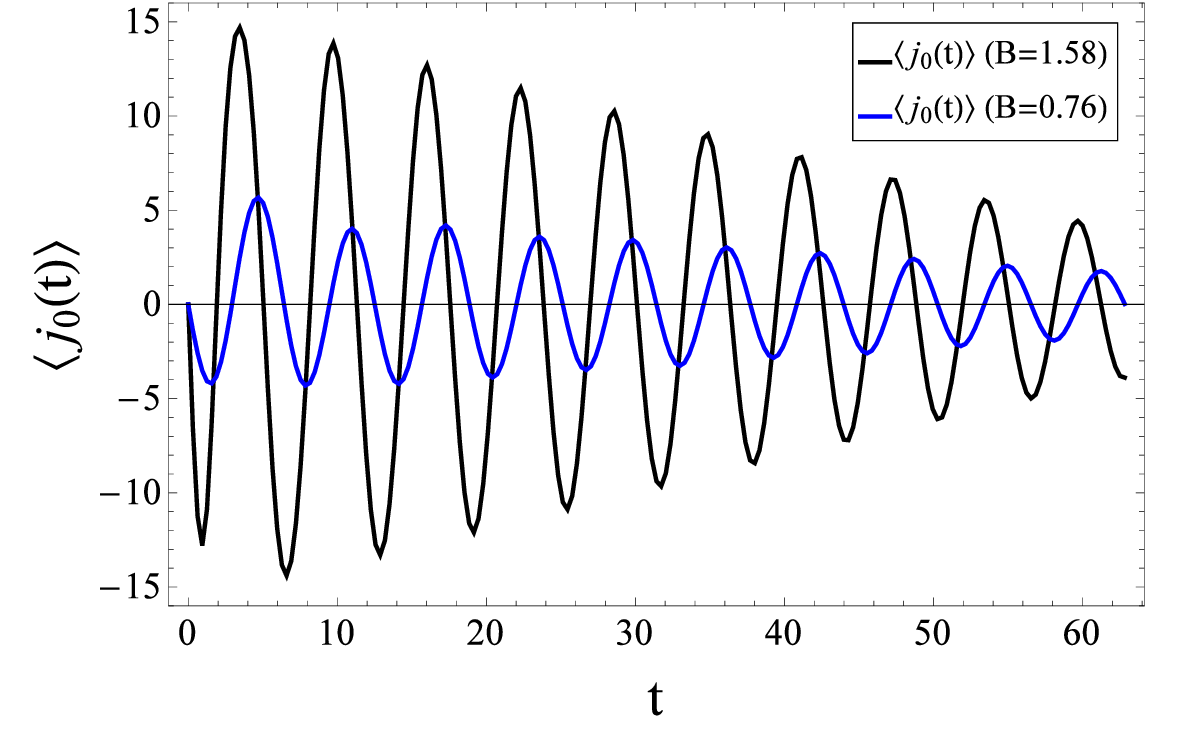}
	\end{tabular}
	\caption{\label{figure7} $B$ dependence for the time evolution of PNA. The horizontal axis is the dimensionless time defined as $t = \omega_{3, {\bf 0}}^r (x^0 -t_0)$. As a reference angular frequency, we choose $\omega_{3, {\bf 0}}^r = 0.35$. We use a set of parameters as $(\tilde{m}_2, T, H_{t_0}, \omega_{3, {\bf 0}})=(3,100,10^{-3}, 0.35)$ for all the lines. In the left plot, the black and magenta lines display the cases $B=1.58$ and $1.01$, respectively. For the right plot, the black and blue lines display the cases $B=1.58$ and $0.76$, respectively.}
\end{figure}

Now we investigate the PNA with the shorter period.  In Fig. \ref{figure6}, we show the temperature ($T$) dependence for the time evolution of PNA. In this regard, the temperature dependence is similar to the one with the longer period. Namely, the amplitude of oscillation becomes larger as the temperature increases. The Fig. \ref{figure7} shows the $B$ dependence. As $B$ parameter decreases, the period of oscillation becomes  longer. However, there were different effects on the amplitude of oscillation. In the left plot, we show the cases that the mass difference $\tilde{m}_2-\tilde{m}_1$ is larger than the frequency $\omega_{3, {\bf 0}}$. 
Since $B^2$, proportional to mass squared difference $\tilde{m}_2^2-\tilde{m}_1^2$, of the magenta line is smaller than that of the black line, the mass difference $\tilde{m}_2-\tilde{m}_1$ of the magenta line is closer to $\omega_{3, {\bf 0}}$. At the beginning ($0<t<22$), the black line of large $B$ has the larger amplitude than that of the magenta line of small $B$. At time $t\sim 22$, the amplitude of the magenta line becomes larger than that of the black line. We also observed that when the mass difference is near to the $\omega_{3, {\bf 0}}$, that is for the case of magenta line, the amplitude grows slowly compared with that of the black line and reaches its maximal value between one and a half period and twice of the period. After taking its maximal value, it slowly decreases. In the right plot, the blue line shows the case that the mass difference $\tilde{m}_2-\tilde{m}_1$ is smaller than the frequency $\omega_{3, {\bf 0}}$. In comparison with the black line, the phase shift of $\frac{\pi}{2}$ was observed in the blue line. The dependence on the parameter $B$ is similar to that of the magenta line. Namely, as $B$ becomes smaller, the amplitude gradually grows at the beginning and slowly decreases at the later time.    

\begin{figure}[t]
	\centering
	\begin{tabular}{c c}
		\includegraphics[width=.5\textwidth]{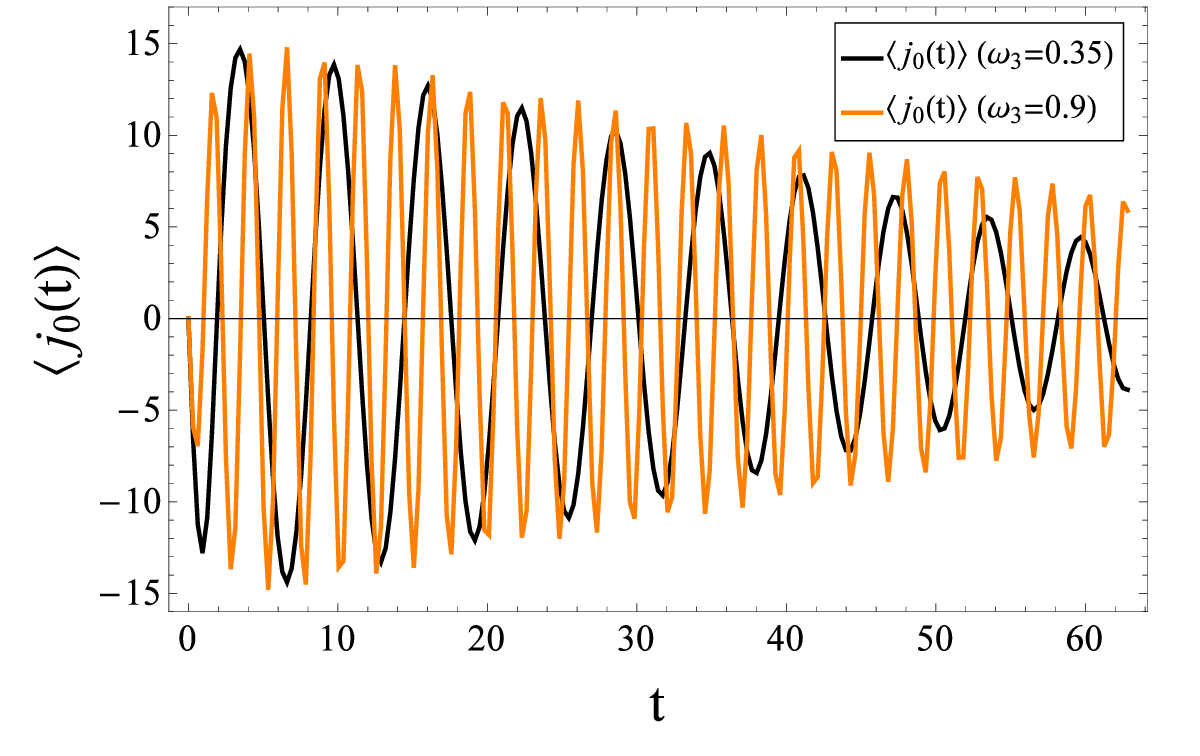}
		&
		\includegraphics[width=.5\textwidth]{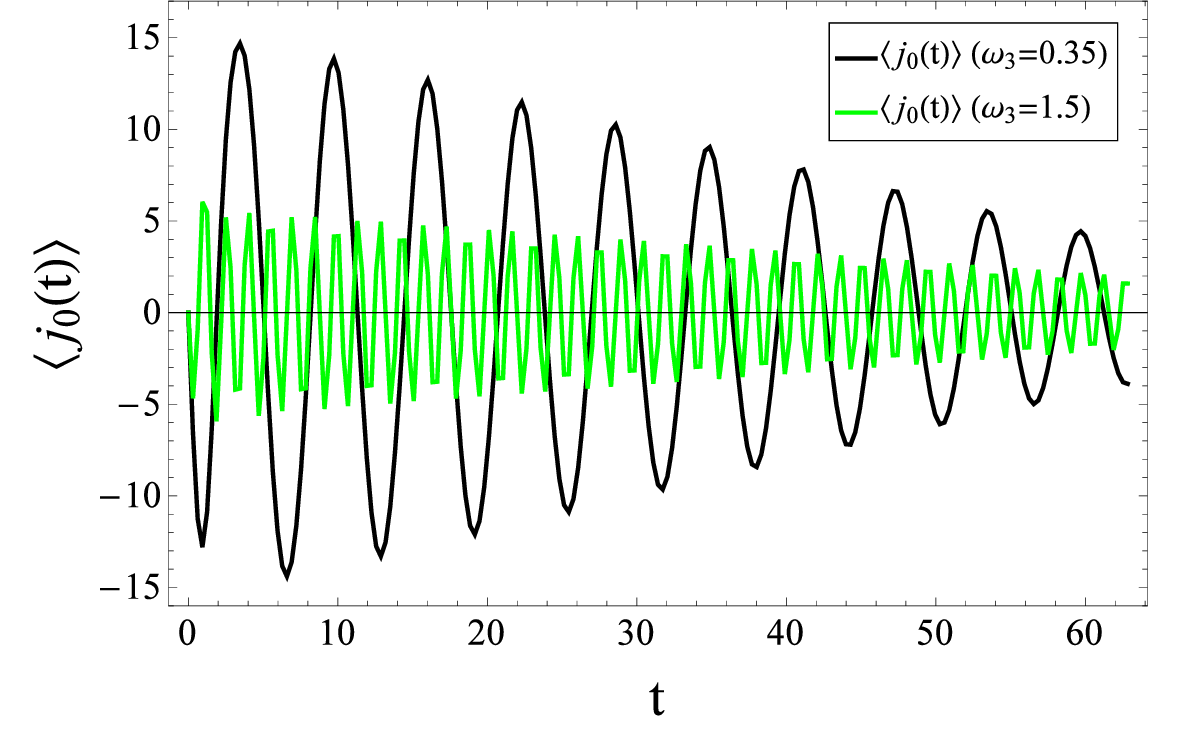}
	\end{tabular}
	\caption{\label{figure8} Dependence on frequency $\omega_{3,{\bf 0}}$ of the time evolution of the PNA. We use the dimensionless time $t = \omega_{3, {\bf 0}}^r (x^0 -t_0)$ for horizontal axis and as a reference angular frequency, we choose $\omega_{3, {\bf 0}}^r = 0.35$. We fix parameters $(\tilde{m}_1,\tilde{m}_2,B,T,H_{t_0})=(2,3,1.58,100,10^{-3})$ for all the lines. The black and orange lines show the cases $\omega_{3, {\bf 0}}=0.35$ and $0.9$, respectively (left plot). The black and green lines show the cases $\omega_{3, {\bf 0}}=0.35$ and $1.5$, respectively (right plot).}
\end{figure}
\begin{figure}[t]
	\centering
	\includegraphics[width=.7\textwidth]{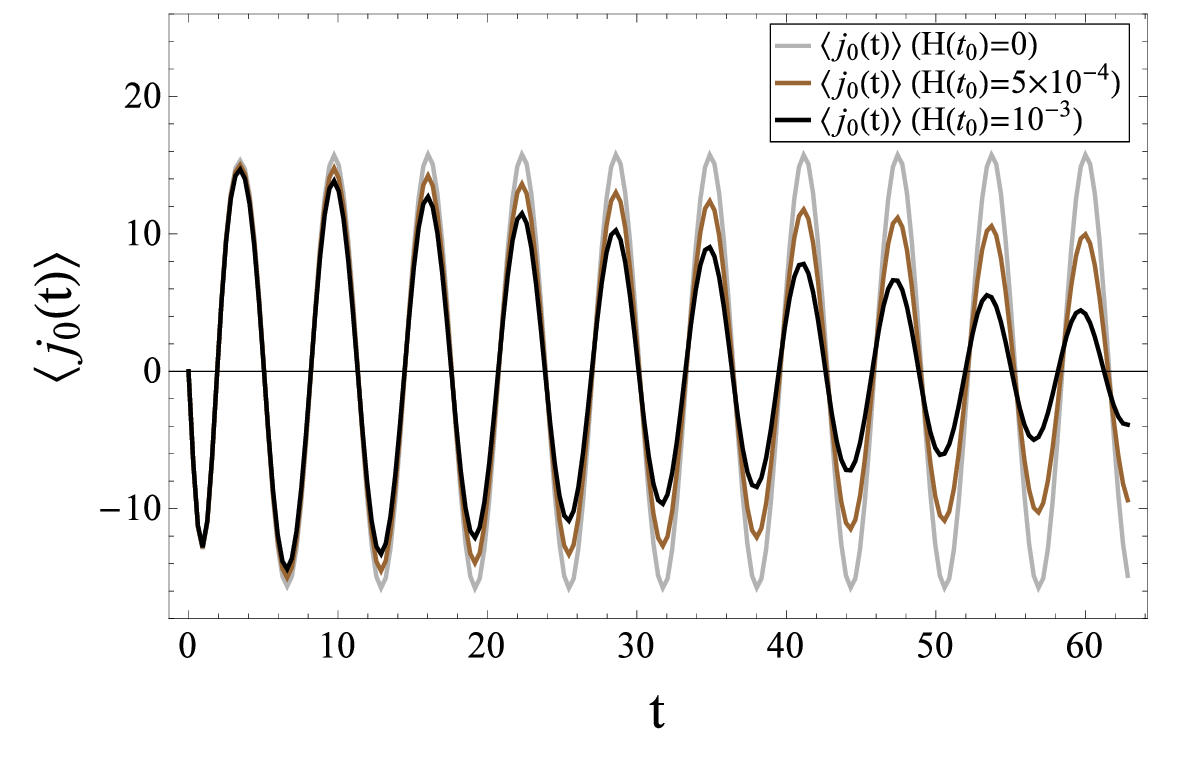}
	\caption{\label{figure9} The expansion rate $H(t_0)$ dependence of the time evolution PNA. In horizontal axes, we use the dimensionless time $t = \omega_{3, {\bf 0}}^r (x^0 -t_0)$ where we choose $\omega_{3, {\bf 0}}^r = 0.35$. We fix parameters as  $(\tilde{m}_1,\tilde{m}_2,B,\omega_{3, {\bf 0}},T)=(2,3,1.58,0.35,100)$ for all the lines. The light gray, brown and black lines display the cases $H(t_0)=0,5\times 10^{-4}$ and $10^{-3}$, respectively.}
\end{figure}
In Fig. \ref{figure8}, we show the dependence on $\omega_{3,{\bf 0}}$. In the left plot, we show the cases that $\omega_{3,{\bf 0}}$'s are smaller than the mass difference as, $\omega_{3, {\bf 0}}^{\text{black}} < \omega_{3, {\bf 0}}^{\text{orange}} < \tilde{m}_2-\tilde{m}_1$.
As $\omega_{3, {\bf 0}}$ increases, the period of the oscillation becomes shorter. There is also a different behavior of the amplitudes as follows. At the beginning, the amplitudes of both black and orange lines increase. After that, in comparison with the black lines, the amplitude of the orange line slowly decreases. In the right plot, the green line shows the case that $\omega_{3, {\bf 0}}$ is larger than the mass difference. We observe that the amplitude of the green line is smaller than that of the black line and the period of the green one is shorter than that of the black one. Figure \ref{figure9} shows the dependence of expansion rate ($H_{t_0}$). In this plot, the PNA gradually decreases as the expansion rate increases.

\subsection{The comparison of two different periods}
\label{sec5C}

\begin{figure}[htbp]
	\centering
	\includegraphics[width=.70\textwidth]{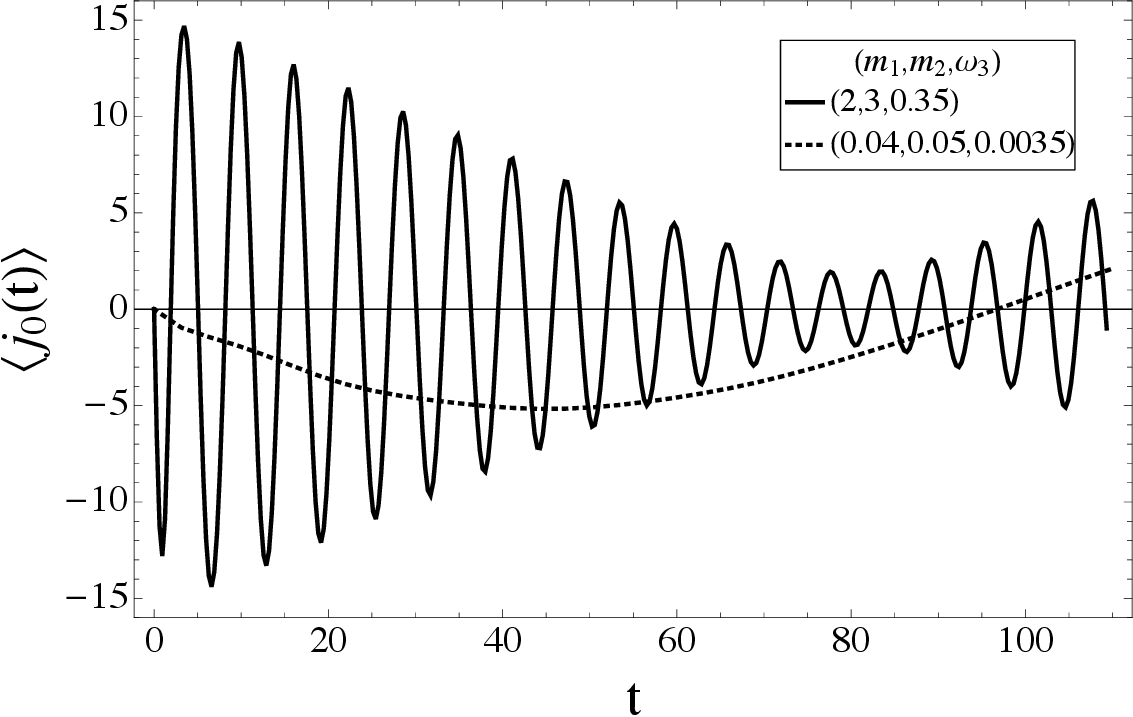}
	\caption{\label{figure10} Comparison two different periods of the time evolution of the PNA. We use the dimensionless time $t = \omega_{3, {\bf 0}}^r (x^0 -t_0)$ for horizontal axis and as a reference we choose $\omega_{3, {\bf 0}}^r = 0.35$. We fix parameters $(T,H_{t_0})=(100,10^{-3})$ for all the lines. The black (shorter period) and black dotted (longer period) lines show the set parameters ($\tilde{m}_1,\tilde{m}_2,B,\omega_{3, {\bf 0}}$) as ($2,3,1.58,0.35$)   and ($0.04,0.05,0.021,0.0035$) , respectively. Noticed that, our approximation will break down after $t=80$.}
\end{figure}

In this subsection, we present a comparison of two different periods of the time evolution of the PNA. In Fig. \ref{figure10}, the black line shows the case of the shorter period and the dotted black line shows the case of the longer one. As can be seen in this figure, the PNA with the shorter period frequently changes the sign and the magnitude also strongly depends on the time. In contrast to the shorter period case, both the sign and magnitude of the longer period case are stable if we restrict to the time range $t=30\sim 60$ in Fig. \ref{figure10}. 

\subsection{The evolution of the PNA with the scale factor of a specific time dependence}
\label{sec5D}

In this subsection, we interpret the numerical simulation in a specific situation. We assume that the  time dependence of the scale factor is given by the one in radiation dominated era. We also specify the unit of the parameters, time and temperature. By doing so, we can clarify implication of the numerical simulation in a more concrete situation. 
	
Specifically, the time dependence of the scale factor is given as follows,	 
\begin{eqnarray}
a (x^0) & = & \sqrt{1 + 2 H_{t_0} (x^0 - t_0)} . \label{sf}
\end{eqnarray}
The above equation is derived as follows. The Einstein's equations without cosmological constant lead to the following equation,
\begin{eqnarray}
\left( \frac{\dot{a}}{a} \right)^2 & = & \frac{8 \pi}{3} G \rho, 
\label{Eeq}
\end{eqnarray}
where $G$ is the Newton's constant. $\rho$ is the energy density for radiation and it is given by,
\begin{eqnarray}
\rho (x^0) & = & \rho_0 a^{- 4} (x^0) , 
\end{eqnarray}
where $\rho_0$ is the initial energy density and we set $a_{t_0}=1$. By setting $x^0 = t_0$ in Eq.(\ref{Eeq}), the initial Hubble parameter is given by,
\begin{eqnarray}
H_{t_0}^2 & = & \frac{8 \pi}{3} G \rho_0 . \label{Ht02}
\end{eqnarray}
Then using Eq.\eqref{Ht02}, Eq.\eqref{Eeq} becomes,
\begin{eqnarray}
\frac{d}{d x^0} \{ a (x^0)^2 \} & = & 2 H_{t_0} . 
\end{eqnarray}
Solving the equation above, one can obtain Eq.\eqref{sf}. 

From the expression in Eq.\eqref{sf}, one needs to specify the unit of the Hubble parameter at $t_0$. Through Eq.\eqref{Ht02}, it is related to the initial energy density $\rho_0$. Assuming $\rho_0$ is given by radiation with an effective degree of freedom $g_{\ast}$ and a temperature $T(t_0)$, one can write $\rho_0$ as follows,
\begin{eqnarray}
	\rho_0  =  g_{\ast} \frac{\pi^2}{30} T^4 (t_0). \label{rho0}
\end{eqnarray}
Hereafter, we assume that the temperature of the radiation $T(t_0)$ is equal to the temperature $T$ in the density operator for the scalar fields. Then one can write the ratio of the initial Hubble parameter and temperature $T$ as follows, 
\begin{eqnarray}
\frac{H_{t_0}}{T}  =  \frac{\pi}{3} \sqrt{\frac{4 \pi g_{\ast}}{5}}
\frac{T}{M_{\text{Pl}}} , \label{Ht0T}
\end{eqnarray}
where $M_{\text{Pl}}$ is the Planck mass, $M_{\text{Pl}}=1.2 \times 10^{19} $ (GeV). Then one can write the temperature $T$ in GeV unit as follows,
\begin{eqnarray}
	T \text{(GeV)}  = \frac{3}{\pi} \sqrt{\frac{5}{4 \pi	g_{\ast}}}  \left( \frac{H_{t_0}}{T} \right)  M_{\text{Pl}} \text{(GeV)}. \label{Ht0Tb}
\end{eqnarray}
In the numerical simulation, the ratio $H_{t_0}/T$ is given. Therefore, for the given ratio and $g_{\ast}$, the temperature $T$ in terms of GeV unit is determined. Then $H_{t_0}$ in GeV unit also becomes,  
\begin{align}
 H_{t_0} \text{(GeV)} = T \text{(GeV)} \times \left( \frac{H_{t_0}}{T} \right) . \label{H0gev}
\end{align}
The masses of the scalar fields $\tilde{m}_{i}$ ($i=1,2,3$) can be also expressed in GeV unit as,
\begin{align}
\tilde{m}_{i} \text{(GeV)} = H_{t_0} \text{(GeV)} \times \left( \frac{\tilde{m}_{i}}{H_{t_0}} \right) , 
\end{align}
where we use the ratios $\frac{\tilde{m}_{i}}{H_{t_0}}$ given in the numerical simulation. 

\begin{table}[t]
	\begin{center}
		\caption{The mass paremeters in GeV unit for both longer and shorter period cases in Fig.\ref{figure10}}
		\begin{tabular}{|c |c|c|} \hline \hline 
			Mass parameter (GeV) & The shorter period & The longer period \\ \hline \hline
			$m_1$ & $2 \times 10^{11}$ & $4 \times 10^9$ \\ \hline 
			$m_2$ & $3 \times 10^{11}$ & $5 \times 10^9$ \\ \hline
			$\omega_{3, {\bf 0}}$  &  $3.5 \times 10^{10}$ & $3.5 \times 10^8$  \\ \hline \hline 
		\end{tabular}
		\label{tb:2}
	\end{center}
\end{table}
As an example, we study the implication of the numerical simulation shown in Fig.\ref{figure10} by specifying the mass parameter in GeV unit. We also determine the unit of time scale. We first determine the temperature in GeV unit using Eq.\eqref{Ht0Tb}. As for the degree of freedom, we can take $g_{\ast}\simeq 100$ which corresponds to the case that all the standard model particles are regarded as radiation. Then, substituting the ratio $H_{t_0}/T=10^{-5}$ adapted in Fig.\ref{figure10} to Eq.\eqref{Ht0Tb} and \eqref{H0gev}, one obtains  $T\sim 10^{13}$ (GeV) and $H(t_0)\sim 10^{8}$ (GeV), respectively. The mass parameters are different between the longer period case (the dotted line) and the shorter period case (the solid line). They are also given in GeV unit shown in Table \ref{tb:2}. 
The time scale $\Delta t=100$ corresponds to $3 \times 10^{- 9} $ (GeV)$^{- 1}$ which is about $2\times 10^{-33}$ (sec).

One can also estimate the size of PNA. Here, we consider the maximum value of the PNA for the longer period case in Fig.\ref{figure10}. We evaluate the ratio of the PNA over entropy density $s$,
\begin{align}
\frac{\left\langle j_0(t\simeq 50)\right\rangle }{s} = - \frac{5 \times 10^{11} (\text{GeV})}{T (\text{GeV})} \times \frac{A_{123}
	(\text{GeV})}{T (\text{GeV})} \times \frac{v_3 (\text{GeV})}{T (\text{GeV})}
\times \frac{45}{2 \pi^2 g_{\ast}}, \label{j0s}
\end{align}
where $s$ is given by,
\begin{equation}
s=g_{\ast} \frac{2 \pi^2}{45} T^3 (\text{GeV}^{3}).
\end{equation}
In Eq.\eqref{j0s}, the first numerical factor, $- 5 \times 10^{11} (\text{GeV})$, is obtained in the following way,
\begin{eqnarray}
   \frac{A_{1 2 3} \left( \text{GeV} \right)}{T
  \left( \text{GeV} \right)} & = & \frac{A_{1 2 3}}{T} , \\
   \frac{v_3 \left( \text{GeV} \right)}{T \left( \text{GeV}
  \right)} & = & \frac{v_3}{T}   , \\
   \frac{- 5 \times 10^{11} \left( \text{GeV} \right)}{T
  \left( \text{GeV} \right)} & = &  \frac{- 5}{T} .
\end{eqnarray}
In the right hand side of the above equations,  $T$ denotes the  temperature in the universal unit of the simulation in Fig.\ref{figure10} where $T=100$ is used. In the left hand side, $T$ denotes the corresponding temperature in GeV unit and it is  $T=10^{13}$ (GeV). 
Substituting the temperature $T=10^{13}$ (GeV) into Eq.\eqref{j0s}, one obtains,
\begin{align}
\frac{\left\langle j_0(t\simeq 50)\right\rangle }{s} = - 1 \times 10^{- 11} \times \left( \frac{A_{123} (\text{GeV})}{10^8 (\text{GeV})}
\right) \times \left( \frac{v_3 (\text{GeV})}{10^{10} (\text{GeV})} \right).
\end{align}
From the equation above, we can achieve the ratio as $10^{-10}$ by taking $A_{123}=10^{8}$ (GeV) and $v_3=10^{11}$ (GeV).


\section{Discussion and conclusion}
\label{sec6}

In this paper, we developed a new mechanism for generating the PNA. This mechanism is realized with the specific model Lagrangian which we have proposed. The model includes a complex scalar. The PNA is associated with U(1) charge of the complex scalar. In addition, we introduce a neutral scalar which interacts with the complex scalar. The U(1) charge is not conserved due to particle number violating interaction. As an another source of particle number violation, the U(1) symmetry breaking mass term for the complex scalar is introduced. The initial value for the condensation of the neutral scalar is non-zero. Using 2PI formalism and specifying the initial condition with density operator, the time-dependent PNA is obtained. 
To include the effect of the time dependence of the scale factor, we approximate it up to the first order of Hubble parameter. 

The results show that the PNA depends on the interaction coupling $A_{123}$ and the initial value of the condensation of the neutral scalar $\hat{\varphi}_{3, t_0}$. It also depends on the mass squared difference of two real scalars which originally form a complex scalar. We found that the interaction coupling $A_{123}$ and the mass squared difference play a key role to give rise to non-vanishing PNA. Even if the initial value of the neutral scalar is non-zero, in the vanishing limit of interaction terms and the mass squared difference, the PNA will vanish. Another important finding is that the contribution to the PNA is divided into four types. The constant scale factor which is the zeroth order of Hubble parameter is the leading contribution. The rests which are the first order term contribute according to their origins. Those are summarized in Table \ref{tb:3}. 
\begin{table}[t]
	\begin{center}
		\caption{The classification of $o(H_{t_0})$ contributions to the PNA}
		\begin{tabular}{|l|p{10cm}|} \hline \hline 
			The effect & The origin \\ \hline \hline
			Dilution & The increase of volume of the universe due to expansion, $\frac{1}{a(x^0)^{3}}-\frac{1}{a_{t_0}^{3}}$\\ \hline 
			Freezing interaction & The decrease of the strength of the cubic interaction $\hat{A}$ as $\hat{A}_{123}-A_{123} $.\\ \hline
			Redshift & The effective energy of particle as indicated in Eq.\eqref{Ome}, $\frac{{\bf k}^{2}}{a (x^0)^2} + \bar{m}_i^2 (x^0)$. \\ \hline \hline 
		\end{tabular}
		\label{tb:3}
	\end{center}
\end{table} 

We have numerically calculated time evolution of the PNA and have investigated its dependence on the temperature, parameter $B$, the angular frequency $\omega_{3, {\bf 0}}$ and the expansion rate of the universe. Starting with the null PNA at the initial time, it is generated by particle number violating interaction. Once the non-zero PNA is generated, it starts to oscillate. The amplitude decreases as the time gets larger. The dumping rate of the amplitude increases as the Hubble parameter becomes larger. The period of the oscillation depends on the angular frequency $\omega_{3, {\bf 0}}$ and the parameter $B$. The former determines the oscillation period for the condensation of the neutral scalar. The latter determines the mass difference of $\phi_1$ and $\phi_2$. 
In the simulation, we focus on the two cases for the oscillation period, one of which corresponds to the longer period case and the other is the shorter period case. 
The longer period is about half of the Hubble time $(1/H_{t_0})$ and the shorter period is one percent of the Hubble time.
The set of parameters $(\omega_{3, {\bf 0}},B)$ which corresponds to the longer period is typically one percent of the values for the shorter period. In both cases, the amplitude gets larger as the temperature increases. For the longer period, as parameter $B$ becomes larger, the amplitude increases. For the shorter period, in order to have large amplitude, the parameter $B$ is taken so that the mass difference $\tilde{m}_2 - \tilde{m}_1$ is near to $\omega_{3, {\bf 0}}$. In other words, when the resonance condition $\omega_{3, {\bf 0}}\simeq \tilde{m}_2 - \tilde{m}_1$ is satisfied, the amplitude becomes large. For the longer period case,
as the angular frequency $\omega_{3, {\bf 0}}$ decreases, the amplitude becomes large.

To show how the mechanism can be applied to a realistic situation,
we study the simulated results for radiation dominated era when the degree of freedom of light particles is assumed to be $g_{\ast}\simeq O(100)$. 
Then when the initial temperature
of the scalar fields is the same as that of the light particles, 
the simulation with $\frac{H_{t_0}}{T}=10^{-5}$  corresponds to the case that 
the temperature of the
universe is $10^{13}$ (GeV) which is slightly lower than GUT scale $\sim
10^{16}$ (GeV) \cite{AMALDI1991447,Kazakov2000}.  The masses of the
scalar fields in Fig.\ref{figure10} are different between the shorter period case and the longer period case as shown in Table \ref{tb:2}.
In the shorter period case, the mass spectrum of the scalar ranges from
 $10^{10}$ (GeV) to  $10^{11}$ (GeV) while
for the longer period case, it is lower than that of the shorter period case by two orders of magnitude.
For the longer period case,  the maximum asymmetry is achieved  at $10^{-33} $ (sec) after the
initial time. For shorter period, it is achieved at about $10^{-34}$ (sec). 
We have estimated the ratio of the PNA over entropy density by substituting the numerical values of the coupling constant ($A_{123}$) and the initial expectation value ($v_3$). 

Compared with the previous works \cite{Hotta2014,PhysRevD.95.095014,cirelli2012,tulin2012}, instead of assuming the non-zero PNA at the initial time, the PNA is created through interactions.  These interactions have the following  unique feature; namely, the interaction between the complex scalars and oscillating condensation of a neutral scalar leads to the PNA. In our work, by assuming the initial condensation of the neutral scalar is away from the equilibrium point, the condensation starts to oscillate. In the expression of the amplitude of PNA, one finds that it is proportional to the CP violating coupling between the scalars and the condensation, the initial condensation of the neutral scalars, and mass difference between mass eigenstates of the two neutral scalars which are originally introduced  as a complex scalar with the particle number violating mass and curvature terms. One of the distinctive feature of the present mechanism from the one which utilizes the PNA created through the heavy particle decays is as follows. In the mechanism which utilizes the heavy particle decays, the temperature must be high enough so that it once brings the heavy particle to the state of the thermal equilibrium. Therefore the temperature of the universe  at reheating era must be as high as the mass of the heavy particle. In contrast to this class of the models, the present model is not restricted by such condition. In place of the condition, the initial condensation must be large enough to explain the asymmetry. 
  
In our model, even for the longer period case, the oscillation period is shorter than the Hubble time $(1/H_{t_0})$. It implies that one of Sakharov conditions for BAU, namely 
non-equilibrium condition is not satisfied. In this respect, we expect that due to the finite life time of the condensation of the neutral scalar, the interaction between the condensation and complex scalars will vanish and eventually the oscillation of the PNA may terminate. The detailed study will be given in the future work. 
The relation between the PNA and the observed BAU should be also studied. 
In particular, we need to consider the mechanism how the created PNA is transferred to the observed BAU.

\begin{acknowledgments}
H.T. and K.I.N. would like to thank the theoretical particle physics group in Hiroshima University for its kind hospitality. A.S.A. would like to thank Summer Institute (SI) 2017 for stimulating discussion where the part of this work has been completed. This work is supported by JSPS KAKENHI Grant Number JP17K05418 (T.M.) and supported in part by JSPS Grant-in-Aid for Scientific Research for Young Scientists (B) 26800151 (K.I.N.). 
\end{acknowledgments}

\appendix

\section{The solution of SDEs for both Green's function and field}
\label{sec_apply}

\subsection{The general solutions of SDEs}
In this subsection, we provide the general solution of SDEs. Let us introduce the following differential equation for a field $\varphi$,
\begin{eqnarray}
\left[\frac{\partial^{2}}{{\partial x^{0}}^{2}}+m^{2}(x^0)\right]\varphi(x^{0}) = S(x^{0}) + E(t_0)\ \delta(x^{0}-t_{0})  + F(T)\  \delta (x^{0}-T), \label{modiffeq}
\end{eqnarray}
where $S$ is an arbitrary function of time $x^0$ and we will find the solution within the time range from $x^0=t_0$ to $x^0=T$. At the boundaries $x^0=t_0$ and $x^0=T$, we introduce the source terms of the form of delta function. The strength of the delta function is denoted as $E(t_0)$ and $F(T)$, respectively. 
One may assume that field vanishes at $x^0 < t_0$, 
\begin{eqnarray}
\varphi(x^{0}<t_0) =0, \quad \frac{\partial \varphi(x^{0})}{\partial x^{0}}\bigg|_{x^{0}<t_0} = 0.   \label{phi0} 
\end{eqnarray}
One integrates Eq.\eqref{modiffeq} with respect to time $x^0$ from $t_0-\epsilon $ to $ t_0+\epsilon$ and obtains initial condition for the first derivative of the field as,
\begin{eqnarray}
\frac{\partial \varphi(x^{0})}{\partial x^{0}}\bigg|_{t_0^{+}}=  E(t_0), \label{intcond}
\end{eqnarray}
where we have used the above assumption and taken limit $\epsilon\rightarrow 0$. $t_0^{+}$ denotes $t_0+0$.

The method of variation of constants has been employed to determine the solution of Eq.\eqref{modiffeq} and it is written as,
\begin{eqnarray}
\varphi(x^{0}) & = & C_{1}(x^{0})\ f(x^{0}) + C_{2}(x^{0})\ g(x^{0}), \label{modsol}
\end{eqnarray}
where  $f(x^0)$ and $g(x^0)$ are two linear independent solutions of the following homogeneous equation,
\begin{eqnarray}
\left[\frac{\partial^{2}}{(\partial x^{0})^{2}}+m^{2}(x^0)\right]\varphi_{\text{homo}}(x^{0})  =  0. 
\end{eqnarray}
Firstly, the following condition is imposed,
\begin{eqnarray}
 \dot{C}_{1}(x^{0})f(x^{0}) + \dot{C}_{2}(x^{0})g(x^{0})=0. \label{con1}
\end{eqnarray}
Eq.\eqref{modsol} becomes the solution of the differential equation Eq.\eqref{modiffeq} when $C_{i}(x^{0})$ satisfies another condition, 
\begin{eqnarray}
 \dot{C}_{1}(x^{0})\dot{f}(x^{0}) + \dot{C}_{2}(x^{0})\dot{g}(x^{0}) = S(x^{0}) + F(T)  \delta (x^{0}-T). \label{con2}
\end{eqnarray}
The remaining task is to find $C_{i}(x^{0})$ which satisfy the conditions in Eqs.\eqref{con1} and \eqref{con2}. It is convenient to write these conditions in matrix form as,
\begin{eqnarray}
\left( \begin{array}{c}
\dot{C}_{1}(x^{0}) \\
\dot{C}_{2}(x^{0})
\end{array} \right)
=\frac{1}{W}
\left( \begin{array}{c c}
g(x^{0}) & -\dot{g}(x^{0}) \\
-f(x^{0}) & \dot{f}(x^{0})
\end{array} \right) 
\left( \begin{array}{c}
S(x^{0}) + F(T)  \delta (x^{0}-T)  \\
0
\end{array} \right), \label{matCi}
\end{eqnarray}
where we have defined,
\begin{eqnarray}
W=\dot{f}(x^{0})g(x^{0})-f(x^{0})\dot{g}(x^{0}). \label{defW}
\end{eqnarray}
and it is a constant with respect to time. 
Integrating Eq.\eqref{matCi} with respect to time from $t_0^{+}$ to $x^{0}$, one obtains,
\begin{align}
C_{1}(x^{0}) &=  \int_{t_0^{+}}^{x^{0}} \frac{1}{W}\left(g(t)\left\{S(t) + F(t)\ \delta(t-T)\right\}\right)dt  + C_{1}(t_0^{+}) \\
C_{2}(x^{0}) &=  -\int_{t_0^{+}}^{x^{0}} \frac{1}{W}\left(f(t)\left\{S(t) + F(t)\ \delta(t-T)\right\}\right)dt + C_{2}(t_0^{+})
\end{align}
Therefore Eq.\eqref{modsol} becomes,
\begin{align}
\varphi(x^{0})  =& \frac{1}{W}\int_{t_0^{+}}^{x^{0}} \left[f(x^0)g(t)-g(x^0)f(t)\right] \left\{S(t) + F(t)\ \delta(t-T)\right\} dt  \nonumber \\ & + C_{1}(t_0^{+}) f(x^0)  + C_{2}(t_0^{+}) g(x^0) \label{varx0}
\end{align}
In this regard, it enables us to define,
\begin{eqnarray}
\bar{K}[x^{0},y^{0}] := \frac{1}{W}\left[f(x^0)g(y^0)-g(x^0)f(y^0)\right]. \label{defKbar}
\end{eqnarray}
Then Eq.\eqref{varx0} alters into
\begin{eqnarray}
\varphi(x^{0}) = \int_{t_0^{+}}^{x^{0}} \bar{K}[x^{0},t] \left\{S(t) + F(t)\ \delta(t-T)\right\} dt + C_{1}(t_0^{+}) f(x^0)
+ C_{2}(t_0^{+}) g(x^0). \label{modsol01}
\end{eqnarray}
To determine $C_{i}(t_0^{+})$, it is required the first derivative of field,
\begin{align}
\frac{\partial}{\partial x^0}\varphi(x^{0})  =& \int_{t_0^{+}}^{x^{0}} \left(\frac{\partial \bar{K}[x^{0},t]}{\partial x^{0}}\right) \left\{S(t) + F(t)\ \delta(t-T)\right\} dt  \nonumber \\ &+ C_{1}(t_0^{+}) \dot{f}(x^0)  + C_{2}(t_0^{+}) \dot{g}(x^0).
\end{align}
Consequently, using definition Eq.\eqref{defKbar}, the term which includes $S(t_0)$ vanishes after integration. From now on, we will denote $\dot{\varphi}(x^{0})$ as $\frac{\partial \varphi(x^{0})}{\partial x^0}$. 

Let us now write $C_{i}(t_0^{+})$ in terms of $\varphi(t_0)$ and $\dot{\varphi}(t_0)$. We consider the following equations,
\begin{eqnarray}
\varphi(t_0^{+}) & = & C_{1}(t_0^{+})\ f(t_0^{+}) + C_{2}(t_0^{+})\ g(t_0^{+}),  \\
\dot{\varphi}(t_0^{+}) & = & C_{1}(t_0^{+})\ \dot{f}(t_0^{+}) + C_{2}(t_0^{+})\ \dot{g}(t_0^{+}). 
\end{eqnarray}
Next, one can write $C_{i}(t_0^{+})$ in matrix form as,
\begin{eqnarray}
\left( \begin{array}{c}
C_{1}(t_0^{+}) \\
C_{2}(t_0^{+})
\end{array} \right)
=-\frac{1}{W}
\left( \begin{array}{c c}
\dot{g}(t_0^{+}) & -g(t_0^{+}) \\
-\dot{f}(t_0^{+}) & f(t_0^{+})
\end{array} \right)
\left( \begin{array}{c}
\varphi(t_0^{+})  \\
\dot{\varphi}(t_0^{+})
\end{array} \right).
\end{eqnarray}
Thus one substitutes these $C_{i}(t_0^{+})$ into Eq.\eqref{modsol01} and obtains,
\begin{align}
\varphi(x^{0})  =& \int_{t_0^{+}}^{x^{0}} \bar{K}[x^{0},t]S(t) dt + \bar{K}[x^{0},T]F(T) \theta(x^{0}-T) - \bar{K'}[x^0,t_0] \varphi(t_0) \nonumber \\
&+ \bar{K}[x^0,t_0] E(t_0), \label{modsol02}
\end{align}
where we have used the initial condition in Eq.\eqref{intcond} and $\bar{K'}[x^0,y^0]$ is defined as,
\begin{align}
\bar{K'}[x^0,y^0]:=\frac{\partial \bar{K}[x^0,y^0]}{\partial y^0}. \label{kprime_bar}
\end{align}
In the next subsection, we will use the obtained solution to provide the solution of SDEs.
 
\subsection{The SDEs for the field}
\label{App2}
Next we move to consider the SDEs for the field in Eqs.\eqref{diffphi01}. It is rewritten as, 
\begin{align}  
	\left[\frac{\partial^{2}}{{\partial x^{0}}^{2}} + \Omega^{2}_{i,x^{0},{\bf k}=0}\right] \hat{\varphi}^{d}_{i,x^{0}} = 
	S_{i,x^0}^{d} , \label{diffphi02} 
\end{align}
where we have defined $S_{i,x^0}^{d} $ as,
\begin{align}
	S_{i,x^0}^{d} &:= c^{da} D_{a b c}  \hat{A}_{i j k} (x^0) \{ \hat{\varphi}_j^b (x^0)
	\hat{\varphi}_k^c (x^0) + \hat{G}^{bc}_{jk} (x, x) \}  
	.	
	\label{defSd} 
\end{align}
To solve Eq.\eqref{diffphi02}, one first sets $E(t_0)=0$ and $ F(T)=0 $ in Eq.\eqref{modiffeq}. Then, the general differential equation is similar to the SDEs for the field. The SDEs of the field in the form of integral equation are given by,
\begin{eqnarray}
	\hat{\varphi}_{i,x^0}^{d} &=& - \bar{K}_{i,x^{0} t_0}'  \hat{\varphi}_{i,t_0}^{d} +\int_{t_0^+}^{x^0} \bar{K}_{i,x^{0} t} S_{i, t}^{d}\ dt
	, \label{solphi2} \\
	\hat{\varphi}_{i,x^0}^{d,\text{free}} &=& - \bar{K}_{i,x^{0} t_0}'  \hat{\varphi}_{i,t_0}^{d} \label{def_field_free} ,\\
	\hat{\varphi}_{i,x^0}^{d,\text{int}} &=& \int_{t_0^+}^{x^0} \bar{K}_{i,x^{0} t} S_{i, t}^{d}\ dt .\label{def_field_int}
\end{eqnarray}
By using the above equations and keeping the solutions up to the first order of the cubic interaction, one obtains Eqs.\eqref{phigen}-\eqref{phihatOA}.

\subsection{The SDEs for the Green's function}
\label{App3}

In this subsection, we consider the SDEs for Green's function in Eqs.\eqref{diffGx01} and \eqref{diffGy01}. They are simply rewritten as,
\begin{align}  
\left[\frac{\stackrel{\rightarrow}{\partial^{2}}}{\partial {x^{0}}^2} + \Omega^{2}_{i,x^{0},{\bf k}}\right] \hat{G}^{ab}_{ij,x^{0}y^{0},{\bf k}} &= 
Q^{ab}_{ij,x^{0}y^{0},{\bf k}} + E^{ac}_{ik,{\bf k}} \hat{G}^{cb}_{kj,t_{0} y^{0},{\bf k}} \delta_{t_{0} x^{0}} 
+ F_{ij}^{ab} \delta_{x^{0}y^{0}}, \label{diffGx02} \\
\hat{G}^{ab}_{ij,x^{0}y^{0},{\bf k}} \left[\frac{\stackrel{\leftarrow}{\partial^{2}}}{\partial {y^{0}}^{2}} + \Omega^{2}_{i,y^{0},{\bf k}}\right]  &= 
R^{ab}_{ij,x^{0}y^{0},{\bf k}} + \hat{G}^{ac}_{ik,x^{0}t_{0},{\bf k}} E^{Tcb}_{kj,{\bf k}} \delta_{t_{0} y^{0}} 
+ F_{ij}^{ab} \delta_{x^{0}y^{0}}, \label{diffGy02} 
\end{align}
where we have defined,
\begin{align}
F_{ij}^{ab} := & -i\delta_{ij}\frac{c^{ab}}{a_{t_{0}}^{3}} \label{def_Fac}, \\
Q^{a b}_{ij,x^0 y^0} ({\bf k}) := & 2 c^{a d} D_{d c e}  \hat{A}_{i k l, x^0} 
\hat{\varphi}^e_{l, x^0} 
\hat{G}^{cb}_{kj, x^0 y^0} ({\bf k}),
\label{defQab}\\
R^{a b}_{ij,x^0 y^0} ({\bf k}) := & 2 \hat{G}^{ac}_{ik, x^0 y^0} ({\bf k}) D_{c e f}  \hat{A}_{k j l, y^0}
\hat{\varphi}^f_{l, y^0} 
	c^{e b},
\label{defRab} 
\end{align}
and $E^{ac}_{ik,{\bf k}}$ is given in Eq.\eqref{defE}. 

\begin{figure}[t]
	\centering
	\includegraphics[width=.50\textwidth]{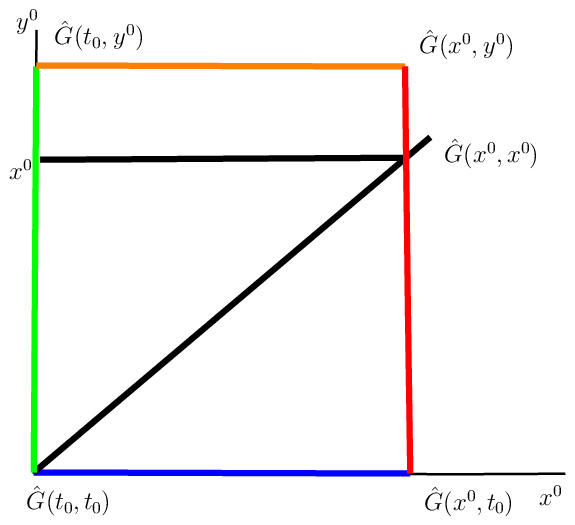}
	\caption{Two paths to obtain $\hat{G}(x^0, y^0)$. We show the paths for the case $x^0<y^0$.}
	\label{figure11}
\end{figure}
In the following, we will obtain SDEs for Green's functions at $(x^0,y^0)$ in the form of integral equation. Starting with the initial condition for Green's function at $(t_0,t_0)$, we obtain two expressions for the Green's function at $(x^0,y^0)$. The two expressions correspond to two paths shown in Figure \ref{figure11} which are used to integrate the differential equation in Eqs.\eqref{diffGx02} and \eqref{diffGy02}. They are given by\footnote{Dot multiplication describes matrix product corresponding their indices.},
\begin{eqnarray}
\hat{G}^{br}_{x^0 y^0} &=& \left\{\bar{K}_{x^0 t_0} \cdot \left(E \cdot \hat{G}_{t_0 t_0}+F\right) 
- \bar{K}'_{x^0 t_0}\cdot \hat{G}_{t_0 t_0} \right\} \cdot \left(E^{T} \cdot \bar{K}_{y^0 t_0}- \bar{K}'_{y^0 t_0} \right) \nonumber \\
& & + \theta(y^0 - x^0) F \cdot \bar{K}_{y^0 x^0} + \int^{y^0}_{t_0^{+}} R_{x^0 t} \cdot \bar{K}_{y^0 t}\ dt \nonumber \\
& & + \int^{x^0}_{t_0^{+}} \bar{K}_{x^0 t} \cdot Q_{t t_0}\ dt  \cdot \left(E^{T} \cdot \bar{K}_{y^0 t_0}- \bar{K}'_{y^0 t_0}\right) , \label{Ghat_b_tot}
\end{eqnarray}
\begin{eqnarray}
\hat{G}_{x^0 y^0}^{go} &=& \left(\bar{K}_{x^0 t_0}\cdot E - \bar{K}'_{x^0 t_0}\right) \cdot 
\left\{\left(\hat{G}_{t_0 t_0}\cdot E^{T} +F\right) \cdot \bar{K}_{y^0 t_0} - \hat{G}_{t_0 t_0} \cdot \bar{K}'_{y^0 t_0} \right\} \nonumber \\
& & + \theta(x^0 - y^0) \bar{K}_{x^0 y^0} \cdot F + \int_{t_0^{+}}^{x^0} \bar{K}_{x^0 t} \cdot Q_{t y^0}\ dt \nonumber \\
& & + \left(\bar{K}_{x^0 t_0}\cdot E - \bar{K}'_{x^0 t_0} \right) \cdot \int^{y^0}_{t_0^{+}} R_{0 t} \cdot \bar{K}_{y^0 t}\ dt , \label{Ghat_a_tot}
\end{eqnarray} 
where the upper indices \lq\lq $ br $\rq\rq and \lq\lq $ go $\rq\rq denote the blue red path and the green orange path, respectively.

Now let us explain how one can derive Eq.\eqref{Ghat_b_tot}. The steps are summarized below.
\begin{itemize}
	\item We first consider the differential equation in Eq.\eqref{diffGx02} which Green's function $\hat{G}(t,t_0)$ on the blue line ($ t_0 \leq t\leq x^0$) satisfies. Using the solution of the general differential equation in Eq.\eqref{modsol02}, we obtain the expression,  
	\begin{eqnarray}
	\hat{G}_{x^0 t_0} = \int_{t_0^{+}}^{x^0} \bar{K}_{x^0 t} \cdot Q_{t t_0}\ dt 
	+  \bar{K}_{x^0 t_0}\cdot \left(E\cdot \hat{G}_{t_0 t_0}+F\right) - \bar{K}'_{x^0 t_0} \cdot \hat{G}_{t_0 t_0},  \label{solG_x0} 
	\end{eqnarray}	 
	where $\hat{G}_{t_0t_0}$ denotes the initial condition. 
	
	\item Next we consider the differential equation in Eq.\eqref{diffGy02} which Green's function $\hat{G}(x^0,t)$ on the red line ($ t_0 \leq t\leq y^0 $) satisfies. Using the solution of the general differential equation in Eq.\eqref{modsol02}, we obtain the expression,
	\begin{eqnarray}
	\hat{G}_{x^0 y^0} &=& \int_{t_0^{+}}^{y^0} R_{x^0 t} \cdot \bar{K}_{y^0 t} \ dt
	+ \hat{G}_{x^0 t_0} \cdot \left(E \cdot \bar{K}_{y^0 t_0} - \bar{K}'_{y^0 t_0} \right)  
	+ \theta(y^0 - x^0) F \cdot \bar{K}_{y^0 x^0},\nonumber \\  \label{solG_x0y0b}
	\end{eqnarray}  
	where $\hat{G}_{x^0t_0}$ denotes the initial condition.
	\item Substituting Eq.\eqref{solG_x0} to Eq.\eqref{solG_x0y0b}, we obtain Eq.\eqref{Ghat_b_tot}.
\end{itemize}
The Eq.\eqref{Ghat_a_tot} is obtained through the steps similar to the above. The difference is as follows. We first integrate the differential equation on the green line with the initial condition at $(t_0,t_0)$ and obtain the expression for $\hat{G}_{t_0y^0}$. Using it as the initial condition, we integrate the differential equation on the orange line and obtain the expression in Eq.\eqref{Ghat_a_tot}.

\subsection{The derivation of free part for Green's function and its path independence}
\label{app_a4}
In the following subsection, we derive the free parts of Green's function which are the zeroth order of cubic interaction. From Eqs.\eqref{Ghat_b_tot} and \eqref{Ghat_a_tot}, we can write them respectively as, 
\begin{eqnarray}
\hat{G}^{br, \text{free}}_{x^0 y^0} &=& \left\{\bar{K}_{x^0 t_0} \cdot \left(E \cdot \hat{G}_{t_0 t_0}+F\right) 
- \bar{K}'_{x^0 t_0}\cdot \hat{G}_{t_0 t_0} \right\} \cdot \left(E^{T} \cdot \bar{K}_{y^0 t_0}- \bar{K}'_{y^0 t_0} \right) \nonumber \\
& & + \theta(y^0 - x^0) F \cdot \bar{K}_{y^0 x^0} , \label{Ghat_b_free} \\
\hat{G}_{x^0 y^0}^{go, \text{free}} &=& \left(\bar{K}_{x^0 t_0}\cdot E - \bar{K}'_{x^0 t_0}\right) \cdot 
\left\{\left(\hat{G}_{t_0 t_0}\cdot E^{T} +F\right) \cdot \bar{K}_{y^0 t_0} - \hat{G}_{t_0 t_0} \cdot \bar{K}'_{y^0 t_0} \right\} \nonumber \\
& & + \theta(x^0 - y^0) \bar{K}_{x^0 y^0} \cdot F . \label{Ghat_a_free}
\end{eqnarray} 
Both of the above expressions satisfy the differential equations in which we turn off the interaction part, namely,
\begin{align}  
\left[\frac{\stackrel{\rightarrow}{\partial^{2}}}{\partial {x^{0}}^2} + \Omega^{2}_{i,x^{0},{\bf k}}\right] \hat{G}^{ab,\text{free}}_{ij,x^{0}y^{0},{\bf k}} &= 
E^{ac}_{ik,{\bf k}} \hat{G}^{cb,\text{free}}_{kj,t_{0} y^{0},{\bf k}} \delta_{t_{0} x^{0}} 
+ F_{ij}^{ab} \delta_{x^{0}y^{0}}, \label{diffGx02_free} \\
\hat{G}^{ab,\text{free}}_{ij,x^{0}y^{0},{\bf k}} \left[\frac{\stackrel{\leftarrow}{\partial^{2}}}{\partial {y^{0}}^{2}} + \Omega^{2}_{i,y^{0},{\bf k}}\right]  &= 
\hat{G}^{ac,\text{free}}_{ik,x^{0}t_{0},{\bf k} } E^{Tcb}_{kj,{\bf k}} \delta_{t_{0} y^{0}} 
+ F_{ij}^{ab} \delta_{x^{0}y^{0}}, \label{diffGy02_free} 
\end{align}
Below we show both expressions in Eqs.\eqref{Ghat_b_free} and \eqref{Ghat_a_free} lead to a single expression. Using Eqs.\eqref{eq:incon},\eqref{defE} and \eqref{def_Fac}, we can rewrite them as follows,
\begin{eqnarray}
\hat{G}^{ab,br,\text{free}}_{ij,x^0 y^0}({\bf k})  &=& \frac{\delta_{ij}}{2\omega_{i}({\bf k})a^{3}_{t_0}}\left[\frac{\sinh \beta \omega_{i}({\bf k})}{\cosh \beta \omega_{i}({\bf k}) - 1}\right] \left( \begin{array}{c c}
1 & 1 \\
1 & 1
\end{array} \right)^{ab}\nonumber \\
& &\times \left[\bar{K}'_{i,x^0 t_0}  \bar{K}'_{i,y^0 t_0}    + \omega^{2}_{i}({\bf k}) \bar{K}_{i,x^0 t_0}\bar{K}_{i,y^0 t_0}\right]
\nonumber \\
& &+ \frac{i \delta_{ij}}{2 a^{3}_{t_0}} \left(\bar{K}'_{i,x^0 t_0} \bar{K}_{i,y^0 t_0} - \bar{K}_{i,x^0 t_0} \bar{K}'_{i,y^0 t_0}\right) 
\begin{array}{l l}
\left( \begin{array}{c c}
-1 & 1 \\
-1 & 1
\end{array} \right)^{ab}
\end{array}
 \nonumber \\
& &- \frac{i \delta_{ij}}{2 a^{3}_{t_0}} \theta(y^0 - x^0) \bar{K}_{i,y^0 x^0}c^{ab} . \label{Ghat_tot_free_br}
\end{eqnarray}
\begin{eqnarray}
\hat{G}^{ab,go,\text{free}}_{ij,x^0 y^0}({\bf k})  &=& \frac{\delta_{ij}}{2\omega_{i}({\bf k})a^{3}_{t_0}}\left[\frac{\sinh \beta \omega_{i}({\bf k})}{\cosh \beta \omega_{i}({\bf k}) - 1}\right] \left( \begin{array}{c c}
1 & 1 \\
1 & 1
\end{array} \right)^{ab}\nonumber \\
& &\times \left[ \bar{K}'_{i,x^0 t_0}  \bar{K}'_{i,y^0 t_0}   + \omega^{2}_{i}({\bf k}) \bar{K}_{i,x^0 t_0}\bar{K}_{i,y^0 t_0}\right]
\nonumber \\
& &+ \frac{i \delta_{ij}}{2 a^{3}_{t_0}} \left(\bar{K}'_{i,x^0 t_0} \bar{K}_{i,y^0 t_0} - \bar{K}_{i,x^0 t_0} \bar{K}'_{i,y^0 t_0}\right) 
\begin{array}{l l}
\left( \begin{array}{c c}
1 & 1 \\
-1 & -1
\end{array} \right)^{ab} 
\end{array}
\nonumber \\
& &- \frac{i \delta_{ij}}{2 a^{3}_{t_0}} \theta(x^0 - y^0) \bar{K}_{i,x^0 y^0} c^{ab}
, \label{Ghat_tot_free_go}
\end{eqnarray}
By using the following relation, 
\begin{align}
\bar{K}'_{i,x^0 t_0} \bar{K}_{i,y^0 t_0} - \bar{K}_{i,x^0 t_0} \bar{K}'_{i,y^0 t_0} = \bar{K}_{i,x^0 y^0}, 
\end{align}
we can show that two expressions are identical to each other. Therefore, they can be summarized into a single expression which is Eq.\eqref{Ghatfree}. 

\subsection{The interaction part of Green's function and its path independence}
\label{app_a5}

Now we move to consider the interaction parts of Green's function. From Eqs.\eqref{Ghat_b_tot} and \eqref{Ghat_a_tot}, we can temporary define them respectively as,
\begin{eqnarray}
	\hat{G}^{br,\text{int}}_{x^0 y^0} &:=& \int^{y^0}_{t_0^{+}} R_{x^0 t} \cdot \bar{K}_{y^0 t}\ dt
	+ \int^{x^0}_{t_0^{+}} \bar{K}_{x^0 t} \cdot Q_{t t_0}\ dt  \cdot \left(E^{T} \cdot \bar{K}_{y^0 t_0}- \bar{K}'_{y^0 t_0}\right), \label{Ghat_b_int} \\
	\hat{G}^{go,\text{int}}_{x^0 y^0} &:=& \int_{t_0^{+}}^{x^0} \bar{K}_{x^0 t} \cdot Q_{t y^0}\ dt 
	+ \left(\bar{K}_{x^0 t_0}\cdot E - \bar{K}'_{x^0 t_0} \right) \cdot \int^{y^0}_{t_0^{+}} R_{t_{0} t} \cdot \bar{K}_{y^0 t}\ dt , \label{Ghat_a_int}
\end{eqnarray}
where $Q$ and $R$ are written in terms of the same $\hat{G}$ and $\hat{\varphi}$ in Eqs.\eqref{defQab} and \eqref{defRab}.
Below, we will show the two expressions are the same to each other. Remind us that $Q$ and $R$ in Eqs.\eqref{Ghat_b_int} and \eqref{Ghat_a_int} are written in terms of $\hat{G}_{x^0 y^0,\text{int}}$ in Eq.\eqref{solG} through the following differential equation, 
\begin{eqnarray}
 Q_{x^0 y^0} & = & \left[ \frac{\stackrel{\rightarrow}{\partial^{2}}}{{\partial x^{0}}^{2}} + \Omega^2_{x^0}
 \right] \hat{G}^{\text{int}}_{x^0 y^0} - E \cdot \hat{G}_{t_0 y^0}^{\text{int}} \delta_{t_0 x^0}, \label{diffGx02b} \\
 R_{x^0 y^0} & = & \hat{G}_{x^0 y^0}^{\text{int}}  \left[ \frac{\stackrel{\leftarrow}{\partial^{2}}}{{\partial y^{0}}^{2}} + \Omega^2_{y^0} \right] - \hat{G}_{x^0 t_0}^{\text{int}}
\cdot E^T \delta_{t_0 y^0}. \label{diffGy02b}
\end{eqnarray}
Substituting these expressions to Eqs.\eqref{Ghat_b_int} and \eqref{Ghat_a_int}, we obtain,
\begin{eqnarray}
\hat{G}^{go,\text{int}}_{x^0 y^0} &=& \hat{G}_{x^0 y^0 }^{\text{int}} + \bar{K}_{x^0 t_0}\cdot \left(  E \cdot \hat{G}^{\text{int}}_{t_0 y^0}
-   \lim_{t\rightarrow t_0} \frac{\partial \hat{G}^{\text{int}}_{t y^0}}{\partial t}   \right) 
 \nonumber \\
& &  + \left(\bar{K}_{x^0 t_0}\cdot E - \bar{K}'_{x^0 t_0} \right)\cdot \hat{G}^{\text{int}}_{t_0 t_0}\cdot \bar{K}'_{y^0 t_0}  \nonumber \\ 
& & +  \left(\bar{K}'_{x^0 t_0} - \bar{K}_{x^0 t_0}\cdot E\right)\cdot \lim_{t\rightarrow t_0} \frac{\partial \hat{G}^{\text{int}}_{t_0 t}}{\partial t}  \cdot \bar{K}_{y^0 t_0}  , \label{Ghat_a_int_n1} \\
&=& \hat{G}^{\text{int}}_{x^0 y^0}, \label{Ghat_a_int_n2}
\end{eqnarray}
\begin{eqnarray}
\hat{G}^{br,\text{int}}_{x^0 y^0} &=& \hat{G}^{\text{int}}_{x^0 y^0} +  \left( \hat{G}^{\text{int}}_{x^0 t_0} \cdot E^{T}  -  \lim_{t\rightarrow t_0} \frac{\partial \hat{G}^{\text{int}}_{x^0 t}}{\partial t} \right)   \cdot \bar{K}_{y^0 t_0} \nonumber \\
& &   +  \bar{K}'_{x^0 t_0} \cdot \hat{G}^{\text{int}}_{t_0 t_0} \cdot \left(E^{T} \cdot \bar{K}_{y^0 t_0}- \bar{K}'_{y^0 t_0}\right)   \nonumber \\
& &+ \bar{K}_{x^0 t_0} \cdot   \lim_{t\rightarrow t_0} \frac{\partial \hat{G}^{\text{int}}_{t t_0}}{\partial t}  \cdot \left(\bar{K}'_{y^0 t_0} - E^{T}\cdot \bar{K}_{y^0 t_0}\right), \label{Ghat_b_int_n1} \\
&= & \hat{G}^{\text{int}}_{x^0 y^0}, \label{Ghat_b_int_n2}
\end{eqnarray}
respectively. To show the  equalities of Eqs.\eqref{Ghat_a_int_n2} and \eqref{Ghat_b_int_n2}, we have used the following relations,
\begin{eqnarray}
 \hat{G}^{\text{int}}_{t_0 t_0} &=& 0, \\
 \lim_{t\rightarrow t_0} \frac{\partial}{\partial t} \hat{G}^{\text{int}}_{t_0 t} &=& 0, \\
 \lim_{t\rightarrow t_0} \frac{\partial \hat{G}^{\text{int}}_{t t_0}}{\partial t}  &=& 0,\\
 E \cdot \hat{G}^{\text{int}}_{t_0 y^0}
 &=&   \lim_{t\rightarrow t_0} \frac{\partial \hat{G}^{\text{int}}_{t y^0}}{\partial t} ,\\
 \hat{G}^{\text{int}}_{x^0 t_0} \cdot E^{T}  &=&  \lim_{t\rightarrow t_0} \frac{\partial \hat{G}^{\text{int}}_{x^0 t}}{\partial t}.
\end{eqnarray}
We complete the proof of equality of two expressions given in Eqs.\eqref{Ghat_b_int} and \eqref{Ghat_a_int}. Since they are identical each other, from now on, we will use $\hat{G}^{br,\text{int}}_{x^0 y^0}$ in Eq.\eqref{Ghat_b_int}. 

To summarize this subsection, let us write the SDEs of Green's functions in the form of integral equations. Omitting the upper and lower indices $i, j$, $a$ and $ b$, the interaction part of Green's function is written as, 
\begin{eqnarray}
\hat{G}^{\text{int}}_{ x^0 y^0} & = & \int_{t_0^+}^{y^0} R_{ x^0 t} \cdot
\bar{K}_{y^0 t}\ dt  - \int_{t_0^+}^{x^0} \bar{K}_{x^0 t} \cdot \left[Q_{t  t_0} \cdot \bar{K}_{y^0 t_0}^{\prime}  - Q_{ t t_0} \cdot E^T \cdot
\bar{K}_{y^0 0}\right]\ dt, \label{eq:266a}
\end{eqnarray}
where $Q$ and $R$ are given in Eqs.\eqref{defQab} and \eqref{defRab}.
Using the above equations and keeping the solutions up to the first order of the cubic interaction, one can obtain Eq.\eqref{GhatOA}.

\section{Derivation for $f (x^0)$ and $g (x^0)$ up to first order of $H(t_0)$}
\label{f_and_g}

$f(x^0)$ and $g(x^0)$ are the solutions of a homogeneous differential equation given in Eqs.\eqref{fx0} and \eqref{gx0}. In this appendix, we derive those solutions for the case that the scale factor is given in Eq.\eqref{SF}. We present them within linear approximation with respect to $H(t_0)$. 


One first considers a couple of general homogeneous differential equations given 
in Eqs.\eqref{fx0} and \eqref{gx0}. As for $\Omega_{i, {\bf k}} (x^0)$, we substitute the expression given in Eq.\eqref{OmeB2}.
The solutions are expanded up to the first order with respect to $H(t_0)$,
\begin{eqnarray}
f_{i,{\bf k}} (x^0) & = & f_{i,{\bf k}}^{(0)} (x^0) + f_{i,{\bf k}}^{(1)} (x^0) , \label{solfx0}\\
g_{i,{\bf k}} (x^0) & = & g_{i,{\bf k}}^{(0)} (x^0) + g_{i,{\bf k}}^{(1)} (x^0)  ,\label{solgx0}
\end{eqnarray}
where $f^{(0)} (x^0)$ and $g^{(0)} (x^0)$ are the solutions which correspond to the zeroth order of $H (t_0)$ while $f^{(1)} (x^0)$ and $g^{(1)} (x^0)$ are the solutions which correspond to the first order correction with respect to $H (t_0)$.

We first compute solution for $f (x^0)$. The differential equations of $f
(x^0)$ in Eq.\eqref{fx0} can be rewritten as,
\begin{eqnarray}
\left[ \frac{1}{\omega^2_{i, {\bf k}} } \frac{\partial^2}{{\partial x^{0}}^{2}} +
\frac{\Omega^2_{i, {\bf k}} (x^0)}{\omega^2_{i, {\bf k}} } \right] f_{i,{\bf k}}^{} (x^0) & = & 0. \label{Eq_Ome}
\end{eqnarray}
Using Eq.\eqref{OmeB2}, the second term in parentheses of above expression is rewritten as,
\begin{eqnarray}
\frac{\Omega^2_{i, {\bf k}} (x^0)}{\omega^2_{i, {\bf k}} } & = & 1 - 2 H (t_0) (x^0
- t_0) \frac{{\bf  k}^{2}}{[a (t_0) \omega_{i, {\bf k}} ]^2}.  \label{FO}
\end{eqnarray}
Then using Eq.(\ref{FO}), Eq.\eqref{Eq_Ome} alters into,
\begin{eqnarray}
\left[ \frac{1}{\omega^2_{i, {\bf k}} } \frac{\partial^2}{{\partial x^{0}}^{2}} +
1 - 2 H (t_0) (x^0 - t_0) \frac{{\bf  k}^2}{[a (t_0) \omega_{i, {\bf k}}
	]^2} \right] f_{i,{\bf k}} (x^0) & = & 0.  \label{DF2}
\end{eqnarray}
One can define several dimensionless parameters as,
\begin{eqnarray}
s & := & \omega_{i, {\bf k}}  x^0, \quad s_0  :=  \omega_{i, {\bf k}}  t_0, \quad h_0 :=  \frac{H (t_0)}{\omega_{i, {\bf k}}}, \quad {\bf  l}  :=  \frac{{\bf  k}}{a (t_0) \omega_{i, {\bf k}}} \label{dp1}
\end{eqnarray}
Then using the above dimensionless parameters, Eq.(\ref{DF2}) is rewritten
as (for simplicity, we omit the lower indices, $i$ and ${\bf k}$),
\begin{eqnarray}
	\left[ \frac{\partial^2}{\partial s^2} + 1 - 2 h_0 (s - s_0) {\bf  l}^2
	\right] f (s) & = & 0. \label{fdif}
\end{eqnarray}
The above equation leads to the following leading equations,
\begin{eqnarray}
\left[ \frac{\partial^2}{\partial s^2} + 1 \right] f^{(0)} (s) & = & 0, 
\label{f0}\\
\left[ \frac{\partial^2}{\partial s^2} + 1 \right] f^{(1)} (s) & = & 2 h_0
(s - s_0) {\bf  l}^2 f^{(0)} (s) .  \label{f1}
\end{eqnarray}
As for the solution of Eq.\eqref{f0}, we choose,
\begin{eqnarray}
f^{(0)} (s) & = & \sin (s) \label{sol_f0}. 
\end{eqnarray}
Next one needs to solve $f^{(1)} (s)$. The solution is written in terms of linear combination of sine and cosine functions,
\begin{eqnarray}
f^{(1)} (s) & = & C_1 (s) \sin (s) + C_2 (s) \cos (s) ,  
\end{eqnarray}
where their coefficients $C_i$ depend on time. Since we can impose the following condition, 
\begin{align}
C_1' (s) \sin (s) + C_2' (s) \cos (s) = 0, \label{B16}
\end{align}
one can show
that $C_i (s)$ satisfy,
\begin{eqnarray}
C_1' (s) \cos (s) - C_2' (s) \sin (s) & = & 2 h_0 (s - s_0) {\bf  l}^2
\sin (s) , \label{B17}
\end{eqnarray}
where $C_i' (s)$ are defined as,
\begin{eqnarray*}
	C_i' (s) & : = & \frac{d C_i (s)}{d s} .
\end{eqnarray*}
From Eqs.\eqref{B16} and \eqref{B17}, one can write $C_i' (s)$ as,
\begin{eqnarray*}
	C_1' (s) & = & h_0 (s - s_0) {\bf  l}^2 \sin (2 s),\\
	C_2' (s) & = & h_0 (s - s_0) {\bf  l}^2 \{ \cos (2 s) - 1 \} .
\end{eqnarray*}
With the initial conditions $C_i (s_0) = 0$, $C_i (s)$ are written  as,
\begin{eqnarray}
C_1 (s) & = & - \frac{h_0 {\bf  l}^2}{2} \left[ (s - s_0) \cos (2 s) -
\frac{1}{2} \{ \sin (2 s) - \sin (2 s_0) \} \right],  \label{c1s}\\
C_2 (s) & = & \frac{h_0 {\bf  l}^2}{2} \left[ (s - s_0) \{ \sin (2 s) -
2 s \} + \frac{1}{2} \{ \cos (2 s) - \cos (2 s_0) \} + (s^2 - s_0^2) \right]
.  \label{c2s}
\end{eqnarray}
Now using Eqs.(\ref{c1s}) and (\ref{c2s}), the solution of $f^{(1)} (s)$ is
written as,
\begin{eqnarray}
f^{(1)} (s) & = & \frac{h_0 {\bf  l}^2}{2} (s - s_0) \left\{ \sin (s) -
\frac{\sin (s - s_0)}{(s - s_0)} \sin (s_0) - (s - s_0) \cos (s) \right\} . \label{f1B}
\end{eqnarray}
To summarize this part, let us write  $f^{(0)} (s)$ and $f^{(1)} (s)$ in terms of original dimensional parameters. They are given by,
\begin{eqnarray}
f^{(0)} (x^0) & = & \sin[\omega_{i, {\bf  k}}  x^0] ,\\
f^{(1)} (x^0) & = &  \frac{H (t_0)\ {\bf  k}^2 (x^0 - t_0) }{2\{ a (t_0)
	\omega_{i, {\bf  k}}  \}^2} \left\{ \sin [\omega_{i, {\bf  k}} x^0]
- \frac{\sin [\omega_{i, {\bf  k}}  (x^0 - t_0)]}{\omega_{i, {\bf  k}}  (x^0 -
	t_0)} \sin [\omega_{i, {\bf  k}}  t_0] \right. \nonumber\\
&  & \left. -  \omega_{i, {\bf  k}} (x^0 - t_0) \cos [\omega_{i, {\bf  k}} x^0] \right\}  . 
\end{eqnarray}

Now we move to compute for $g (x^0)$. In this regard, $g(s)$ satisfies the same equation in Eq.\eqref{fdif} which $f(s)$ satisfies. The difference is that $g^{0}(s)$ is a cosine function, 
\begin{eqnarray}
g^{(0)} (s) & = & \cos (s) \label{sol_g0}. 
\end{eqnarray}
Applying the same procedure which we have used for the derivation of $f^{1}(s)$, we obtain,
\begin{eqnarray}
g^{(1)} (s) & = & \frac{h_0 {\bf  l}^2}{2} (s - s_0) \left[ \cos (s) -
\frac{\sin (s - s_0)}{(s - s_0)} \cos (s_0) + (s - s_0) \sin (s) \right] . \label{g1B}
\end{eqnarray}
Finally, one rewrites $g^{(0)} (s)$ and $g^{(1)} (s)$ in terms of original variables. They are given by, 
\begin{eqnarray}
g^{(0)} (x^0) & = & \cos [\omega_{i, {\bf  k}}  x^0], \\
g^{(1)} (x^0) & = & \frac{H (t_0)\ {\bf  k}^2  (x^0 - t_0)}{2 \{ a (t_0) \omega_{i, {\bf  k}} \}^2}
\left\{ \cos [\omega_{i, {\bf  k}} x^0] -
\frac{\sin [\omega_{i, {\bf  k}} (x^0 - t_0)]}{\omega_{i, {\bf  k}} (x^0 -
	t_0)} \cos [\omega_{i, {\bf  k}} t_0] \right.\nonumber \\
&  & \left. + \omega_{i, {\bf  k}} (x^0 - t_0) \sin [\omega_{i, {\bf  k}} x^0] \right\} .
\end{eqnarray}

\section{Derivation of $\bar{K}_i [x^0, y^0]$ up to first order of $H(t_0)$}
\label{sec_kbar}
In this appendix, we present $\bar{K}_{i, x^0 y^0, {\bf k}}$ given in Eq.\eqref{defKbar2} within the linear approximation with respect to $H(t_0)$.
For simplicity, momentum index ${\bf k}$ is suppressed. $\bar{K}_{i, x^0 y^0, {\bf k}}$ is also expanded up to the first order with respect to $H (t_0)$, namely,
\begin{eqnarray}
\bar{K}_i [x^0, y^0] & = & \bar{K}^{(0)}_i [x^0, y^0] + \bar{K}^{(1)}_i
[x^0, y^0],  \label{KBC2}
\end{eqnarray}
where we have defined,
\begin{eqnarray}
\bar{K}^{(0)}_i [x^0, y^0] & : = & \frac{1}{W_{i}} \{ f^{(0)}_{i} (x^0) g^{(0)}_{i}
(y^0) - f^{(0)}_{i} (y^0) g^{(0)}_{i} (x^0) \},  \label{KB0}\\
\bar{K}^{(1)}_i [x^0, y^0] & : = & \frac{1}{W_{i}} \{ f^{(0)}_{i} (x^0) g^{(1)}_{i}
(y^0) - f^{(0)}_{i} (y^0) g^{(1)}_{i} (x^0) + f^{(1)}_{i} (x^0) g^{(0)}_{i} (y^0) 
\nonumber\\
&  &  - f^{(1)}_{i} (y^0) g^{(0)}_{i} (x^0) \} .  \label{KB1}
\end{eqnarray}
One can show that $W$ is written in terms of the zeroth order solutions $f^{(0)}_{i} $, $g^{(0)}_{i} $ and their derivatives.
\begin{eqnarray*}
	W_{i} & = & \dot{f}^{(0)}_{i} (t_0) g^{(0)}_{i} (t_0) - f^{(0)}_{i} (t_0) \dot{g}^{(0)}_{i} (t_0) =\omega_{i, {\bf  k}}, 
\end{eqnarray*}
because $f^{1}(t_0)$, $\dot{f}^{(1)} (t_0)$,  $g^{1}(t_0)$and $\dot{g}^{(1)} (t_0)$ vanish.
Substituting the zeroth order function and the first order function of $f$ and $g$ in Eqs.\eqref{sol_f0}, \eqref{f1B}, \eqref{sol_g0} and \eqref{g1B}, Eqs.\eqref{KB0} and \eqref{KB1} alters into,
\begin{eqnarray}
	\bar{K}^{(0)}_i [x^0, y^0] & = & \frac{\sin (s - u)}{\omega_{i, {\bf  k}} } \nonumber  \\
	\bar{K}^{(1)}_i [x^0, y^0] & = & \frac{h_0 {\bf l}^2}{2 \omega_{i, {\bf  k}}
	} \{ u + s - 2 s_0 \} [\sin (s - u) + (u - s) \cos (s - u)] \nonumber 
\end{eqnarray}
where we have defined the following dimensionless parameters as,
\begin{eqnarray*}
	u & : = & \omega_{i, {\bf  k}}  y^0,\\
	s_0 & : = & u_0 = \omega_{i, {\bf  k}}  t_0.
\end{eqnarray*}
and $s,s_0,h_0$ and ${\bf  l}$ are defined in Eq.\eqref{dp1}, respectively. 
Therefore, $\bar{K}_i [x^0, y^0]$ in Eq.\eqref{KBC2} is written in terms of original parameters as,
\begin{eqnarray}
\bar{K}_i [x^0, y^0]  &=& \bar{K}^{(0)}_i [x^0, y^0] + \bar{K}^{(1)}_i[x^0, y^0]  \nonumber \\
\bar{K}^{(0)}_i [x^0, y^0] &= & \frac{\sin [\omega_{i, {\bf  k}} (x^0 - y^0)]}{\omega_{i, {\bf  k}}}, \label{k0} \\
\bar{K}^{(1)}_i[x^0, y^0] &=& \frac{H (t_0)}{2} \frac{{\bf  k}^2}{\omega_{i, {\bf  k}}^2 a (t_0)^2}
(x^0 + y^0 - 2 t_0) \nonumber \\
& & \times \left( \frac{\sin [\omega_{i, {\bf  k}} (x^0 -
	y^0)]}{\omega_{i, {\bf  k}}} - (x^0 - y^0) \cos [\omega_{i, {\bf  k}}
(x^0 - y^0)] \right). \nonumber \\ \label{k1} 
\end{eqnarray}

We define $\dot{\bar{K}}_i $ and $\dot{\bar{K}}'_i$ as,
\begin{eqnarray}
\dot{\bar{K}}_i [x^0, y^0]&:=&\frac{\partial \bar{K}_{i}[x^0,y^0]}{\partial x^0}, \\
\dot{\bar{K}}'_i [x^0, y^0]&:=&\frac{\partial^2 \bar{K}_{i}[x^0,y^0]}{\partial x^0 \partial y^0}.
\end{eqnarray}
Then $\bar{K}'_i $ in Eq.\eqref{kprime_bar}, $\dot{\bar{K}}_i $ and $\dot{\bar{K}}'_i$ are given in terms of original parameters by,
\begin{eqnarray}
\bar{K}'_i [x^0, y^0] & = & \bar{K}^{(0)\prime}_i [x^0, y^0] +
\bar{K}^{(1)\prime}_i [x^0, y^0] ,\nonumber \\
\bar{K}^{(0)\prime}_i [x^0, y^0] &=&  - \cos (s - u)
 =  - \cos [\omega_{i, {\bf  k}} (x^0 - y^0)] , \label{kpri0} \\
\bar{K}^{(1)\prime}_i [x^0, y^0] & = & \frac{h_0 {\bf l}^2}{2} [(u - s) \cos (s - u) + \{ 1 + (u + s -
2 s_0) (u - s) \} \sin (s - u)], \nonumber \\
& =& H (t_0) \frac{{\bf  k}^2}{2
	\omega_{i, {\bf  k}}^3 a (t_0)^2} [\omega_{i, {\bf  k}} (y^0 - x^0)
\cos [\omega_{i, {\bf  k}} (x^0 - y^0)]\nonumber \\
&  &  + \{ 1 + \omega_{i, {\bf  k}}^2 (y^0 + x^0 - 2 t_0)
(y^0 - x^0) \} \sin [\omega_{i, {\bf  k}} (x^0 - y^0)]], \label{kpri1}
\end{eqnarray}
\begin{eqnarray}
\dot{\bar{K}}_i [x^0, y^0] & = & \dot{\bar{K}}^{(0)}_i [x^0, y^0] +
\dot{\bar{K}}^{(1)}_i [x^0, y^0] ,\nonumber \\
\dot{\bar{K}}^{(0)}_i [x^0, y^0] & = & \cos (s - u) = \cos [\omega_{i, {\bf  k}} (x^0 - y^0)], \label{kdot0} \\
\dot{\bar{K}}^{(1)}_i [x^0, y^0] & = & \frac{h_0 {\bf l}^2}{2} [(u - s) \cos (s - u) + \{ 1 - (u + s -
2 s_0) (u - s) \} \sin (s - u)] \nonumber \\
& =& H (t_0) \frac{{\bf  k}^2}{2
	\omega_{i, {\bf  k}}^3 a (t_0)^2} [\omega_{i, {\bf  k}} (y^0 - x^0)
\cos [\omega_{i, {\bf  k}} (x^0 - y^0)]\nonumber \\
&  &  + \{ 1 - \omega_{i, {\bf  k}}^2 (y^0 + x^0 - 2 t_0)
(y^0 - x^0) \} \sin [\omega_{i, {\bf  k}} (x^0 - y^0)]] , \label{kdot1}
\end{eqnarray}
\begin{eqnarray}
\dot{\bar{K}}'_i [x^0, y^0] & = & \dot{\bar{K}}^{(0)\prime}_i [x^0, y^0] +
\dot{\bar{K}}^{(1)\prime}_i [x^0, y^0] ,\nonumber \\
\dot{\bar{K}}^{(0)\prime}_i [x^0, y^0] & = &   \omega_{i, {\bf  k}}  \sin (s - u) = \omega_{i, {\bf  k}} \sin [\omega_{i, {\bf  k}} (x^0 - y^0)], \label{kdotpri0} \\
\dot{\bar{K}}^{(1)\prime}_i [x^0, y^0] & = & \frac{h_0 {\bf l}^2}{2} \omega_{i, {\bf  k}}  (u + s - 2 s_0) [(u
- s) \cos (s - u) - \sin (s - u)]\nonumber \\
&=& H (t_0) \frac{{\bf  k}^2}{2 \omega_{i, {\bf  k}} a (t_0)^2} (y^0 +
x^0 - 2 t_0) \nonumber \\
& &\times \{ \omega_{i, {\bf  k}} (y^0 - x^0) \cos [\omega_{i,
	{\bf  k}} (x^0 - y^0)] - \sin [\omega_{i, {\bf  k}} (x^0 - y^0)] \}. \nonumber \\   \label{kdotpri1}
\end{eqnarray}

\section{Calculation of the expectation value of PNA}
\label{sec_calJ}

\subsection{Time integration and momentum integration of $\langle j_0 (x^0) \rangle_{1 \text{st}}$}

In this section, we provide both time and momentum integrations in the expression of the expectation value of the PNA. Let us first consider Eq.\eqref{J01st}. It can be rewritten as,
\begin{align}
\langle j_0 (x^0) \rangle_{1\text{st}} = \langle j_0 (x^0) \rangle_{1\text{st}, A} + \langle j_0 (x^0) \rangle_{1\text{st}, B}
, \label{J01st_T}
\end{align}
where $\langle j_0 (x^0) \rangle_{1\text{st}, A}$ and $\langle j_0 (x^0) \rangle_{1\text{st}, B}$ are given by,
\begin{eqnarray}
\langle j_0 (x^0) \rangle_{1\text{st}, A} & = & \frac{1}{a_{t_0}^3} 
\hat{\varphi}_{3, t_0} A_{1 2 3} \left\{ 1 - 3 (x^0 - t_0) H (t_0)   \right\} \int \frac{d^3  {\bf k}}{(2 \pi)^3} 
\left[ \frac{1}{2} \left( \frac{1}{\omega_{2, {\bf k}}} +
\frac{1}{\omega_{1, {\bf k}}} \right)  \right. \nonumber\\
&  & \times \left( \coth \frac{\beta
	\omega_{2, {\bf k}}}{2} - \coth \frac{\beta \omega_{1,
		{\bf k}}}{2} \right) \int_{t_0}^{x^0} \cos \omega_{3, {\bf 0}} t \cos
(\omega_{1, {\bf k}} - \omega_{2, {\bf k}}) (x^0 - t) d t
\nonumber\\
&  & + \frac{1}{2} \left( \frac{1}{\omega_{2, {\bf k}}} -
\frac{1}{\omega_{1, {\bf k}}} \right) \left( \coth \frac{\beta
	\omega_{2, {\bf k}}}{2} + \coth \frac{\beta \omega_{1,
		{\bf k}}}{2} \right) \nonumber\\
&  &\left. \times \int_{t_0}^{x^0} \cos \omega_{3, {\bf 0}} t \cos
(\omega_{1, {\bf k}} + \omega_{2, {\bf k}}) (x^0 - t) d t \right] ,
\label{J01stA} \\
\langle j_0 (x^0) \rangle_{1 \text{st}, B} & = & - \frac{3}{2} \frac{H
	(t_0)}{a_{t_0}^3}  \hat{\varphi}_{3, t_0} A_{1 2 3} \int \frac{d^3 
	{\bf k}}{(2 \pi)^3} \left[ \frac{1}{2} \left( \frac{1}{\omega_{2,
		{\bf k}}} + \frac{1}{\omega_{1, {\bf k}}} \right) \left( \coth
\frac{\beta \omega_{2, {\bf k}}}{2} - \coth \frac{\beta \omega_{1,
		{\bf k}}}{2} \right) \right. \nonumber\\
&  & \times \int_{t_0}^{x^0} (t - t_0) \cos \omega_{3,
	{\bf 0}} (t - t_0) \cos (\omega_{1, {\bf k}} - \omega_{2,
	{\bf k}}) (x^0 - t) d t \nonumber\\
&  & + \frac{1}{2} \left( \frac{1}{\omega_{2, {\bf k}}} -
\frac{1}{\omega_{1, {\bf k}}} \right) \left( \coth \frac{\beta
	\omega_{2, {\bf k}}}{2} + \coth \frac{\beta \omega_{1,
		{\bf k}}}{2} \right) \nonumber\\
&  &\left. \times \int_{t_0}^{x^0} (t - t_0) \cos \omega_{3,
	{\bf 0}} (t - t_0) \cos (\omega_{1, {\bf k}} + \omega_{2,
	{\bf k}}) (x^0 - t) d t \right].  \label{J01stB}
\end{eqnarray}
$\langle j_0 (x^0) \rangle_{1\text{st}, A}$ is the part which includes the PNA with constant scale factor and dilution effect while $\langle j_0 (x^0) \rangle_{1\text{st}, B}$ is the part which includes the freezing interaction effect. One can derive Eqs.\eqref{J01stA} and \eqref{J01stB} by substituting Eqs.\eqref{k0}, \eqref{kpri0}, \eqref{kdot0} and \eqref{kdotpri0} into Eq.\eqref{J01st}. 

Below we first carry out time integration of $\langle j_0 (x^0) \rangle_{1\text{st}, A}$. One defines the new variable of integration as, 
\begin{eqnarray}
s &:=& x^0 - t \label{new_s}, \\
x^1 &:= & x^0 - t_0. \label{new_x1}
\end{eqnarray}
Then one obtains that,
\begin{eqnarray}
\langle j_0 (x^1+t_0) \rangle_{1\text{st}, A} & = & \frac{\omega_{3, {\bf 0}}}{2 a_{t_0}^3}
\hat{\varphi}_{3, t_0} A_{1 2 3} (\tilde{m}_1^2 - \tilde{m}_2^2) \left\{ 1 - 3 x^1 H (t_0)   \right\} \int \frac{d^3 
	{\bf k}}{(2 \pi)^3} \frac{1}{\omega_{1, {\bf k}} \omega_{2,
		{\bf k}}} \nonumber\\
&  & \times \left[ \left( \frac{2 e^{- \beta \omega_{2 ,{\bf k}}}}{1 - e^{-
		\beta \omega_{2 ,{\bf k}}}} - \frac{2 e^{- \beta \omega_{1, {\bf k}}}}{1 - e^{- \beta
		\omega_{1, {\bf k}}}} \right) \frac{\frac{\sin [\omega_{3, {\bf 0}} x^1]}{(\omega_{1, {\bf k}} - \omega_{2, {\bf k}})} -^{}
	\frac{\sin [(\omega_{1, {\bf k}} - \omega_{2, {\bf k}}) x^1]}{\omega_{3, {\bf 0}}}}{\omega_{3, {\bf 0}}^2 - (\omega_{1,
		{\bf k}} - \omega_{2, {\bf k}})^2} \right. \nonumber\\
&  & \left. + \left( \frac{1 + e^{- \beta \omega_{2, {\bf k}}}}{1 - e^{-
		\beta \omega_{2, {\bf k}}}} + \frac{1 + e^{- \beta \omega_{1,
			{\bf k}}}}{1 - e^{- \beta \omega_{1, {\bf k}}}} \right)
\frac{\frac{\sin [\omega_{3, {\bf 0}} x^1]}{\omega_{1, {\bf k}} + \omega_{2, {\bf k}}} - \frac{\sin
		[(\omega_{1, {\bf k}} + \omega_{2, {\bf k}}) x^1]}{\omega_{3, {\bf 0}} }}{\omega_{3, {\bf 0}}^2 - (\omega_{1,
		{\bf k}} + \omega_{2, {\bf k}})^2}  \right].  \label{J01stAR}
\end{eqnarray}
By taking first time derivative of the above expression, setting $x^{1}=0$, using the initial expectation value of $\hat{\varphi}_{3, t_0} =v_3$ and setting $H(t_0)=0$, we obtain Eq.\eqref{tder}.

We move now to compute $\langle j_0 (x^0) \rangle_{1\text{st}, B}$. To perform time integration for $\langle j_0 (x^0) \rangle_{1\text{st}, B}$, one introduces the new variable of integration, 
\begin{eqnarray}
s : =  t - t_0, \label{new2_s}
\end{eqnarray}
and obtains,
\begin{eqnarray}
\langle j_0 (x^1+t_0) \rangle_{1\text{st}, B} & = & -  \frac{3 H
	(t_0) \omega_{3,
		{\bf 0}}}{4 a_{t_0}^3}  \hat{\varphi}_{3, t_0} A_{1 2 3} (\tilde{m}_1^2 - \tilde{m}_2^2) \int \frac{d^3 
	{\bf k}}{(2 \pi)^3} \frac{1}{\omega_{1, {\bf k}} \omega_{2,
		{\bf k}}} \left[ \left( \frac{\coth \frac{\beta \omega_{2,
			{\bf k}}}{2} - \coth \frac{\beta \omega_{1,
			{\bf k}}}{2}}{\omega^-_{12, {\bf k}}} \right) \right. \nonumber\\
&  & \times \left( \frac{x^1 \sin \omega_{3, {\bf 0}} x^1}{(\omega_{3, {\bf 0}}^2 - {\omega^-_{12, {\bf k}}}^2)} +
\frac{\frac{\omega_{3, {\bf 0}}^2 + {\omega^-_{12,
				{\bf k}}}^2}{\omega_{3, {\bf 0}}} \{ \cos \omega_{3, {\bf 0}}
	x^1 - \cos [\omega^-_{12, {\bf k}} x^1] \}}{(\omega_{3,
		{\bf 0}}^2 - {\omega^-_{12, {\bf k}}}^2)^2} \right) \nonumber\\
&  & + \left( \frac{\coth \frac{\beta \omega_{2, {\bf k}}}{2} + \coth
	\frac{\beta \omega_{1, {\bf k}}}{2}}{\omega^+_{12, {\bf k}}}
\right) \left( \frac{x^1 \sin \omega_{3, {\bf 0}} x^1}{(\omega_{3, {\bf 0}}^2 - {\omega^+_{12, {\bf k}}}^2)} \right.
\nonumber\\
&  &\left. \left. + \frac{\frac{\omega_{3, {\bf 0}}^2 + {\omega^+_{12,
				{\bf k}}}^2}{\omega_{3, {\bf 0}} } \{ \cos \omega_{3, {\bf 0}}
	x^1 - \cos [\omega^+_{12, {\bf k}} x^1] \}}{(\omega_{3,
		{\bf 0}}^2 - {\omega^+_{12, {\bf k}}}^2)^2} \right) \right], \label{J01stBR}
\end{eqnarray}
where we have defined $\omega^{\pm}_{12, {\bf k}}$ as,
\begin{eqnarray}
	\omega^{\pm}_{12, {\bf k}} & := & \omega_{1, {\bf k}} \pm
	\omega_{2, {\bf k}} .
\end{eqnarray}


The next task is to integrate Eqs.\eqref{J01stAR} and \eqref{J01stBR} with respect to spatial momentum. Using those equations, Eq.\eqref{J01st_T} leads to the following expression,
\begin{eqnarray}
\langle j_0 (x^1+t_0) \rangle_{1\text{st}} & = & - \frac{\hat{\varphi}_{3, t_0}
	A_{1 2 3}}{a_{t_0}^3} \frac{(\tilde{m}_1^2 - \tilde{m}_2^2)}{\omega_{3,
		{\bf 0}}}   \left[ \{ 1 - 3 x^1 H (t_0) \} \left\lbrace  J_1 (x^1, \tilde{m}_1, \tilde{m}_2, \omega^{}_{3,
	{\bf 0}}) \right. \right. \nonumber \\ & & \left. \left. + J_2 (x^1, \tilde{m}_1, \tilde{m}_2, \omega^{}_{3,
	{\bf 0}}) \right\rbrace + \frac{3}{4} H (t_0)   \left\lbrace J_3 (x^1, \tilde{m}_1, \tilde{m}_{2 }, \omega^{}_{3,
	{\bf 0}}) +  J_4 (x^1, \tilde{m}_1, \tilde{m}_{2 }, \omega^{}_{3, {\bf 0}})\right\rbrace \right], \nonumber \\
\end{eqnarray}
where auxiliary functions $J_i (x^1, \tilde{m}_1, \tilde{m}_{2 },\omega_{3, {\bf 0}})$ ($i=1,\ldots ,4$) are defined as,
\begin{eqnarray}
	J_1 (x^1, \tilde{m}_1, \tilde{m}_2, \omega_{3, {\bf 0}}) &
	:= & - \frac{\sin [\omega_{3, {\bf 0}} x^1]}{2} \int \frac{d^3  {\bf k}}{(2 \pi)^3}
	\frac{1}{\omega_{2, {\bf k}} \omega_{1, {\bf k}}} \frac{\left(
		\frac{1 + e^{- \beta \omega_{2, {\bf k}}}}{1 - e^{- \beta \omega_{2,
					{\bf k}}}} + \frac{1 + e^{- \beta \omega_{1, {\bf k}}}}{1 - e^{-
				\beta \omega_{1, {\bf k}}}} \right)}{1 - \frac{(\omega^{+}_{12, {\bf k}})^2}{\omega_{3, {\bf 0}}^2}} 
	\frac{1}{\omega^{+}_{12, {\bf k}}} \nonumber  \\ & &
	+ \frac{1}{2 \omega_{3, {\bf 0}}} \int \frac{d^3  {\bf k}}{(2 \pi)^3}
	\frac{1}{\omega_{2, {\bf k}} \omega_{1, {\bf k}}}  \frac{\left(
		\frac{1 + e^{- \beta \omega_{2, {\bf k}}}}{1 - e^{- \beta \omega_{2,
					{\bf k}}}} + \frac{1 + e^{- \beta \omega_{1, {\bf k}}}}{1 - e^{-
				\beta \omega_{1, {\bf k}}}} \right) \sin [\omega^{+}_{12, {\bf k}} x^1]}{1 - \frac{(\omega^{+}_{12, {\bf k}})^2}{\omega_{3, {\bf 0}}^2}}, \nonumber \\ 
\end{eqnarray}
\begin{eqnarray}
	J_2 (x^1, \tilde{m}_1, \tilde{m}_2, \omega_{3, {\bf 0}}) &
	:= & - \frac{\sin [\omega_{3, {\bf 0}} x^1]}{2 (\tilde{m}_1^2 - \tilde{m}_2^2)}   \int \frac{d^3 
		{\bf k}}{(2 \pi)^3} \frac{\omega^{+}_{12, {\bf k}}}{\omega_{1, {\bf k}} \omega_{2, {\bf k}}}
	\frac{\left( \frac{1 + e^{- \beta \omega_{2, {\bf k}}}}{1 - e^{- \beta
				\omega_{2, {\bf k}}}} - \frac{1 + e^{- \beta \omega_{1,
					{\bf k}}}}{1 - e^{- \beta \omega_{1, {\bf k}}}} \right)}{1 -
		\frac{(\omega^{-}_{12, {\bf k}})^2}{\omega_{3,
				{\bf 0}}^2}}\nonumber \\ & & + \frac{1}{2 \omega_{3, {\bf 0}}} \int \frac{d^3  {\bf k}}{(2 \pi)^3}
	\frac{1}{\omega_{2, {\bf k}} \omega_{1, {\bf k}}}  \frac{\left(
		\frac{1 + e^{- \beta \omega_{2, {\bf k}}}}{1 - e^{- \beta \omega_{2,
					{\bf k}}}} - \frac{1 + e^{- \beta \omega_{1, {\bf k}}}}{1 - e^{-
				\beta \omega_{1, {\bf k}}}} \right) \sin [\omega^{-}_{12, {\bf k}} x^1]}{1 - \frac{(\omega^{-}_{12, {\bf k}})^2}{\omega_{3, {\bf 0}}^2}}, \nonumber \\
\end{eqnarray}
\begin{eqnarray}
	J_3 (x^1, \tilde{m}_1, \tilde{m}_{2 } ,
	\omega_{3, {\bf 0}}) & : = & \int \frac{d^3  {\bf k}}{(2 \pi)^3} 
	\frac{1}{\omega_{1, {\bf k}} \omega_{2, {\bf k}}}
	\frac{{\omega^+_{12, {\bf k}}}^2 \left( \coth \frac{\beta \omega_{2,
				{\bf k}}}{2} - \coth \frac{\beta \omega_{1, {\bf k}}}{2}
		\right)}{{\omega^+_{12, {\bf k}}}^2 - \frac{(\tilde{m}_1^2 -
			\tilde{m}_2^2)^2}{\omega_{3, {\bf 0}}^2}} \left[ \frac{x^1 \sin [\omega_{3, {\bf 0}} x^1]}{\omega^-_{12, {\bf k}}} \right. \nonumber \\ & & \left.  + \frac{\left( {\omega^+_{12, {\bf k}}}^2 +
		\frac{(\tilde{m}_1^2 - \tilde{m}_2^2)^2}{\omega_{3, {\bf 0}}^2} \right)
		\left( \frac{\cos [\omega_{3, {\bf 0}} x^1]}{\omega^-_{12,
				{\bf k}}} - \frac{\cos \left[ \frac{(\tilde{m}_1^2 - \tilde{m}_2^2)
				x^1}{\omega^+_{12, {\bf k}}} \right]}{\omega^-_{12,
				{\bf k}}} \right)}{\omega_{3, {\bf 0}} \left( {\omega^+_{12,
				{\bf k}}}^2 - \frac{(\tilde{m}_1^2 - \tilde{m}_2^2)^2}{\omega_{3,
				{\bf 0}}^2} \right)} \right], 
\end{eqnarray}
\begin{eqnarray}
	J_4 (x^1, \tilde{m}_1, \tilde{m}_{2 } ,\omega_{3, {\bf 0}}) & := & \int \frac{d^3  {\bf k}}{(2
		\pi)^3} \frac{1}{\omega_{1, {\bf k}} \omega_{2, {\bf k}}}
	\frac{\left( \frac{1 + e^{- \beta \omega_{2, {\bf k}}}}{1 - e^{- \beta
				\omega_{2, {\bf k}}}} + \frac{1 + e^{- \beta \omega_{1,
					{\bf k}}}}{1 - e^{- \beta \omega_{1, {\bf k}}}} \right)}{1 -
		\frac{{\omega^+_{12, {\bf k}}}^2}{\omega_{3, {\bf 0}}^2}} \left[
	\frac{x^1 \sin [\omega_{3, {\bf 0}} x^1]}{\omega^+_{12,
			{\bf k}}}  \right. \nonumber\\
	&  & \left. + \left( 1 + \frac{{\omega^+_{12, {\bf k}}}^2}{\omega_{3,
			{\bf 0}}^2} \right) \frac{\{ \cos [\omega_{3, {\bf 0}} x^1] - \cos [\omega^+_{12, {\bf k}} x^1] \}}{\omega_{3,
			{\bf 0}} \omega^+_{12, {\bf k}} \left( 1 - \frac{{\omega^+_{12,
					{\bf k}}}^2}{\omega_{3, {\bf 0}}^2} \right)} \right].
\end{eqnarray}
We carry out the momentum integration of the above expressions numerically. 


\subsection{Time integration and momentum integration of $\langle j_0 (x^0) \rangle_{2 \text{nd}}$}
Below we consider time and momentum integrations of the second part of PNA which is Eq.\eqref{J02nd}.  Substituting Eqs.\eqref{k0}, \eqref{k1}, \eqref{kpri0}-\eqref{kdotpri1} into Eq.\eqref{J02nd} and performing time integration, we obtain,
\begin{eqnarray}
& &\langle j_0 (x^1+t_0) \rangle_{2 \text{nd}} \nonumber\\
& = & - \frac{2 \hat{\varphi}_{3, t_0} A_{1 2 3} }{a_{t_0}^3}   \int
\frac{d^3  {\bf k}}{(2 \pi)^3} \left[ \left\{ \frac{1}{2 \omega_2 }
\coth \frac{\beta \omega_2}{2} \right. \right. \frac{\sqrt{\Delta_1'
		\Delta_2'}}{2}  \left[ \frac{1}{\omega_2^2} \left\{ \frac{\omega_{12}^{- 2}
	\cos [\omega_{12}^+ x^1]}{\omega_3^2 - \omega_{12}^{- 2}} \right. \right.
\nonumber\\
&  & - \frac{\omega_{12}^{+ 2} \cos [\omega_{12}^- x^1]}{\omega_3^2 -
	\omega_{12}^{+ 2}} - \cos [\omega_3 x^1] \cos [2 \omega_2 x^1] \left(
\frac{\omega_{12}^{- 2} }{\omega_3^2 - \omega_{12}^{- 2}} -
\frac{\omega_{12}^{+ 2}}{\omega_3^2 - \omega_{12}^{+ 2}} \right) + \omega_3
\sin [\omega_3 x^1] \sin [2 \omega_2 x^1] \nonumber\\
&  & \left. \times \left( \frac{\omega_{12}^- }{\omega_3^2 - \omega_{12}^{-
		2}} + \frac{\omega_{12}^+}{\omega_3^2 - \omega_{12}^{+ 2}} \right) \right\}
+ \frac{(\tilde{m}_1^2 - \tilde{m}_2^2)}{\omega_1} \left\{ 4 \omega_3 x^1
\sin [\omega_3 x^1] \left( \frac{\omega_{12}^+}{(\omega_3^2 - \omega_{12}^{+
		2})^2} - \frac{\omega_{12}^- }{(\omega_3^2 - \omega_{12}^{- 2})^2} \right)
\right. \nonumber\\
&  & - \frac{2 \omega_{12}^+  \{ \cos [\omega_3 x^1] - \cos [\omega_{12}^+
	x^1] \}}{(\omega_3^2 - \omega_{12}^{+ 2})^2} 
+ \frac{2 \omega_{12}^-  \{ \cos [\omega_3 x^1] - \cos [\omega_{12}^-
	x^1] \}}{(\omega_3^2 - \omega_{12}^{- 2})^2} - \left( (x^1)^2 -
\frac{1}{\omega_1 \omega_2} \right) \frac{\omega_{12}^+ \cos [\omega_{12}^+
	x^1]}{\omega_3^2 - \omega_{12}^{+ 2}} \nonumber\\
&  & + 8 \omega_3^2  \left( \frac{\omega_{12}^+  \{ \cos [\omega_3 x^1] -
	\cos [\omega_{12}^+ x^1] \}}{(\omega_3^2 - \omega_{12}^{+ 2})^3} -
\frac{\omega_{12}^-  \{ \cos [\omega_3 x^1] - \cos [\omega_{12}^- x^1]
	\}}{(\omega_3^2 - \omega_{12}^{- 2})^3} \right) \nonumber\\
&  & \left. + \left( (x^1)^2 + \frac{1}{\omega_1 \omega_2} \right)
\frac{\omega_{12}^- \cos [\omega_{12}^- x^1]}{\omega_3^2 - \omega_{12}^{-
		2}} - \frac{\cos [\omega_3 x^1]}{\omega_1 \omega_2} \left(
\frac{\omega_{12}^+ }{\omega_3^2 - \omega_{12}^{+ 2}} +
\frac{\omega_{12}^-}{\omega_3^2 - \omega_{12}^{- 2}} \right) \right\}
\nonumber\\
&  & - \frac{\frac{x^1}{\omega_1} (\tilde{m}_1^2 - \tilde{m}_2^2)}{\omega_1
	\omega_2} \left\{ \left( \frac{\omega_{12}^+ {\color[HTML]{000000}}
}{\omega_3^2 - \omega_{12}^{+ 2}} + \frac{\omega_{12}^-
	{\color[HTML]{000000}} }{\omega_3^2 - \omega_{12}^{- 2}} \right) \omega_3
\sin [\omega_3 x^1] - \frac{\omega_{12}^{+ 2} \sin [\omega_{12}^+
	x^1]}{\omega_3^2 - \omega_{12}^{+ 2}} - \frac{\omega_{12}^{- 2} \sin
	[\omega_{12}^- x^1]}{\omega_3^2 - \omega_{12}^{- 2}} \right\} \nonumber\\
&  & - \frac{1}{\omega_1} \left( \frac{\omega_1}{\omega_2} +
\frac{\omega_2}{\omega_1} \right) \left\{ \frac{(\omega_3^2 + \omega_{12}^{+
		2}) \omega_{12}^- {\color[HTML]{000000}}}{(\omega_3^2 - \omega_{12}^{+
		2})^2} \{ \cos [\omega_3 x^1] - \cos [\omega_{12}^+ x^1] \} +
\frac{(\omega_3^2 + \omega_{12}^{- 2}) \omega_{12}^+ {\color[HTML]{000000}}
}{(\omega_3^2 - \omega_{12}^{- 2})^2} \right. \nonumber\\
&  & \left. \left. \left. \times \{ \cos [\omega_3 x^1] - \cos
[\omega_{12}^- x^1] \} + \left( \frac{\omega_{12}^- {\color[HTML]{000000}}
}{\omega_3^2 - \omega_{12}^{+ 2}} + \frac{\omega_{12}^+
	{\color[HTML]{000000}} }{\omega_3^2 - \omega_{12}^{- 2}} \right) \omega_3
x^1 \sin [\omega_3 x^1] \right\} \right] \right\} \nonumber\\
&  &  - \{ 1 \leftrightarrow 2\ \text{for lower indices} \}],  \label{RGresult}
\end{eqnarray}
where we have introduced $\Delta_{i, {\bf k}}'$ defined by,
\begin{eqnarray}
\Delta_{i, {\bf k}}' & = & \frac{H (t_0)}{2} 
\frac{{\bf k}^2}{\omega_{i, {\bf k}}^2 a (t_0)^2}.   \label{DD}
\end{eqnarray}
For notational simplicity, $\omega_{i,{\bf k}}$ is denoted by $\omega_{i}$ ($i=1,2$) and $\omega_{12,{\bf k}}^{\pm}$ is denoted by $\omega_{12}^{\pm}$ in Eq.\eqref{RGresult}. 


The next task is to integrate Eq.\eqref{RGresult} with respect to spacial momentum.
Eq.(\ref{RGresult}) leads to the following expression,
\begin{eqnarray}
 \langle j_0 (x^1 + t_0) \rangle_{2 \text{nd}} 
& = & \frac{\hat{\varphi}_{3, t_0} A_{1 2 3}}{a_{t_0}^3} \frac{H (t_0)}{4}
(\tilde{m}_1^2 - \tilde{m}_2^2) \left[  J_{11} (x^1, \tilde{m}_1, \tilde{m}_{2 } ,\omega_{3, {\bf 0}}) + J_{12} (x^1, \tilde{m}_1, \tilde{m}_{2 } ,\omega_{3, {\bf 0}}) \right.  \nonumber\\
&  & \left. +  J_{13} (x^1, \tilde{m}_1, \tilde{m}_{2 } ,\omega_{3, {\bf 0}})
+ J_{14} (x^1, \tilde{m}_1, \tilde{m}_{2 } ,\omega_{3, {\bf 0}}) + J_{15}
(x^1, \tilde{m}_1, \tilde{m}_{2 } ,\omega_{3, {\bf 0}}) \right] , \nonumber \\
\end{eqnarray}
where we have defined auxiliary functions $J_i (x^1, \tilde{m}_1, \tilde{m}_{2 } ,\omega_{3, {\bf 0}})$
$(i = 11, \ldots, 15)$ as,
\begin{eqnarray}
J_{11} (x^1, \tilde{m}_1, \tilde{m}_{2 } ,\omega_{3, {\bf 0}}) & := & \int
\frac{d^3  {\bf k}}{(2 \pi)^3} \frac{{\bf k}^2}{a (t_0)^2} 
\frac{x^1}{(\omega_1 \omega_2)^3} \left[ \left( \coth \frac{\beta
	\omega_2}{2} + \coth \frac{\beta \omega_1}{2} \right) \right. \nonumber\\
&  & \times \left( \frac{\omega_{12}^+ {\color[HTML]{000000}} \omega_3 \sin
	[\omega_3 x^1]}{\omega_3^2 - \omega_{12}^{+ 2}} - \frac{\omega_{12}^{+ 2}
	\sin [\omega_{12}^+ x^1]}{\omega_3^2 - \omega_{12}^{+ 2}} \right) + \left(
\coth \frac{\beta \omega_2}{2} - \coth \frac{\beta \omega_1}{2} \right)
\nonumber\\
&  & \times \left. \left( \frac{\omega_{12}^- {\color[HTML]{000000}}
	\omega_3 \sin [\omega_3 x^1]}{\omega_3^2 - \omega_{12}^{- 2}} -
\frac{\omega_{12}^{- 2} \sin [\omega_{12}^- x^1]}{\omega_3^2 -
	\omega_{12}^{- 2}} \right) \right],  \label{Jm}
\end{eqnarray}
\begin{eqnarray}
J_{12} (x^1, \tilde{m}_1, \tilde{m}_{2 } ,\omega_{3, {\bf 0}}) & := &
(\tilde{m}_1^2 - \tilde{m}_2^2)^{- 1} \int \frac{d^3  {\bf k}}{(2
	\pi)^3}  \frac{{\bf k}^2}{a (t_0)^2}  \frac{1}{\omega_1 \omega_2} 
\left[ \left( \coth \frac{\beta \omega_1}{2} \frac{1}{\omega_1^3} - \coth
\frac{\beta \omega_2}{2} \frac{1}{\omega_2^3} \right) \right. \nonumber\\
&  & \times \left( \frac{\omega_{12}^{- 2} \cos [\omega_{12}^+
	x^1]}{\omega_3^2 - \omega_{12}^{- 2}} - \frac{\omega_{12}^{+ 2} \cos
	[\omega_{12}^- x^1]}{\omega_3^2 - \omega_{12}^{+ 2}} \right) + \left( \coth
\frac{\beta \omega_2}{2} \frac{\cos [2 \omega_2 x^1]}{\omega_2^3} \right.
\nonumber\\
&  & \left. - \coth \frac{\beta \omega_1}{2} \frac{\cos [2 \omega_1
	x^1]}{\omega_1^3} \right) \left( \frac{\omega_{12}^{- 2} }{\omega_3^2 -
	\omega_{12}^{- 2}} - \frac{\omega_{12}^{+ 2}}{\omega_3^2 - \omega_{12}^{+
		2}} \right) \cos [\omega_3 x^1] \nonumber\\
&  & - \left( \coth \frac{\beta \omega_2}{2} \frac{\sin [2 \omega_2
	x^1]}{\omega_2^3} + \coth \frac{\beta \omega_1}{2} \frac{\sin [2 \omega_1
	x^1]}{\omega_1^3}  \right) \frac{\omega_{12}^- \omega_3 \sin [\omega_3
	x^1]}{\omega_3^2 - \omega_{12}^{- 2}} \nonumber\\
&  & \left. - \left( \coth \frac{\beta \omega_2}{2} \frac{\sin [2 \omega_2
	x^1]}{\omega_2^3} - \coth \frac{\beta \omega_1}{2} \frac{\sin [2 \omega_1
	x^1]}{\omega_1^3} \right) \frac{\omega_{12}^+ \omega_3 \sin [\omega_3
	x^1]}{\omega_3^2 - \omega_{12}^{+ 2}}  \right] ,\nonumber \\  \label{Jr}
\end{eqnarray}
\begin{eqnarray}
J_{13} (x^1, \tilde{m}_1, \tilde{m}_{2 } ,\omega_{3, {\bf 0}}) & := &
(\tilde{m}_1^2 - \tilde{m}_2^2)^{- 1} \int \frac{d^3  {\bf k}}{(2
	\pi)^3}  \frac{{\bf k}^2}{a (t_0)^2}  \frac{(\omega_1^2 +
	\omega_2^2)}{(\omega_1 \omega_2)^3} \left[  \left( \coth \frac{\beta
	\omega_2}{2} + \coth \frac{\beta \omega_1}{2} \right) \right. \nonumber\\
&  & \times \left( \frac{(\omega_3^2 + \omega_{12}^{+ 2}) \omega_{12}^-
	{\color[HTML]{000000}}  \{ \cos [\omega_3 x^1] - \cos [\omega_{12}^+ x^1]
	\}}{(\omega_3^2 - \omega_{12}^{+ 2})^2} + \frac{\omega_{12}^-
	{\color[HTML]{000000}} \omega_3 x^1 \sin [\omega_3 x^1]}{\omega_3^2 -
	\omega_{12}^{+ 2}} \right) \nonumber\\
&  & + 2 \left( \frac{e^{- \beta \omega_2}}{1 - e^{- \beta \omega_2}} -
\frac{e^{- \beta \omega_1}}{1 - e^{- \beta \omega_1}} \right) \left(
\frac{(\omega_3^2 + \omega_{12}^{- 2}) \omega_{12}^+ {\color[HTML]{000000}} 
	\{ \cos [\omega_3 x^1] - \cos [\omega_{12}^- x^1] \}}{(\omega_3^2 -
	\omega_{12}^{- 2})^2} \right. \nonumber\\
&  & \left. \left. + \frac{\omega_{12}^+ {\color[HTML]{000000}} \omega_3
	x^1 \sin [\omega_3 x^1]}{\omega_3^2 - \omega_{12}^{- 2}} \right) \right] ,
\label{Jg}
\end{eqnarray}
\begin{eqnarray}
J_{14} (x^1, \tilde{m}_1, \tilde{m}_{2 } ,\omega_{3, {\bf 0}}) & := & - \int
\frac{d^3  {\bf k}}{(2 \pi)^3} \frac{{\bf k}^2}{a (t_0)^2}
\frac{1}{(\omega_1 \omega_2)^2} \left( \coth \frac{\beta \omega_1}{2} +
\coth \frac{\beta \omega_2}{2} \right) \nonumber\\
&  & \times \left[ 4 \omega_3 x^1 \sin [\omega_3 x^1]
\frac{\omega_{12}^+}{(\omega_3^2 - \omega_{12}^{+ 2})^2} - \frac{2
	\omega_{12}^+  \{ \cos [\omega_3 x^1] - \cos [\omega_{12}^+ x^1]
	\}}{(\omega_3^2 - \omega_{12}^{+ 2})^2} \right. \nonumber\\
&  & + \frac{8 \omega_3^2 \omega_{12}^+  \{ \cos [\omega_3 x^1] - \cos
	[\omega_{12}^+ x^1] \}}{(\omega_3^2 - \omega_{12}^{+ 2})^3} - \left(
{\color[HTML]{000000}(x^1)^2} - \frac{1}{\omega_1 \omega_2} \right)
\frac{\omega_{12}^+ \cos [\omega_{12}^+ x^1]}{\omega_3^2 - \omega_{12}^{+
		2}} \nonumber\\
&  & \left. - \frac{\cos [\omega_3 x^1]}{\omega_1 \omega_2} 
\frac{\omega_{12}^+ }{\omega_3^2 - \omega_{12}^{+ 2}} \right] , \label{Jb1}
\end{eqnarray}
\begin{eqnarray}
J_{15} (x^1, \tilde{m}_1, \tilde{m}_{2 } ,\omega_{3, {\bf 0}}) & := & - 2
\int \frac{d^3  {\bf k}}{(2 \pi)^3} \frac{{\bf k}^2}{a (t_0)^2}
\frac{1}{(\omega_1 \omega_2)^2} \left( \frac{e^{- \beta \omega_1}}{1 - e^{-
		\beta \omega_1}} - \frac{e^{- \beta \omega_2}}{1 - e^{- \beta \omega_2}}
\right) \nonumber\\
&  & \times \left[ 4 \omega_3 x^1 \sin [\omega_3 x^1] \frac{\omega_{12}^-
}{(\omega_3^2 - \omega_{12}^{- 2})^2} - \frac{2 \omega_{12}^-  \{ \cos
	[\omega_3 x^1] - \cos [\omega_{12}^- x^1] \}}{(\omega_3^2 - \omega_{12}^{-
		2})^2} \right. \nonumber\\
&  & + \frac{8 \omega_3^2 \omega_{12}^-  \{ \cos [\omega_3 x^1] - \cos
	[\omega_{12}^- x^1] \}}{(\omega_3^2 - \omega_{12}^{- 2})^3} - \left(
{\color[HTML]{000000}(x^1)^2} + \frac{1}{\omega_1 \omega_2} \right)
\frac{\omega_{12}^- \cos [\omega_{12}^- x^1]}{\omega_3^2 - \omega_{12}^{-
		2}} \nonumber\\
&  & \left. + \frac{\cos [\omega_3 x^1]}{\omega_1 \omega_2} 
\frac{\omega_{12}^-}{\omega_3^2 - \omega_{12}^{- 2}} \right].  \label{Jb2}
\end{eqnarray}

We carry out the momentum integration of the above expressions numerically.

%

\end{document}